\documentstyle[aps,amsfonts,graphicx,pstricks,axodraw,eqsecnum]{revtex}

\newcommand{\elle}{\ell\hspace{-0.16cm}/}
\newcommand{\ka}{k\!\!\!/}
\newcommand{\erre}{r\!\!\!/}
\newcommand{\qu}{q\!\!\!/}
\newcommand{\pslush}{p\hspace{-0.16cm}/}
\newcommand{\D}{\displaystyle}

\begin{document}
\begin{titlepage}
\begin{flushright}
FTUV/011016 \\
IFIC/01-56 \\
\end{flushright}

\vspace{0.6cm}

\begin{center}
\Large{{\bf Gauge-Independent Off-Shell Fermion Self-Energies \\ 
at Two Loops: The Cases of QED and QCD}}\\
\end{center}

\vspace{0.5 cm}

\begin{center}
\large Daniele Binosi and Joannis Papavassiliou
\end{center}

\begin{center}
{\em
Departamento de F\'\i sica Te\'orica and IFIC, Centro Mixto, \\
Universidad de Valencia-CSIC,\\
E-46100, Burjassot, Valencia, Spain\\}
\end{center}

\vspace{0.1cm}

\begin{abstract}

We   use   the   pinch    technique   formalism   to   construct   the
gauge-independent  off-shell  two-loop fermion  self-energy,  both for
Abelian  (QED) and  non-Abelian  (QCD) gauge  theories.   The new  key
observation   is   that  all   contributions   originating  from   the
longitudinal  parts  of gauge  boson  propagators,  by  virtue of  the
elementary  tree-level  Ward identities  they  trigger,  give rise  to
effective vertices, which do not exist in the original Lagrangian; all
such vertices cancel diagrammatically inside physical quantities, such
as current  correlation functions or $S$-matrix  elements.  We present
two  different, but  complementary derivations:  First,  we explicitly
track down the  aforementioned cancellations inside two-loop diagrams,
resorting to nothing more than basic algebraic manipulations.  Second,
we present  an absorptive derivation, exploiting the  unitarity of the
$S$-matrix, and the Ward identities imposed on tree-level and one-loop
physical amplitudes by gauge invariance, in the case of QED, or by the
underlying  Becchi-Rouet-Stora  symmetry, in  the  case  of QCD.   The
propagator-like  sub-amplitude   defined  by  means   of  this  latter
construction  corresponds  precisely to  the  imaginary  parts of  the
effective self-energy obtained  in the former case; the  real part may
be obtained from a (twice  subtracted) dispersion relation.  As in the
one-loop  case,  the final  two-loop  fermion self-energy  constructed
using   either  method   coincides  with   the   conventional  fermion
self-energy computed in the Feynman gauge.

\bigskip 
\bigskip 


\end{abstract}

\vspace{1cm}

\begin{flushleft}
E-mail: Daniele.Binosi@uv.es; Joannis.Papavassiliou@uv.es
\end{flushleft}
\end{titlepage}

\section*{Introduction}

It is well known that off-shell Green's functions depend in general on
the  gauge-fixing  procedure  used  to  quantize the  theory,  and  in
particular on  the gauge-fixing parameter (GFP) chosen  within a given
scheme.  A  celebrated exception  to this general  fact is  the vacuum
polarization  of the  photon in  QED,  which is  both gauge  invariant
({\it i.e.} transverse)  and GFP-independent  to all orders  in perturbation
theory.   In  contrast,  the   fermion  self-energy  $\Sigma  (p)$  is
GFP-dependent already  at the one-loop  level.  The dependence  on the
GFP is  in general non-trivial and  affects the properties  of a given
Green's  function.   In the  framework  of  the  covariant gauges  for
example, depending on  the choice of the GFP  $\xi$, one may eliminate
the ultraviolet divergence of the one-loop electron propagator $\Sigma
(p,\xi)$  by  choosing  the  Landau  gauge $\xi=0$,  or  the  infrared
divergence  appearing after on-shell  renormalization by  choosing the
Yennie-Fried   gauge  $\xi=3$.   The   situation  becomes   even  more
complicated  in the  case  of non-Abelian  gauge  theories, where  all
Green's  functions  depend  on  the  GFP.   Of  course,  when  forming
observables  the gauge  dependences  of the  Green's functions  cancel
among  each  other order  by  order  in  perturbation theory,  due  to
powerful  field-theoretical  properties, a  fact  which reduces  their
seriousness.  However, these dependences  pose a major difficulty when
one  attempts  to   extract  physically  meaningful  information  from
individual  Green's functions.   This  is  the case  in  at least  two
important  situations, which  both lie  beyond the  confines  of fixed
order perturbation theory:  first, the Schwinger-Dyson (SD) equations,
which constitute one of the few methods for obtaining non-perturbative
information in the  continuum; second, resonant transition amplitudes,
where the taming of  the physical kinematic singularity necessitates a
resummation,  which amounts  to  a non-trivial  reorganization of  the
perturbative series.
  
In the context of  the two non-perturbative situations mentioned above
the fermion propagator is of particular interest.  In the former case,
SD  equation  involving fermion  propagators  (self-energies) have  an
extended  range of  applications,  most  of which  are  linked to  the
mechanism  of  dynamical mass  generation,  which  is explored  by
looking for non-trivial solutions to gap equations.  The study of such
equations has been  particularly popular in QED \cite{Johnson:1964da},
and  even  more  so  in  QCD \cite{Lane:1974he},  where  it  has  been
intimately  associated  with  the  mechanism that  breaks  the  chiral
symmetry.   Similar  equations  are  relevant in  QED$_3$,  where  the
infrared regime of the theory is probed for a non-trivial fixed point
\cite{Dorey:1990sz}, for
technicolor models \cite{Appelquist:1988ee}, 
gauged Nambu--Jona-Lasinio models \cite{Leung:1989hw},  
and more recently  color superconductivity
\cite{Hong:2000fh}.  A similar quest  takes place in top-color models,
where the  mass of the top  quark is generated through  a gap equation
involving a strongly interacting massive gauge field
\cite{Bardeen:1990ds}.   The usual  problem  with the  SD approach  in
general \cite{Cornwall:1974vz,Marciano:1978su}
and  the  gap equations  in  particular \cite{Atkinson:1987av} is  that
sooner  or later  one is  forced  to choose  a gauge,  resorting to  a
variety of arguments, but gauge  choices cast in general doubts on the
robustness  of the conclusions  thusly reached.   In the  latter case,
{\it i.e.}  the   resonant  production   of  fermions,  and   in  particular
top-quarks, one makes use of the resumed off-shell quark self-energy.
Even  though exactly at  the resonance  the gauge  dependences cancel,
infinitesimally away  from it they  persist, giving rise  to artifacts
obscuring the  notion of the  running width and the  implementation of
perturbative unitarity in the resulting Born-improved amplitudes
\cite{Pilaftsis:1990zt}.

It  is  known   that  gauge-invariant  and  GFP-independent  effective
off-shell  Green's functions can  be constructed  by resorting  to the
pinch  technique  (PT)  \cite{Cornwall:1982zr}.   The  PT  reorganizes
systematically a  given physical amplitude  into sub-amplitudes, which
have   the  same  kinematic   properties  as   conventional  $n$-point
functions,  (propagators,  vertices,  boxes),  but, in  addition,  are
endowed  with  desirable physical  properties.   Most importantly,  at
one-loop order (i) are independent of the GFP; (ii)
satisfy naive,  (ghost-free) tree-level Ward  identities (WI), instead
of the usual Slavnov-Taylor identities
\cite{Slavnov:1972fg};  (iii) contain only
physical thresholds and satisfy  very  special  unitarity  relations
\cite{Papavassiliou:1995fq}; (iv) 
coincide  with  the conventional $n$-point  functions when the latter are 
computed in the background field method Feynman  gauge  
\cite{Denner:1994nn}.
These  properties  are realized  diagrammatically,  by exploiting  the
elementary WI's  of the theory in order  to enforce crucial
cancellations.  The extension of the PT to two-loops has only recently
been accomplished in the case  of massless Yang-Mills theories such as
QCD
\cite{Papavassiliou:2000az}.  
The studies presented so far in the literature
have mainly focused on the  general
construction of the  effective gauge-independent gluon self-energy, 
but  little
has been  said about the  fermion propagator 
\cite{Papavassiliou:1995yi,Watson:1999vv,Yamada:2001px}.

Throughout the two-loop analysis
of \cite{Papavassiliou:2000az}
it has been assumed that one can work  without loss  of  generality in 
the covariant  (renormalizable) Feynman gauge, {\it i.e.}
begin the analysis by  choosing   the Feynman gauge when  
writing  down  the   Feynman  diagrams contributing  to the  $S$-matrix.
Of course, there is no doubt that
the  entire S-matrix written in  the
Feynman gauge
is equal  to the same entire $S$-matrix written in any other gauge.
What is less obvious is that 
all relevant cancellations proceed without  need of carrying out
integrations over the virtual loop momenta, thus maintaining the kinematic
identity of the various  Green's functions intact,
a point of crucial importance within the PT philosophy.  
As has been shown by explicit calculations 
(see for example \cite{Papavassiliou:1995yi} ), 
this is indeed 
the case at one-loop.
Assuming that this important property persists at two loops,
the  highly non-trivial issue which was
resolved in \cite{Papavassiliou:2000az} 
was  how the splitting of the  three-gluon vertices appearing in
the two-loop diagrams should proceed.   
We shall not review this point further,
given  that it has been  exhaustively treated in \cite{Papavassiliou:2000az}; 
here it should suffice to
say that no such splitting should take place for the internal  three-gluon
vertices appearing  inside the  two-loop fermion propagator,
or any other diagram for that matter. 
Therefore,  one of  the 
conclusions  presented in \cite{Papavassiliou:2000az}, under the 
aforementioned assumption,
was that  the gauge-independent  two-loop
quark  propagator in  the presence  of QCD interactions {\it coincides} with
the conventional one computed in the Feynman gauge.   
In  this  paper  we  will  verify 
this assumption in the cases of  
QED and QCD for the two-loop
fermion self-energy. 

In  the  first  part of  this  paper  we  will  track down  the  gauge
cancellations  systematically,  and   provide  a  simple  diagrammatic
algorithm which allows one to follow easily their implementation. 
\footnote{
Some aspects of the
cancellation mechanism  described  in this paper are similar in  spirit to
that  presented in \cite{Feng:1996vg}; 
however, we  do not resort to ``color 
orientation'' techniques.}.  
 The key
observation, which will be  used as the only guiding principle throughout   
the   intermediate    steps,   is  that all contributions
originating from the longitudinal parts of gauge boson propagators, by
virtue of the  WI they trigger,  give rise to {\it unphysical} effective
vertices,  {\it i.e.} vertices which do not exist in  the   original 
Lagrangian. All  such  vertices   cancel  {\it diagrammatically} inside
ostensibly gauge-invariant  quantities, such as current correlation functions
or $S$-matrix elements 
\footnote{ Given that the vertices involved are unphysical, one might
be tempted to directly discard all such contributions by hand,
instead of cancelling them 
algebraically against each other, as we do in the paper. 
In the case we consider here this seemingly
ad hoc procedure would furnish the
correct answer, but it is not known to us if it 
would work in general}. 
The final calculational  recipe resulting from 
this analysis is  that one can use directly the  covariant Feynman gauge, 
which of course happens to be the simplest operationally.  
It is important to
emphasize  that exactly the same result is obtained 
even in  the context  of
the non-covariant  axial gauges  for example
\cite{Dokshitzer:1980hw}, 
where the  Feynman gauge cannot be  reached a priori  by simply fixing
appropriately the value of the  gauge fixing parameter.  Thus, even if
one  uses a bare  gluon propagator  of the  general axial  gauge form,
after the  aforementioned cancellations  have taken place  one arrives
effectively to the answer written in the covariant Feynman gauge. Also
notice that in calculating the final answer (something we shall not do
here)  one  never  has  to  carry  out any  of  the  tricky  integrals
characteristic  of  the  axial  gauges,  {\it  {\it i.e.}}   integrals  with
unphysical poles of the form $n\! \cdot\! k$
\cite{Leibbrandt:1984pj}.

The second part  of the paper is devoted  to the absorptive derivation
of  the  same  results.   The  absorptive  construction  exploits  the
unitarity and analyticity  properties of physical amplitudes, together
with the  fundamental WI satisfied by {\it  entire} physical processes
dictated     by      the     Becchi-Rouet-Stora     (BRS)     symmetry
\cite{Becchi:1976nq}.  The salient points  of this general method have
been presented in detail in \cite{Papavassiliou:1996zn}.  Here we will
apply it to the case  of the two-loop quark self-energy containing QED
or QCD corrections.

The paper is  organized as follows.  In Sec.~I we  review the one loop
construction in both  the QED and QCD.  This will allow  us to fix the
notation  and  introduce  in  a simplified  setting  the  diagrammatic
algorithm used throughout the paper. In particular we will discuss how
the  gauge  cancellations are  achieved  both  in current  correlation
functions as well  as in physical on-shell processes,  such as $\gamma
Q\to\gamma Q$ or  $G Q\to GQ$, where  $Q$ is a quark and  $G$ a gluon.
In Sec.~II we tackle the two  loop case.  The procedure is carried out
in full  detail, beginning from the same  current correlation function
as in  the one-loop  case.  By means  of a systematic,  albeit lenghty
analysis,  we demonstrate explicitly  all relevant  cancellations, and
finally  define  the GFP-independent  two-loop  electron (quark)  self
energy for  QED and QCD.  We then  turn to the description  of how one
may construct  the PT effective Green's functions  using unitarity and
analyticity arguments.  Thus, we  first review the one loop absorptive
construction  in  both  the  Abelian and  non-Abelian  gauge  theories
(Sec.~III), and introduce the notation  which will be used in Sec.~IV,
where we  will carry  out in detail  the full  absorptive construction
both in the QED and QCD  frameworks. Finally, in Sec.~V we present our
concluding remarks.

\section{The one-loop case.}

Before venturing  into the  intricacies of the  two-loop construction,
which is the  main topic of this paper, we will  first present the one
loop case, in  an attempt to fix the ideas and  the notation.  In this
section  we will  explain in  detail the  method which  gives  rise to
effective,  gauge-independent  fermion  self-energies. In  particular,
after  setting  up  the  diagrammatic  notation  which  will  be  used
throughout the  paper, we will  illustrate how the procedure  works in
the  case of  QED and  QCD.  The  results for  the one-loop  case have
already been  presented in \cite{Papavassiliou:2000az},  albeit from a
slightly  different point of  view; here  we will  recast them  in the
diagrammatic language introduced below,  thus setting up the stage for
the two-loop derivation.   It turns out that in  the one-loop case the
difference   between   the  Abelian   (QED)   and  non-Abelian   (QCD)
constructions  is purely  group-theoretical, and  therefore  a unified
presentation  will be  followed; this  will  seize being  the case  at
two-loops.

We will assume that the theory has been  gauge-fixed by introducing in the
gauge-invariant Lagrangian a gauge-fixing term of the form 
$\frac{1}{2\xi}(\partial_{\mu}A^{\mu})^2$, {\it i.e.}
 a  linear, covariant gauge;
the parameter $\xi$ is the GFP. This gauge-fixing term gives rise to a bare
gauge-boson propagator of the form
\begin{equation}
\Delta_{\mu\nu}(\ell,\xi ) = 
-\frac{\D i}{\D \ell^2}
\left[\ g_{\mu\nu} - (1-\xi) \frac{\D \ell_\mu
\ell_\nu}{\D \ell^2}\right]
\end{equation}
which explicitly depends on $\xi$.  The trivial  color factor $\delta_{ab}$
appearing in the (gluon) propagator  has been suppressed. The form of
$\Delta_{\mu\nu}(\ell,\xi )$ for the  special choice   $\xi =1$ (Feynman gauge)
will be of central importance in what follows; we will denote it by
$\Delta_{\mu\nu}^{F}(\ell)$, {\it i.e.}
\begin{equation}
\Delta_{\mu\nu}(\ell, 1 )\equiv \Delta_{\mu\nu}^{F}(\ell)
=  -\frac{\D i}{\D \ell^2}\, g_{\mu\nu}\,  . 
\end{equation}
$\Delta_{\mu\nu}(\ell,\xi )$  and $\Delta_{\mu\nu}^{F}(\ell)$ will  be
denoted graphically   as  follows:\footnote{For  convenience,   in our
diagrammatic notation we  will remove  all factors of $i$
appearing in the  fermionic/bosonic propagators; they can be  easily
recovered as a  global coefficient multiplying  the Feynman
diagram under consideration.}

\begin{center}
\begin{picture}(0,20)(100,-15)

\Gluon(-5,-5)(30,-5){2.5}{7} \Photon(160,-5)(195,-5){2}{6}

\Text(35,-5)[l]{\normalsize{$\equiv   i  \Delta_{\mu\nu}(\ell,\xi)$,}}
\Text(200,-5)[l]{\normalsize{$\equiv i \Delta_{\mu\nu}^F(\ell)$.}}

\end{picture}
\end{center}
For the diagrammatic proofs that   will follow, in  addition  to the
propagators   $\Delta_{\mu\nu}(\ell)$  and   $\Delta_{\mu\nu}^F(\ell)$
introduced  above,   we   will need  six   auxiliary  propagator-like
structures, as shown below:

\begin{center}
\begin{picture}(0,80)(100,-75)

\Photon(-5,-5)(30,-5){2}{6}
\Photon(-5,-3.5)(30,-3.5){2}{6}
\Text(11,-5)[c]{\rotatebox{19}{\bf{\big /}}}
\Text(10.4,-5)[c]{\rotatebox{19}{\bf{\big /}}}
\Text(14,-5)[c]{\rotatebox{19}{\bf{\big /}}}
\Text(14.6,-5)[c]{\rotatebox{19}{\bf{\big /}}}
\Text(35,-5)[l]{\normalsize{$\equiv\ \frac{\D \ell_\mu
\ell_\nu}{\D \ell^4}$}}

\Photon(-5,-35)(30,-35){2}{6}
\Photon(-5,-33.5)(30,-33.5){2}{6}
\Text(12.2,-35)[c]{\rotatebox{19}{\bf{\big /}}}
\Text(12.8,-35)[c]{\rotatebox{19}{\bf{\big /}}}
\Vertex(-5,-35){2}
\Text(35,-35)[l]{\normalsize{$\equiv\ \frac{\D \ell_\mu}{\D \ell^4}$}}

\Photon(-5,-65)(30,-65){2}{6}
\Photon(-5,-63.5)(30,-63.5){2}{6}
\Vertex(-5,-65){2}
\Vertex(30,-65){2}
\Text(35,-65)[l]{\normalsize{$\equiv\ \frac{\D 1}{\D \ell^4}$}}

\Photon(160,-5)(195,-5){2}{6}
\Text(176,-5)[c]{\rotatebox{19}{\bf{\big /}}}
\Text(175.4,-5)[c]{\rotatebox{19}{\bf{\big /}}}
\Text(179,-5)[c]{\rotatebox{19}{\bf{\big /}}}
\Text(179.6,-5)[c]{\rotatebox{19}{\bf{\big /}}}
\Text(200,-5)[l]{$\equiv\ \frac{\D \ell_\mu
\ell_\nu}{\D \ell^2}$}

\Photon(160,-35)(195,-35){2}{6}
\Vertex(160,-35){2}
\Text(200,-35)[l]{\normalsize{$\equiv\ \frac{\D \ell_\mu}{\D \ell^2}$}}
\Text(177.2,-35)[c]{\rotatebox{19}{\bf{\big /}}}
\Text(177.8,-35)[c]{\rotatebox{19}{\bf{\big /}}}

\Photon(160,-65)(195,-65){2}{6}
\Vertex(160,-65){2}
\Vertex(195,-65){2}
\Text(200,-65)[l]{\normalsize{$\equiv\ \frac{\D 1}{\D \ell^2}$}}

\end{picture}
\end{center}
All of these six structures will arise from  algebraic manipulations of the
original $\Delta_{\mu\nu}(\ell)$. For example, in terms of the above notation
we have the following simple relation (we will set
$\lambda  \equiv \xi-1$):

\begin{center}
\begin{picture}(0,10)(60,20)

\Gluon(5,25)(40,25){2.5}{7}
\Text(45,25)[l]{\normalsize{$\equiv$}}
\Photon(60,25)(95,25){2}{6}
\Text(100,25)[l]{\normalsize{$+\ \lambda$}}
\Photon(120,25.75)(155,25.75){2}{6}
\Photon(120,24.25)(155,24.25){2}{6}
\Text(136,25)[c]{\rotatebox{19}{\bf{\big /}}}
\Text(135.4,25)[c]{\rotatebox{19}{\bf{\big /}}}
\Text(139,25)[c]{\rotatebox{19}{\bf{\big /}}}
\Text(139.6,25)[c]{\rotatebox{19}{\bf{\big /}}}

\end{picture}
\end{center}
We next turn to the study of the gauge-dependence of the fermion self-energy
(electron in QED, quarks in QCD). The inverse electron propagator  of order $n$
in the perturbative expansion has the form (again suppressing color)
\begin{equation}
S_n^{-1}(p,\xi) = \pslush -m - \Sigma^{(n)}(p,\xi)
\end{equation}
where $\Sigma^{(n)} (p,\xi)$ is the $n-th$ order  self-energy. Clearly $
\Sigma^{(0)} = 0$, and $S_0^{-1}(p) = \pslush -m$. The quantity $\Sigma^{(n)}
(p,\xi)$ depends  explicitly on $\xi$ already for $n=1$. In particular
\begin{eqnarray}
\Sigma^{(1)}(p,\xi) &=& \int  [d\ell]
\gamma^{\mu} S_0(p+\ell) \gamma^{\nu} 
\Delta_{\mu\nu}(\ell,\xi) \nonumber\\
 &=&
\Sigma^{(1)}_F (p)
+ \lambda \Sigma^{(1)}_L (p)
\label{STOT}
\end{eqnarray}
with
\begin{equation}
\Sigma^{(1)}_F (p) \equiv \Sigma^{(1)}(p,1) = 
\int  [d\ell]
\gamma^{\mu} S_0 (p+\ell) \gamma^{\nu} 
\Delta_{\mu\nu}^F (\ell)
\label{SF}
\end{equation}
and 
\begin{eqnarray}
\Sigma^{(1)}_L (p) 
&=& - S_0^{-1}(p) 
\int  \frac{[d\ell]}{\ell^4} \, S_0(p+\ell)\gamma^{\nu} \ell_{\nu} \nonumber\\
&=& - \int \frac{[d\ell]}{\ell^4} \,   \ell_{\mu}\gamma^{\mu} 
S_0(p+\ell)\,\,  S_0^{-1}(p)  
\nonumber\\     
&=& 
  S_0^{-1}(p) \, \int \frac{[d\ell]}{\ell^4} S_0(p+\ell) \,\, S_0^{-1}(p) 
 - S_0^{-1}(p) \int  \frac{[d\ell]}{\ell^4} .
\label{SL}
\end{eqnarray}
In the above formulas  $ [d\ell] \equiv g^2 
\mu^{2\epsilon}\frac{d^D \ell}{(2\pi)^D}$,  
with $D=4-2\epsilon$ the dimension of space-time, $\mu$ the 't Hooft
mass\footnote{Throughout the paper we use   $\int  \frac{[d\ell]}{\ell^2} = 0 $ and 
$\int  \frac{[d\ell] \ell_{\alpha}\ell_{\beta}}{\ell^4} =
g_{\alpha\beta} D^{-1} \int \frac{[d\ell]}{\ell^2} =
0$,  valid in dimensional regularization. In addition, integrals odd in the
integration  variable are considered to vanish.  Notice however that nowhere
will we use the slightly subtler  $\int \frac{[d\ell]}{\ell^4} = 0 $,  which is often
employed in the literature.},  and $g$ the gauge coupling ($ g\equiv e$ for
QED, and $ g\equiv g_s$ for QCD). The subscripts ``F'' and ``L'' stand for
``Feynman'' and  ``Longitudinal'', respectively.  Notice that $\Sigma^{(1)}_L$ 
is proportional to  $S_0^{-1}(p)$ and thus vanishes ``on-shell''.  The most
direct way to arrive at the results of Eq.(\ref{SL}) is to employ the
fundamental WI \footnote{ A formal derivation of the gauge dependence may be 
obtained by resorting to the so called 
Nielsen identities \cite{Nielsen:1975ph}.} 
\begin{equation} 
\elle=S_0^{-1}(p+\ell)-S_0^{-1}(p),
\label{WIzero} 
\end{equation} 
which is triggered every time the longitudinal
momenta of $\Delta_{\mu\nu}(\ell,\xi)$ gets contracted with the appropriate
$\gamma$ matrix appearing in the  vertices. 
Diagrammatically, this elementary WI gets translated to

\begin{center}
\begin{picture}(0,40)(60,20)

\Line(0,25)(50,25)
\Photon(24.3,50)(24.3,25){2}{5}
\Photon(25.7,50)(25.7,25){2}{5}
\Text(25,37.2)[c]{\rotatebox{-71}{\bf{\big /}}}
\Text(25,37.8)[c]{\rotatebox{-71}{\bf{\big /}}}
\Text(55,25)[l]{$\equiv$}

\Line(70,25)(100,25)
\Photon(100.7,25)(100.7,50){2}{5}
\Photon(99.3,25)(99.3,50){2}{5}
\Vertex(100,25){2}

\Text(120,25)[c]{$-$}

\Line(140,25)(170,25)
\Photon(140.7,25)(140.7,50){-2}{5}
\Photon(139.3,25)(139.3,50){-2}{5}
\Vertex(140,25){2}

\end{picture}
\end{center} 
Then, the diagrammatic representation of
Eq.(\ref{STOT}), Eq.(\ref{SF}), and Eq.(\ref{SL}) will be given by

\begin{equation}
\begin{picture}(0,110)(130,-45)

\Line(-15,25)(55,25)
\GlueArc(20,25)(20,0,180){2.5}{9}
\Text(60,25)[l]{$\equiv$}

\Line(70,25)(140,25)
\PhotonArc(105,25)(20,0,180){2}{7.5}
\Text(145,25)[l]{$-\ \lambda$}

\Line(170,25)(230,25)
\PhotonArc(190,25)(20,0,180){2}{7.5}
\PhotonArc(190,25)(18.8,0,180){2}{7.5}
\Text(190.9,44.5)[c]{\rotatebox{19}{\bf{\big /}}}
\Text(190.3,44.5)[c]{\rotatebox{19}{\bf{\big /}}}

\Text(60,-30)[l]{$=$}
\Line(70,-30)(140,-30)
\PhotonArc(105,-30)(20,0,180){2}{7.5}
\Text(145,-30)[l]{$+\ \lambda$}

\Line(170,-30)(230,-30)
\PhotonArc(200,-40)(30,20,160){2}{7.5}
\PhotonArc(200,-40)(28.8,20,160){2}{7.5}
\Vertex(170,-30){2}
\Vertex(230,-30){2}
\Text(237,-30)[l]{$-\ \lambda$}

\Line(270,-30)(320,-30)
\PhotonArc(270,-20)(10,180,540){2}{9}
\PhotonArc(270,-20)(8.8,180,540){2}{9}

\Vertex(270,-30){2}
\Vertex(170,25){2}

\label{GR}
\end{picture}
\end{equation}

When considering physical amplitudes,  the characteristic structure of the
longitudinal parts established above   allows for their cancellation against
identical contributions  originating from diagrams which are kinematically
different from fermion self-energies,  such as vertex-graphs or boxes, {\it
without} the need for integration over the internal virtual momenta.  This last
property is important because  in this way the original  kinematical identity
is guaranteed to be maintained; instead, loop integrations generally mix the
various kinematics.   Diagrammatically, the action of the WI  is very
distinct:  it always gives rise to unphysical effective vertices, {\it i.e.}
vertices
which do not appear in the original Lagrangian; all such vertices   cancel in
the full, gauge-invariant amplitude.

To actually pursue these special 
cancellations explicitly one may choose among a variety
of gauge invariant quantities.  For example, one may consider the current
correlation function  $I_{\mu\nu}$ defined as (in momentum space)
\begin{eqnarray} 
I_{\mu\nu}(q) &=& i \int d^{4}x e^{iq\cdot x}
\langle 0  | T \left[J_{\mu}(x) J_{\nu}(0) \right] |0 \rangle
\nonumber\\
 &=& (g_{\mu\nu} q^2 - q_{\mu}q_{\nu}) I(q^2) \, ,
\label{IQED}
\end{eqnarray} 
where the current $J_{\mu} (x)$ is given by $J_{\mu}(x) =\ :\!\bar{Q} (x)
\gamma_{\mu} Q (x)\!:$. Of course, $I_{\mu\nu}(q)$  coincides with the photon
vacuum polarization of QED. Equivalently, one may study   physical on-shell
processes such as   $\gamma e \to \gamma e$, or $\gamma Q \to \gamma Q$, or  
$G Q \to G Q$, where  $G$ is a gluon. Of these three processes the first two are the most economical, 
since in the latter the algebra is more complicated due
to the appearance of three gluon vertices.

To see explicitly the mechanism enforcing  these cancellations in the   QED and
QCD cases, we first  consider the one-loop photonic or gluonic corrections,
respectively, to the quantity  $I_{\mu\nu}$. Clearly either set of corrections
is GFP-independent, since the current $J_{\mu}(x)$ is invariant under both the
$U(1)$ and the $SU(3)$ gauge transformations  
\begin{eqnarray}
Q (x) &\to& \exp\{ -i\theta(x)\} Q(x)  \qquad  
Q (x) \to \exp\{-i\theta_a(x) T^a\} Q (x)
\nonumber\\
\bar{Q}(x) &\to& \exp\{ i\theta(x)\} \bar{Q}(x) \qquad \quad \!
\bar{Q}(x) \to  \exp\{i\theta_a(x) T^a \} \bar{Q}(x)
\end{eqnarray}
where $T^a = \frac{1}{2}\lambda^a$, with  $\lambda^a$ the Gell-Mann matrices. 

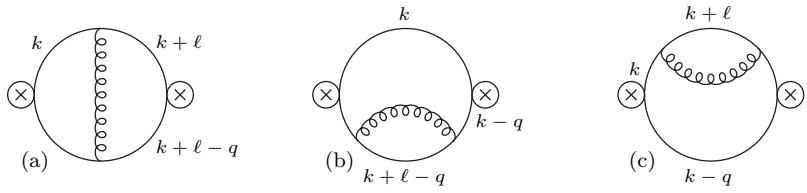
\begin{figure}

\begin{picture}(100,100)(-60,0)

\Text(20,25)[l]{\footnotesize{(a)}}
\Text(135,25)[l]{\footnotesize{(b)}}
\Text(250,25)[l]{\footnotesize{(c)}}

\Text(24,70)[l]{\scriptsize{$k$}}
\Text(71,70)[l]{\scriptsize{$k+\ell$}}
\Text(71,30)[l]{\scriptsize{$k+\ell-q$}}

\Text(165,81)[c]{\scriptsize{$k$}}
\Text(192,40)[l]{\scriptsize{$k-q$}}
\Text(165,19)[c]{\scriptsize{$k+\ell-q$}}

\Text(280,81)[c]{\scriptsize{$k+\ell$}}
\Text(250,60)[l]{\scriptsize{$k$}}
\Text(280,19)[c]{\scriptsize{$k-q$}}

\CArc(20,50)(5,0,360)
\Line(18,52)(22,48)
\Line(22,52)(18,48)
\CArc(50,50)(25,0,360)
\CArc(80,50)(5,0,360)
\Line(78,52)(82,48)
\Line(82,52)(78,48)

\CArc(135,50)(5,0,360)
\Line(133,52)(137,48)
\Line(137,52)(133,48)
\CArc(165,50)(25,0,360)
\CArc(195,50)(5,0,360)
\Line(193,52)(197,48)
\Line(197,52)(193,48)

\CArc(250,50)(5,0,360)
\Line(248,52)(252,48)
\Line(252,52)(248,48)
\CArc(280,50)(25,0,360)
\CArc(310,50)(5,0,360)
\Line(308,52)(312,48)
\Line(312,52)(308,48)

\Gluon(50,25)(50,75){2}{9}

\GlueArc(165,25)(19.2,22.5,157.5){2}{9}

\GlueArc(280,75)(19.2,202.5,337.5){2}{9}

\end{picture}

\caption{One loop diagram contributing to the QED/QCD fermion self-energy.}

\label{fig1}
\end{figure}

The relevant diagrams are those shown in Fig.\ref{fig1}. To see the
appearance of the unphysical vertices,  
we carry out the manipulations presented
in Eq.(\ref{STOT}), Eq.(\ref{SF}), and Eq.(\ref{SL}), or, equivalently, in
 Eq.(\ref{GR}), 
this time  embedded inside $I_{\mu\nu}(q)$. Thus, from  diagrams (b) and
(c) we arrive at
\begin{center}
\begin{picture}(100,50)(30,25)

\Text(-55,50)[l]{(b)+(c) $\to\ 2\lambda$}

\CArc(20,50)(5,0,360)
\Line(18,52)(22,48)
\Line(22,52)(18,48)
\CArc(50,50)(25,0,360)
\CArc(80,50)(5,0,360)
\Line(78,52)(82,48)
\Line(82,52)(78,48)
\PhotonArc(50,25)(19.2,22.5,157.5){2}{9}
\PhotonArc(50,25)(18,22.5,157.5){2}{9}
\Text(48.5,44.2)[c]{\rotatebox{19}{\bf{\big /}}}
\Text(47.9,44.2)[c]{\rotatebox{19}{\bf{\big /}}}
\Text(51.5,44.2)[c]{\rotatebox{19}{\bf{\big /}}}
\Text(52.1,44.2)[c]{\rotatebox{19}{\bf{\big /}}}

\Text(90,50)[l]{$=-\,2\lambda$}
\hspace{-1.3 cm}

\CArc(160,50)(5,0,360)
\Line(158,52)(162,48)
\Line(162,52)(158,48)
\CArc(190,50)(25,0,360)
\CArc(220,50)(5,0,360)
\Line(218,52)(222,48)
\Line(222,52)(218,48)
\PhotonArc(215,25)(25.7,90,180){-2}{7.5}
\PhotonArc(215,25)(24.3,90,180){-2}{7.5}
\Vertex(215,50){2}
\Text(196,41)[c]{\rotatebox{65}{\bf{\big /}}}
\Text(196.45,41.45)[c]{\rotatebox{65}{\bf{\big /}}}

\end{picture}
\end{center}
We thus see that since the action of the
elementary WI of  Eq.(\ref{WIzero}) amounts to the cancellation of internal 
propagators, 
its diagrammatic  
consequence  
is that of introducing an unphysical effective vertex, 
describing an interaction of the form $\gamma\gamma \bar{Q} Q$ or
$\gamma G \bar{Q} Q$, depending on whether we consider photonic or 
gluonic corrections. 
This type of vertex
may be depicted by means of a Feynman rule of the form

\begin{center}
\begin{picture}(0,35)(20,10)

\CArc(20,10)(5,0,360)
\Line(18,12)(22,8)
\Line(22,12)(18,8)

\Line(-5,15)(45,15)
\Gluon(20,15)(20,40){2}{5}
\Vertex(20,15){2}

\Text(50,15)[l]{$\equiv i\gamma_\mu$}
\Text(30,10)[c]{\footnotesize{$\mu$}}

\end{picture}
\end{center}
being $\mu$ the index of the external current.

To see how the above unphysical contributions cancel inside  $I_{\mu\nu}$
we turn to diagram (a). The action of 
the WI may be
translated to the following diagrammatic picture

\begin{center}
\begin{picture}(100,50)(85,25)

\Text(-30,50)[l]{(a) $\to\ \lambda$}

\CArc(20,50)(5,0,360)
\Line(18,52)(22,48)
\Line(22,52)(18,48)
\CArc(50,50)(25,0,360)
\CArc(80,50)(5,0,360)
\Line(78,52)(82,48)
\Line(82,52)(78,48)
\Photon(49.25,25)(49.25,75){2}{9}
\Photon(50.75,25)(50.75,75){2}{9}
\Text(50,51.5)[c]{\rotatebox{-71}{\bf{\big /}}}
\Text(50,52.1)[c]{\rotatebox{-71}{\bf{\big /}}}
\Text(50,48.5)[c]{\rotatebox{-71}{\bf{\big /}}}
\Text(50,47.9)[c]{\rotatebox{-71}{\bf{\big /}}}
\Text(90,50)[l]{$=\ \ \lambda$}
\hspace{-1.5 cm}

\Text(155,25)[lb]{\footnotesize{$(\alpha)$}}

\CArc(160,50)(5,0,360)
\Line(158,52)(162,48)
\Line(162,52)(158,48)
\CArc(190,50)(25,0,360)
\CArc(220,50)(5,0,360)
\Line(218,52)(222,48)
\Line(222,52)(218,48)
\PhotonArc(215,25)(25.7,90,180){-2}{7.5}
\PhotonArc(215,25)(24.3,90,180){-2}{7.5}
\Vertex(215,50){2}
\Text(196,41)[c]{\rotatebox{65}{\bf{\big /}}}
\Text(196.45,41.45)[c]{\rotatebox{65}{\bf{\big /}}}
\Text(230,50)[l]{$-\ \ \lambda$}
\hspace{-1.6 cm}

\Text(295,25)[lb]{\footnotesize{$(\beta)$}}

\CArc(300,50)(5,0,360)
\Line(298,52)(302,48)
\Line(302,52)(298,48)
\CArc(330,50)(25,0,360)
\CArc(360,50)(5,0,360)
\Line(358,52)(362,48)
\Line(362,52)(358,48)
\PhotonArc(305,25)(25.7,0,90){2}{7.5}
\PhotonArc(305,25)(24.3,0,90){2}{7.5}
\Vertex(305,50){2}
\Text(324,41)[c]{\rotatebox{-25}{\bf{\big /}}}
\Text(323.6,41.4)[c]{\rotatebox{-25}{\bf{\big /}}}

\end{picture}
\end{center}
It is then elementary to establish that the two diagrams on the 
right-hand side 
of the
above diagrammatic  equation add up.  Indeed, for $(\alpha)$ we have
(suppressing the integral measure)
\begin{equation}
(\alpha)\sim \lambda{\rm Tr}\,\left[\gamma^\mu\frac 1{\ka-m} \gamma^\nu \frac
1{\ka+\elle-\qu-m}\gamma^\rho\frac 1{\ka-\qu-m}\right]\frac{\ell_\rho}{\ell^4},
\end{equation}
whereas from $(\beta)$, 
taking the trace counter-clockwise and using the fact that 
 $I_{\mu\nu}$ is symmetric under the exchange
$\mu\leftrightarrow\nu$, we obtain
\footnote{Throughout the paper we will make extensive  
use of suitable shiftings of the
integration variables together with  
various rearrangements of seemingly distinct
diagrams.}
 \begin{eqnarray}
(\beta) & \sim & -\lambda{\rm Tr}\,\left[\gamma^\nu\frac 1{\ka+\elle-m} 
\gamma^\mu \frac
1{\ka-\qu-m}\gamma^\rho\frac 1{\ka+\elle-\qu-m}\right]\frac{\ell_\rho}{\ell^4} 
\nonumber \\ 
& \stackrel{\stackrel{\stackrel{\scriptstyle{k+\ell\to k}}
{\scriptstyle{\ell\to-\ell}}}{\mu\leftrightarrow\nu}}{=} & 
\lambda{\rm Tr}\,\left[\gamma^\mu\frac 1{\ka-m} 
\gamma^\nu \frac
1{\ka+\elle-\qu-m}\gamma^\rho\frac 1{\ka-\qu-m}\right]\frac{\ell_\rho}{\ell^4}. 
\end{eqnarray} 

Summing the two equations above then, it is clear how  the gauge dependent part
of the one loop amplitude cancel all together.
Having proved that the GFP-dependent contributions coming from the 
original graphs containing $\Sigma^{(1)}(p,\xi)$, {\it i.e.} 
Fig.\ref{fig1}(b) and Fig.\ref{fig1}(c) cancel exactly against
equal but opposite {\it propagator-like} contributions coming from
Fig.\ref{fig1}(a), one is left with the ``pure'' GFP-independent
one-loop fermion self-energy, $\widehat{\Sigma}^{(1)}(p)$. Clearly,
it concides with the $\Sigma^{(1)}_F (p)$ of Eq.(\ref{SF}), {\it i.e.}
\cite{Papavassiliou:1995yi}
\begin{equation}
\widehat{\Sigma}^{(1)}(p) = \Sigma^{(1)}_F (p).
\label{GFPI}
\end{equation}

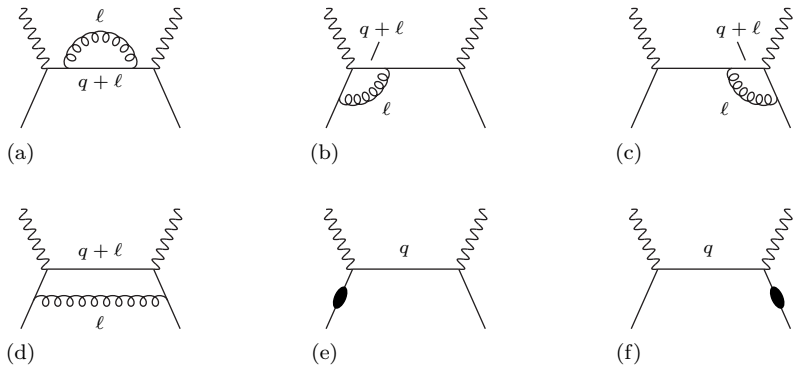
\begin{figure}

\begin{picture}(100,100)(-60,0)

\Text(20,18)[c]{\footnotesize{(a)}}
\Text(135,18)[c]{\footnotesize{(b)}}
\Text(250,18)[c]{\footnotesize{(c)}}

\Text(50,45)[c]{\scriptsize{$q+\ell$}}
\Text(50,70)[c]{\scriptsize{$\ell$}}

\Text(165,65)[r]{\scriptsize{$q+\ell$}}
\Text(160,35)[r]{\scriptsize{$\ell$}}
\Line(152,53)(155,60)

\Text(300,65)[r]{\scriptsize{$q+\ell$}}
\Text(288,35)[r]{\scriptsize{$\ell$}}
\Line(293,53)(290,60)

\Line(20,27.5)(30,50)
\Photon(20,72.5)(30,50){2}{6}
\Line(30,50)(70,50)
\Line(70,50)(80,27.5)
\Photon(70,50)(80,72.5){-2}{6}
\GlueArc(50,50)(12,0,180){2}{8}

\Line(135,27.5)(145,50)
\Photon(135,72.5)(145,50){2}{6}
\Line(145,50)(185,50)
\Line(185,50)(195,27.5)
\Photon(185,50)(195,72.5){-2}{6}
\GlueArc(145,50)(12,246,360){2}{6}

\Line(250,27.5)(260,50)
\Photon(250,72.5)(260,50){2}{6}
\Line(260,50)(300,50)
\Line(300,50)(310,27.5)
\Photon(300,50)(310,72.5){-2}{6}
\GlueArc(300,50)(12,180,294){2}{6}

\end{picture}

\begin{picture}(100,100)(-60,-25)

\Text(20,18)[c]{\footnotesize{(d)}}
\Text(135,18)[c]{\footnotesize{(e)}}
\Text(250,18)[c]{\footnotesize{(f)}}

\Text(50,57)[c]{\scriptsize{$q+\ell$}}
\Text(50,30)[c]{\scriptsize{$\ell$}}

\Text(165,57)[c]{\scriptsize{$q$}}

\Text(280,57)[c]{\scriptsize{$q$}}

\Line(20,27.5)(30,50)
\Photon(20,72.5)(30,50){2}{6}
\Line(30,50)(70,50)
\Line(70,50)(80,27.5)
\Photon(70,50)(80,72.5){-2}{6}
\Gluon(24.7,38)(75.3,38){2}{10}

\Line(135,27.5)(145,50)
\Photon(135,72.5)(145,50){2}{6}
\Line(145,50)(185,50)
\Line(185,50)(195,27.5)
\Photon(185,50)(195,72.5){-2}{6}
\GOval(140.3,39.25)(2,4)(66){0}

\Line(250,27.5)(260,50)
\Photon(250,72.5)(260,50){2}{6}
\Line(260,50)(300,50)
\Line(300,50)(310,27.5)
\Photon(300,50)(310,72.5){-2}{6}
\GOval(305,39.25)(2,4)(-66){0}

\end{picture}

\caption{One loop QED/QCD correction to the process $\gamma Q\to\gamma Q$ 
$(p_1+p_2=q=p_3+p_4)$. Diagrams (e) and (f)
corresponds to the (one loop) renormalization of the external legs.}
\label{ftre}

\end{figure}

Next, we will consider the physical process $\gamma Q\to\gamma Q$, in order to
analyze how the  procedure outlined above works in the
case of an $S$-matrix element.  
The
one-loop diagrams for the process under consideration are listed in
Fig.\ref{ftre}. We will isolate the parts of the above diagrams
proportional to $\lambda$, using again the WI of Eq.(\ref{GR}),    
together with the fact that 
the external particles are on their mass shell. 
We emphasize that the point of this 
exercise is not to prove the GFP-independence of the $S$-matrix,
but rather to recognize that the GFP-cancellations proceed in a very
special way: 
the $\lambda$-dependent parts of vertices [(b),(c)] 
and boxes [(d)]
do not maintain the same kinematic identity as their parent graphs;
instead, they reduce to simpler kinematic structures, which are
precisely those stemming from the original propagator diagram (a),
and finally cancel algebraically against each-other.

To see this in detail, we begin with diagram (a); 
applying the identity of Eq.(\ref{GR}) in a symmetric way,
{\it i.e.} allowing the longitudinal parts to act once on the left and once on
the right vertex of the diagram, one obtains the following  GFP dependent
part 

\begin{center}
\begin{picture}(0,50)(100,0)

\Text(-40,25)[l]{(a)$\to\ -\ \frac{\D \lambda}{\D 2}$}  
\Text(85,25)[l]{$+\ \lambda$}  
\Text(185,25)[l]{$-\ \frac{\D \lambda}{\D 2}$}

\PhotonArc(35,25)(8,90,450){2}{9}
\PhotonArc(35,25)(6.8,90,450){2}{9}
\Line(25,25)(38,25)
\Line(45,25)(65,25)
\Line(15,2.5)(25,25)
\Photon(15,47.5)(25,25){2}{6}
\Line(65,25)(75,2.5)
\Photon(65,25)(75,47.5){-2}{6}
\Vertex(25,25){2}

\Line(115,2.5)(125,25)
\Photon(115,47.5)(125,25){2}{6}
\Line(165,25)(175,2.5)
\Photon(165,25)(175,47.5){-2}{6}
\Line(125,25)(165,25)
\PhotonArc(145,18)(20,22,158){2}{7.5}
\PhotonArc(145,18)(18.8,22,158){2}{7.5}
\Vertex(125.5,25){2}
\Vertex(164.5,25){2}

\Line(215,2.5)(225,25)
\Photon(215,47.5)(225,25){2}{6}
\Line(265,25)(275,2.5)
\Photon(265,25)(275,47.5){-2}{6}
\PhotonArc(255,25)(8,-90,270){2}{9}
\PhotonArc(255,25)(6.8,-90,270){2}{9}
\Line(225,25)(245,25)
\Line(252,25)(265,25)
\Vertex(265,25){2}

\end{picture}
\end{center}
Similarly, we find

\begin{center}
\begin{picture}(0,110)(160,-60)

\Text(-60,25)[l]{(b) + (c) $\to\  \lambda$}  
\Text(85,25)[l]{$-\ \lambda$}  
\Text(185,25)[l]{$+\ \lambda$}
\Text(285,25)[l]{$+\ \lambda$}

\PhotonArc(35,25)(8,90,450){2}{9}
\PhotonArc(35,25)(6.8,90,450){2}{9}
\Line(25,25)(38,25)
\Line(45,25)(65,25)
\Line(15,2.5)(25,25)
\Photon(15,47.5)(25,25){2}{6}
\Line(65,25)(75,2.5)
\Photon(65,25)(75,47.5){-2}{6}
\Vertex(25,25){2}

\Line(115,2.5)(125,25)
\Photon(115,47.5)(125,25){2}{6}
\Line(165,25)(175,2.5)
\Photon(165,25)(175,47.5){-2}{6}
\Line(125,25)(165,25)
\PhotonArc(145,18)(20,22,158){2}{7.5}
\PhotonArc(145,18)(18.8,22,158){2}{7.5}
\Vertex(125.5,25){2}
\Vertex(164.5,25){2}

\Line(215,2.5)(225,25)
\Photon(215,47.5)(225,25){2}{6}
\Line(265,25)(275,2.5)
\Photon(265,25)(275,47.5){-2}{6}
\PhotonArc(255,25)(8,-90,270){2}{9}
\PhotonArc(255,25)(6.8,-90,270){2}{9}
\Line(225,25)(245,25)
\Line(252,25)(265,25)
\Vertex(265,25){2}

\Line(315,2.5)(325,25)
\Photon(315,47.5)(325,25){2}{6}
\Line(365,25)(375,2.5)
\Photon(365,25)(375,47.5){-2}{6}
\Line(325,25)(365,25)
\PhotonArc(355,45)(35.6,216,295.5){2}{9}
\PhotonArc(355,45)(34.4,216,295.5){2}{9}
\Vertex(325,25){2}
\Text(354.7,10)[c]{\rotatebox{19}{\bf{\big /}}}
\Text(355.3,10)[c]{\rotatebox{19}{\bf{\big /}}}

\Text(-40,-35)[l]{(d) $\to\  -\lambda$}

\Photon(15,-12.5)(25,-35){2}{6}
\Line(15,-57.5)(25,-35)
\Photon(65,-35)(75,-12.5){-2}{6}
\Line(65,-35)(75,-57.5)
\Line(25,-35)(65,-35)
\PhotonArc(55,-15)(35.6,216,295.5){2}{9}
\PhotonArc(55,-15)(34.4,216,295.5){2}{9}
\Vertex(25,-35){2}
\Text(54.7,-50)[c]{\rotatebox{19}{\bf{\big /}}}
\Text(55.3,-50)[c]{\rotatebox{19}{\bf{\big /}}}

\end{picture}
\end{center}
Thus we see that the GFP dependent part 
of the box diagram (d) completely cancels against the last 
diagram appearing in the above equation, 
in such a way that, after adding everything up, we get

\begin{center}
\begin{picture}(0,50)(50,0)

\Text(-120,25)[l]{(a) + (b) + (c) + (d) $\to\ \frac{\D \lambda}{\D 2}$}  
\Text(85,25)[l]{$+\ \frac{\D \lambda}{\D 2}$}  

\PhotonArc(35,25)(8,90,450){2}{9}
\PhotonArc(35,25)(6.8,90,450){2}{9}
\Line(25,25)(38,25)
\Line(45,25)(65,25)
\Line(15,2.5)(25,25)
\Photon(15,47.5)(25,25){2}{6}
\Line(65,25)(75,2.5)
\Photon(65,25)(75,47.5){-2}{6}
\Vertex(25,25){2}

\Line(115,2.5)(125,25)
\Photon(115,47.5)(125,25){2}{6}
\Line(165,25)(175,2.5)
\Photon(165,25)(175,47.5){-2}{6}
\PhotonArc(155,25)(8,-90,270){2}{9}
\PhotonArc(155,25)(6.8,-90,270){2}{9}
\Line(125,25)(145,25)
\Line(152,25)(165,25)
\Vertex(165,25){2}

\end{picture}
\end{center}

The remaining tadpole-like contributions will actually cancel
against the GFP-dependent parts of the diagrams (e) and (f),
representing the renormalization of the
external legs. 
The renormalization constant reads 
\begin{equation}
Z_2^{1/2}=1+\frac12\delta Z_2+\dots
\label{rrel}
\end{equation}
where
 $\delta Z_2$ represents 
 the one loop counter-term in the perturbative expansion; 
in the on-shell renormalization scheme it is defined as 
\begin{equation}
\delta Z_2=\frac{\partial}{\partial\pslush}
\Sigma^{(1)}(p,\xi)\Bigg |_{\mbox{}_{\D \pslush\!=m}} =
\frac{\partial}{\partial\pslush} 
\left(\Sigma^{(1)}_F (p) +
\lambda  \Sigma^{(1)}_L (p)\right)\Bigg |_{\mbox{}_{\D \pslush\!=m}}
\label{renor}
\end{equation}
We next focus our attention on the $\Sigma^{(1)}_L$
part 
of Eq.(\ref{renor}). 
As can be seen from Eq.(\ref{SL}),
$\Sigma_L^{(1)}(p)$ is of the form
\begin{equation}
\Sigma_L^{(1)}(p)\equiv
\left(\pslush-m\right)g(\pslush)\left(\pslush-m\right)+
\left(\pslush-m\right)c,
\end{equation}
where $c$ is a momentum independent constant (see footnote 4); thus
\begin{equation}
\Sigma_L^{(1)}(p)\big |_{\mbox{}_{\D \pslush\!=m}}=0, \qquad \qquad 
\delta Z_{2L}= \lambda c.
\end{equation}
Diagrammatically then, Eq.(\ref{GR}) implies that 

\begin{center}
\begin{picture}(0,30)(100,10)

\Line(0,25)(40,25)
\GOval(20,25)(2,4)(0){0}

\Text(45,24)[l]{$=$}

\Line(60,25)(100,25)
\GOval(80,25)(2,4)(0){0}
\Text(80,32)[c]{$\scriptstyle{\lambda=0}$}

\Text(105,25)[l]{$+\ \lambda$}

\Line(130,25)(170,25)
\PhotonArc(170,35)(8,180,540){2}{9}
\PhotonArc(170,35)(6.8,180,540){2}{9}
\Vertex(170,25){2}

\end{picture}
\end{center}
where the right-hand side is evaluated at $\pslush=m$.
Thus, recalling the extra $1/2$ factor appearing in Eq.(\ref{rrel}),
the GFP dependent part of the wave function renormalization diagrams 
is given by

\begin{center}
\begin{picture}(0,50)(50,0)

\Text(-70,25)[l]{(e) + (f) $\to\ -\frac{\D \lambda}{\D 2}$}  
\Text(85,25)[l]{$-\ \frac{\D \lambda}{\D 2}$}  

\PhotonArc(35,25)(8,90,450){2}{9}
\PhotonArc(35,25)(6.8,90,450){2}{9}
\Line(25,25)(38,25)
\Line(45,25)(65,25)
\Line(15,2.5)(25,25)
\Photon(15,47.5)(25,25){2}{6}
\Line(65,25)(75,2.5)
\Photon(65,25)(75,47.5){-2}{6}
\Vertex(25,25){2}

\Line(115,2.5)(125,25)
\Photon(115,47.5)(125,25){2}{6}
\Line(165,25)(175,2.5)
\Photon(165,25)(175,47.5){-2}{6}
\PhotonArc(155,25)(8,-90,270){2}{9}
\PhotonArc(155,25)(6.8,-90,270){2}{9}
\Line(125,25)(145,25)
\Line(152,25)(165,25)
\Vertex(165,25){2}

\end{picture}
\end{center}
which completes the proof that
the GFP independence of the $S$-matrix element is implemented
in the kinematically distinct way advocated above. Again, the
remaining pure GFP-independent fermion self-energy which 
survives is simply the one given in Eq.(\ref{GFPI}).

\section{The two-loop case.}

In the previous section we set up the 
general method for treating the GFP-dependent
contributions associated with the longitudinal
momenta  of the gauge bosons propagators inside Feynman graphs,
studied at the one-loop level, 
the special cancellation mechanism that this implies.
Thus we have been able to
define the one-loop GFP-independent fermion self-energy
$\widehat{\Sigma}^{(1)}(p)$.
In this section we proceed to the main subject 
of this paper, namely the two-loop
definition of the GFP-independent fermion self-energy.

At the two-loop level the presence of up to three gauge boson propagators in
the internal fermion loop will 
give rise to 
${\cal O}(\lambda^3)$,
${\cal O}(\lambda^2)$, and ${\cal O}(\lambda)$ GFP-dependent pieces.  
Of course, the gauge cancellations
proceed independently at each order in $\lambda$, a fact which 
facilitates the identification of diagrams (or parts of diagrams)
which can mix with each-other. 
Notice however that occasionally we will deviate from 
this elementary rule of thumb, in order to exploit  
the fact that one can identify massive
cancellations between different diagrams {\it before} separating out the
different orders in the GFP $\lambda$; this happens for example in the
Abelian-like part of the gauge cancellation, as we will see below.   
We will now proceed
to the detailed analysis of the two-loop  construction, showing first how it
works in the Abelian (QED) case, and concentrating then on the non-Abelian case
(QCD).

\subsection{Abelian case.}

\begin{figure}

\begin{picture}(100,100)(-10,0)

\Text(20,25)[c]{\footnotesize{(a)}}
\Text(135,25)[c]{\footnotesize{(b)}}
\Text(250,25)[c]{\footnotesize{(c)}}
\Text(365,25)[c]{\footnotesize{(d)}}

\Text(20,65)[l]{\scriptsize{$k$}}
\Text(50,81)[c]{\scriptsize{$k+\ell$}}
\Text(78,65)[l]{\scriptsize{$k+\ell+r$}}

\Text(135,65)[l]{\scriptsize{$k$}}
\Text(165,81)[c]{\scriptsize{$k+\ell$}}
\Text(193,65)[l]{\scriptsize{$k+\ell+r$}}

\Text(250,65)[l]{\scriptsize{$k$}}
\Text(308,65)[l]{\scriptsize{$k+r$}}
\Text(223,35)[l]{\scriptsize{$k-\ell-q$}}

\Text(358,70)[l]{\scriptsize{$k-\ell$}}
\Text(423,65)[l]{\scriptsize{$k+r$}}
\Text(352,35)[l]{\scriptsize{$k-q$}}

\CArc(20,50)(5,0,360)
\Line(18,52)(22,48)
\Line(22,52)(18,48)
\CArc(50,50)(25,0,360)
\CArc(80,50)(5,0,360)
\Line(78,52)(82,48)
\Line(82,52)(78,48)
\Gluon(40,27)(40,73){2}{9}
\Gluon(60,27)(60,73){2}{9}

\CArc(135,50)(5,0,360)
\Line(133,52)(137,48)
\Line(137,52)(133,48)
\CArc(165,50)(25,0,360)
\CArc(195,50)(5,0,360)
\Line(193,52)(197,48)
\Line(197,52)(193,48)
\Gluon(149,31)(179,71){2}{9}
\Gluon(180,30)(167,48){2}{4}
\Gluon(162,55)(150,70){2}{3}

\CArc(250,50)(5,0,360)
\Line(248,52)(252,48)
\Line(252,52)(248,48)
\CArc(280,50)(25,0,360)
\CArc(310,50)(5,0,360)
\Line(308,52)(312,48)
\Line(312,52)(308,48)
\Gluon(290,27)(290,73){2}{9}
\GlueArc(263,25)(19.2,0,112){2}{7.5}

\CArc(365,50)(5,0,360)
\Line(363,52)(367,48)
\Line(367,52)(363,48)
\CArc(395,50)(25,0,360)
\CArc(425,50)(5,0,360)
\Line(423,52)(427,48)
\Line(427,52)(423,48)
\Gluon(405,27)(405,73){2}{9}
\GlueArc(378,75)(19.2,248,360){2}{7.5}

\end{picture}

\begin{picture}(100,100)(-10,-25)

\Text(20,25)[c]{\footnotesize{(e)}}
\Text(135,25)[c]{\footnotesize{(f)}}
\Text(250,25)[c]{\footnotesize{(g)}}
\Text(365,25)[c]{\footnotesize{(h)}}

\Text(20,65)[l]{\scriptsize{$k$}}
\Text(73,70)[l]{\scriptsize{$k+\ell+r$}}
\Text(62,23)[l]{\scriptsize{$k+r-q$}}

\Text(135,65)[l]{\scriptsize{$k$}}
\Text(188,70)[l]{\scriptsize{$k+r$}}
\Text(177,23)[l]{\scriptsize{$k+\ell+r-q$}}

\Text(250,65)[l]{\scriptsize{$k$}}
\Text(308,65)[l]{\scriptsize{$k+r$}}
\Text(292,23)[l]{\scriptsize{$k+r-\ell-q$}}

\Text(365,65)[l]{\scriptsize{$k$}}
\Text(410,77)[l]{\scriptsize{$k+\ell+r$}}
\Text(407,23)[l]{\scriptsize{$k+r-q$}}

\CArc(20,50)(5,0,360)
\Line(18,52)(22,48)
\Line(22,52)(18,48)
\CArc(50,50)(25,0,360)
\CArc(80,50)(5,0,360)
\Line(78,52)(82,48)
\Line(82,52)(78,48)

\CArc(135,50)(5,0,360)
\Line(133,52)(137,48)
\Line(137,52)(133,48)
\CArc(165,50)(25,0,360)
\CArc(195,50)(5,0,360)
\Line(193,52)(197,48)
\Line(197,52)(193,48)

\CArc(250,50)(5,0,360)
\Line(248,52)(252,48)
\Line(252,52)(248,48)
\CArc(280,50)(25,0,360)
\CArc(310,50)(5,0,360)
\Line(308,52)(312,48)
\Line(312,52)(308,48)

\CArc(365,50)(5,0,360)
\Line(363,52)(367,48)
\Line(367,52)(363,48)
\CArc(395,50)(25,0,360)
\CArc(425,50)(5,0,360)
\Line(423,52)(427,48)
\Line(427,52)(423,48)

\Gluon(40,27)(40,73){2}{9}
\GlueArc(67,75)(19.2,180,292){2}{7.5}

\Gluon(155,27)(155,73){2}{9}
\GlueArc(182,25)(19.2,68,180){2}{7.5}

\Gluon(280,25)(280,39){2}{3}
\Gluon(280,48)(280,75){2}{6}
\GlueArc(280,25)(19.2,22.5,157.5){2}{9}

\Gluon(395,61)(395,75){2}{3}
\Gluon(395,25)(395,52){2}{6}
\GlueArc(395,75)(19.2,202.5,337.5){2}{9}

\end{picture}

\begin{picture}(100,100)(-10,-50)

\Text(20,25)[c]{\footnotesize{(i)}}
\Text(135,25)[c]{\footnotesize{(l)}}
\Text(250,25)[c]{\footnotesize{(m)}}
\Text(365,25)[c]{\footnotesize{(n)}}

\Text(50,81)[c]{\scriptsize{$k$}}
\Text(73,34.5)[l]{\scriptsize{$k+\ell-q$}}
\Text(50,19)[c]{\scriptsize{$k+\ell+r-q$}}

\Text(165,19)[c]{\scriptsize{$k-q$}}
\Text(190,65.5)[l]{\scriptsize{$k+\ell$}}
\Text(165,81)[c]{\scriptsize{$k+\ell+r$}}

\Text(280,81)[c]{\scriptsize{$k$}}
\Text(306,35)[l]{\scriptsize{$k+r-q$}}
\Text(225,35)[l]{\scriptsize{$k+\ell-q$}}

\Text(395,20)[c]{\scriptsize{$k-q$}}
\Text(418,70)[l]{\scriptsize{$k+r$}}
\Text(355,70)[l]{\scriptsize{$k+\ell$}}

\CArc(20,50)(5,0,360)
\Line(18,52)(22,48)
\Line(22,52)(18,48)
\CArc(50,50)(25,0,360)
\CArc(80,50)(5,0,360)
\Line(78,52)(82,48)
\Line(82,52)(78,48)

\CArc(135,50)(5,0,360)
\Line(133,52)(137,48)
\Line(137,52)(133,48)
\CArc(165,50)(25,0,360)
\CArc(195,50)(5,0,360)
\Line(193,52)(197,48)
\Line(197,52)(193,48)

\CArc(250,50)(5,0,360)
\Line(248,52)(252,48)
\Line(252,52)(248,48)
\CArc(280,50)(25,0,360)
\CArc(310,50)(5,0,360)
\Line(308,52)(312,48)
\Line(312,52)(308,48)

\CArc(365,50)(5,0,360)
\Line(363,52)(367,48)
\Line(367,52)(363,48)
\CArc(395,50)(25,0,360)
\CArc(425,50)(5,0,360)
\Line(423,52)(427,48)
\Line(427,52)(423,48)

\GlueArc(50,25)(18,21,159){2}{7.5}
\GlueArc(50,25)(26.2,31.7,148.5){2}{8.5}

\GlueArc(165,75)(18,201,339){2}{7.5}
\GlueArc(165,75)(26.2,211.7,328){2}{8.5}

\GlueArc(265,25)(18,0.5,117.5){2}{7}
\GlueArc(295,25)(19.2,161,179){2}{1.5}
\GlueArc(295,25)(19.2,63,135){2}{4.5}

\GlueArc(410,75)(18,180.5,297.5){2}{7}
\GlueArc(380,75)(19.2,339,359){2}{1.5}
\GlueArc(380,75)(19.2,243,315){2}{4.5}

\end{picture}

\begin{picture}(100,100)(-10,-75)

\Text(20,25)[c]{\footnotesize{(o)}}
\Text(135,25)[c]{\footnotesize{(p)}}
\Text(250,25)[c]{\footnotesize{(q)}}
\Text(365,25)[c]{\footnotesize{(r)}}

\Text(50,81)[c]{\scriptsize{$k$}}
\Text(73,34.5)[l]{\scriptsize{$k+r-q$}}
\Text(-5,34.5)[l]{\scriptsize{$k+\ell-q$}}

\Text(165,19)[c]{\scriptsize{$k-q$}}
\Text(190,65.5)[l]{\scriptsize{$k+r$}}
\Text(123,65.5)[l]{\scriptsize{$k+\ell$}}

\Text(250,60)[l]{\scriptsize{$k$}}
\Text(280,81)[c]{\scriptsize{$k-\ell$}}
\Text(280,19)[c]{\scriptsize{$k+r-q$}}

\Text(364,60)[l]{\scriptsize{$k$}}
\Text(395,81)[c]{\scriptsize{$k+r$}}
\Text(423,60)[l]{\scriptsize{$k+\ell$}}

\CArc(20,50)(5,0,360)
\Line(18,52)(22,48)
\Line(22,52)(18,48)
\CArc(50,50)(25,0,360)
\CArc(80,50)(5,0,360)
\Line(78,52)(82,48)
\Line(82,52)(78,48)

\CArc(135,50)(5,0,360)
\Line(133,52)(137,48)
\Line(137,52)(133,48)
\CArc(165,50)(25,0,360)
\CArc(195,50)(5,0,360)
\Line(193,52)(197,48)
\Line(197,52)(193,48)

\CArc(250,50)(5,0,360)
\Line(248,52)(252,48)
\Line(252,52)(248,48)
\CArc(280,50)(25,0,360)
\CArc(310,50)(5,0,360)
\Line(308,52)(312,48)
\Line(312,52)(308,48)

\CArc(365,50)(5,0,360)
\Line(363,52)(367,48)
\Line(367,52)(363,48)
\CArc(395,50)(25,0,360)
\CArc(425,50)(5,0,360)
\Line(423,52)(427,48)
\Line(427,52)(423,48)

\GlueArc(30,30)(14,342.5,467.5){2}{6}
\GlueArc(70,30)(14,74,196){2}{6}

\GlueArc(185,70)(14,522.5,647.5){2}{6}
\GlueArc(145,70)(14,254,376.5){2}{6}

\GlueArc(280,25)(19.2,22.5,157.5){2}{9}
\GlueArc(280,75)(19.2,202.5,337.5){2}{9}

\Gluon(395,25)(395,50){2}{6}
\Gluon(375,65)(395,50){-2}{5}
\Gluon(395,50)(415,65){-2}{5}
\GCirc(395,50){1}{1}

\end{picture}

\begin{picture}(100,100)(-10,-100)

\Text(20,25)[c]{\footnotesize{(s)}}
\Text(135,25)[c]{\footnotesize{(t)}}
\Text(250,25)[c]{\footnotesize{(u)}}
\Text(365,25)[c]{\footnotesize{(w)}}

\Text(7,40)[l]{\scriptsize{$k-q$}}
\Text(50,19)[c]{\scriptsize{$k-r-q$}}
\Text(75,65)[l]{\scriptsize{$k+\ell-r$}}

\Text(165,81)[c]{\scriptsize{$k$}}
\Text(183,25)[l]{\scriptsize{$k-\ell-q$}}
\Text(135,18)[l]{\scriptsize{$k+r-\ell-q$}}

\Text(280,19)[c]{\scriptsize{$k-q$}}
\Text(299,74)[l]{\scriptsize{$k+\ell$}}
\Text(248,74)[l]{\scriptsize{$k+r$}}

\Text(366,65)[l]{\scriptsize{$k$}}
\Text(420,65)[l]{\scriptsize{$k-\ell$}}
\Text(410,50)[l]{\scriptsize{$r$}}
\put(377,40){\rotateleft{\scriptsize{$r-\ell$}}}

\Line(145,21)(150,28)

\CArc(20,50)(5,0,360)
\Line(18,52)(22,48)
\Line(22,52)(18,48)
\CArc(50,50)(25,0,360)
\CArc(80,50)(5,0,360)
\Line(78,52)(82,48)
\Line(82,52)(78,48)

\CArc(135,50)(5,0,360)
\Line(133,52)(137,48)
\Line(137,52)(133,48)
\CArc(165,50)(25,0,360)
\CArc(195,50)(5,0,360)
\Line(193,52)(197,48)
\Line(197,52)(193,48)

\CArc(250,50)(5,0,360)
\Line(248,52)(252,48)
\Line(252,52)(248,48)
\CArc(280,50)(25,0,360)
\CArc(310,50)(5,0,360)
\Line(308,52)(312,48)
\Line(312,52)(308,48)

\CArc(365,50)(5,0,360)
\Line(363,52)(367,48)
\Line(367,52)(363,48)
\CArc(395,50)(25,0,360)
\CArc(425,50)(5,0,360)
\Line(423,52)(427,48)
\Line(427,52)(423,48)

\Gluon(50,75)(50,50){2}{5}
\Gluon(30,35)(50,50){2}{5}
\Gluon(50,50)(70,35){2}{5}
\GCirc(50,50){1}{1}

\Gluon(165,25)(165,42.5){2}{3}
\GlueArc(165,25)(19.2,22.5,157.5){2}{9}
\GCirc(165,43.2){1}{1}

\GlueArc(280,75)(19.2,202.5,337.5){2}{9}
\Gluon(280,75)(280,57.5){-2}{3}
\GCirc(280,57){1}{1}

\GlueArc(395,50)(9.5,90,270){2}{5}
\GlueArc(395,50)(9.5,270,450){2}{5}
\Gluon(395,25)(395,40){2}{3}
\Gluon(395,75)(395,62){-2}{3}
\GCirc(395,41){1}{1}
\GCirc(395,61){1}{1}

\end{picture}

\begin{picture}(100,100)(-10,-125)

\Text(20,25)[c]{\footnotesize{(x)}}
\Text(135,25)[c]{\footnotesize{(y)}}
\Text(250,25)[c]{\footnotesize{(z$_1$)}}
\Text(365,25)[c]{\footnotesize{(z$_2$)}}

\Text(50,81)[c]{\scriptsize{$k$}}
\Text(50,19)[c]{\scriptsize{$k+\ell-q$}}
\Text(50,60)[c]{\scriptsize{$r$}}
\Text(50,31)[c]{\scriptsize{$r-\ell$}}

\Text(165,81)[c]{\scriptsize{$k-\ell$}}
\Text(165,19)[c]{\scriptsize{$k-q$}}
\Text(165,70)[c]{\scriptsize{$r$}}
\Text(165,40)[c]{\scriptsize{$r-\ell$}}

\Text(251,65)[l]{\scriptsize{$k$}}
\Text(305,65)[l]{\scriptsize{$k-\ell$}}
\Text(294,50)[l]{\scriptsize{$r$}}
\put(263,40){\rotateleft{\scriptsize{$r-\ell$}}}

\Text(395,81)[c]{\scriptsize{$k$}}
\Text(395,19)[c]{\scriptsize{$k+\ell-q$}}
\Text(395,60)[c]{\scriptsize{$r$}}
\Text(395,31)[c]{\scriptsize{$r-\ell$}}

\CArc(20,50)(5,0,360)
\Line(18,52)(22,48)
\Line(22,52)(18,48)
\CArc(50,50)(25,0,360)
\CArc(80,50)(5,0,360)
\Line(78,52)(82,48)
\Line(82,52)(78,48)

\CArc(135,50)(5,0,360)
\Line(133,52)(137,48)
\Line(137,52)(133,48)
\CArc(165,50)(25,0,360)
\CArc(195,50)(5,0,360)
\Line(193,52)(197,48)
\Line(197,52)(193,48)

\CArc(250,50)(5,0,360)
\Line(248,52)(252,48)
\Line(252,52)(248,48)
\CArc(280,50)(25,0,360)
\CArc(310,50)(5,0,360)
\Line(308,52)(312,48)
\Line(312,52)(308,48)

\CArc(365,50)(5,0,360)
\Line(363,52)(367,48)
\Line(367,52)(363,48)
\CArc(395,50)(25,0,360)
\CArc(425,50)(5,0,360)
\Line(423,52)(427,48)
\Line(427,52)(423,48)

\GlueArc(50,45)(9.5,0,180){2}{4}
\GlueArc(50,45)(9.5,180,360){2}{4}
\GlueArc(59.5,25)(20,50,90){2}{2}
\GlueArc(40.5,25)(20,90,130){2}{2}
\GCirc(40.5,45){1}{1}
\GCirc(59.5,45){1}{1}

\GlueArc(165,55)(9.5,0,180){2}{4}
\GlueArc(165,55)(9.5,180,360){2}{4}
\GlueArc(174.5,75)(20,270,310){2}{2}
\GlueArc(155.5,75)(20,230,270){2}{2}
\GCirc(155.5,55){1}{1}
\GCirc(174.5,55){1}{1}

\DashCArc(280,50)(11,0,360){1}
\Gluon(280,25)(280,39){2}{3}
\Gluon(280,75)(280,61){-2}{3}

\DashCArc(395,45)(11,0,360){1}
\GlueArc(406,25)(20,52,90){2}{2}
\GlueArc(384,25)(20,90,128){2}{2}

\end{picture}

\begin{picture}(100,0)(-10,-60)

\Text(20,25)[c]{\footnotesize{(z$_3$)}}
\Text(135,25)[c]{\footnotesize{(z$_4$)}}
\Text(250,25)[c]{\footnotesize{(z$_5$)}}
\Text(365,25)[c]{\footnotesize{(z$_6$)}}

\Text(50,81)[c]{\scriptsize{$k-\ell$}}
\Text(50,19)[c]{\scriptsize{$k-q$}}
\Text(50,70)[c]{\scriptsize{$r$}}
\Text(50,40)[c]{\scriptsize{$r-\ell$}}

\Text(136,65)[l]{\scriptsize{$k$}}
\Text(190,65)[l]{\scriptsize{$k-\ell$}}
\Text(179,50)[l]{\scriptsize{$r$}}
\put(147,40){\rotateleft{\scriptsize{$r-\ell$}}}

\Text(280,81)[c]{\scriptsize{$k$}}
\Text(280,19)[c]{\scriptsize{$k+\ell-q$}}
\Text(280,60)[c]{\scriptsize{$r$}}
\Text(280,31)[c]{\scriptsize{$r-\ell$}}

\Text(395,81)[c]{\scriptsize{$k-\ell$}}
\Text(395,19)[c]{\scriptsize{$k-q$}}
\Text(395,70)[c]{\scriptsize{$r$}}
\Text(395,40)[c]{\scriptsize{$r-\ell$}}

\CArc(20,50)(5,0,360)
\Line(18,52)(22,48)
\Line(22,52)(18,48)
\CArc(50,50)(25,0,360)
\CArc(80,50)(5,0,360)
\Line(78,52)(82,48)
\Line(82,52)(78,48)

\CArc(135,50)(5,0,360)
\Line(133,52)(137,48)
\Line(137,52)(133,48)
\CArc(165,50)(25,0,360)
\CArc(195,50)(5,0,360)
\Line(193,52)(197,48)
\Line(197,52)(193,48)

\CArc(250,50)(5,0,360)
\Line(248,52)(252,48)
\Line(252,52)(248,48)
\CArc(280,50)(25,0,360)
\CArc(310,50)(5,0,360)
\Line(308,52)(312,48)
\Line(312,52)(308,48)

\CArc(365,50)(5,0,360)
\Line(363,52)(367,48)
\Line(367,52)(363,48)
\CArc(395,50)(25,0,360)
\CArc(425,50)(5,0,360)
\Line(423,52)(427,48)
\Line(427,52)(423,48)

\DashCArc(50,55)(11,0,360){1}
\GlueArc(61,75)(20,270,308){2}{2}
\GlueArc(39,75)(20,232,270){2}{2}

\CArc(165,50)(11,0,360)
\Gluon(165,25)(165,39){2}{3}
\Gluon(165,75)(165,61){-2}{3}

\CArc(280,45)(11,0,360)
\GlueArc(291,25)(20,52,90){2}{2}
\GlueArc(269,25)(20,90,128){2}{2}

\CArc(395,55)(11,0,360)
\GlueArc(406,75)(20,270,308){2}{2}
\GlueArc(384,75)(20,232,270){2}{2}

\end{picture}

\vspace{-1.5 cm}
\caption{Two loops diagram contributing to the QCD fermion self-energy.}

\label{fig3}
\end{figure}
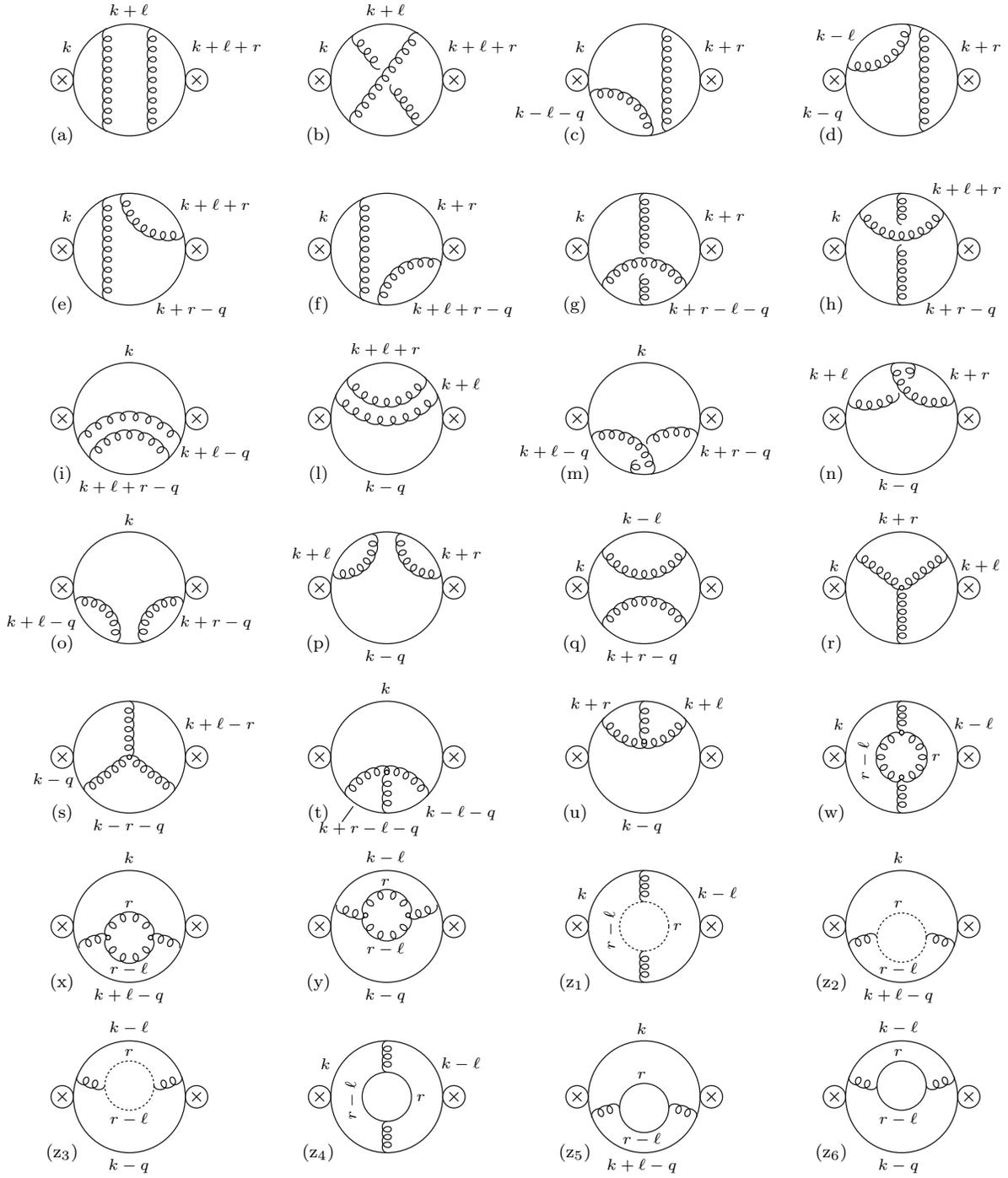

The diagrams contributing to this part of the amplitude are the ones denoted
(a),$\dots$,(q) in Fig.\ref{fig3}\footnote{Diagrams (z$_4$), (z$_5$) and
(z$_6$) can be put in the Feynman gauge right from the start, due to the
transversality of the fermionic sector of the photon/gluon propagator (see also
Sec.\ref{nac}).}. In this part of the cancellation 
one can carry out massive cancellations dealing with 
${\cal O}(\lambda^2)$ and ${\cal O}(\lambda)$ diagrams at the same time, 
by pinching with only one propagator, while letting the other one untouched.
We will now consider in detail a 
couple of these diagrams showing how the 
procedure
works in the two-loop case.\\ From the box diagram (a) we get for example the following  
equation

\begin{center}
\begin{picture}(0,120)(190,-45)

\Text(-28,50)[l]{(a) $\to\ \lambda$}

\CArc(20,50)(5,0,360)
\Line(18,52)(22,48)
\Line(22,52)(18,48)
\CArc(50,50)(25,0,360)
\CArc(80,50)(5,0,360)
\Line(78,52)(82,48)
\Line(82,52)(78,48)
\Photon(40,27)(40,73){-2}{9}
\Photon(41.5,27)(41.5,73){-2}{9}
\Gluon(60,27)(60,73){2.5}{9}
\Text(41,51.5)[c]{\rotatebox{-71}{\bf{\big /}}}
\Text(41,52.1)[c]{\rotatebox{-71}{\bf{\big /}}}
\Text(41,48.5)[c]{\rotatebox{-71}{\bf{\big /}}}
\Text(41,47.9)[c]{\rotatebox{-71}{\bf{\big /}}}

\Text(90,50)[l]{$=\  \lambda$}
\Text(190,50)[l]{$-\ \lambda$}
\Text(90,-15)[l]{$=\ \lambda$}
\Text(115,-40)[lb]{\footnotesize{$(\alpha)$}}
\Text(190,-15)[l]{$-\ \lambda$}
\Text(215,-40)[lb]{\footnotesize{$(\beta)$}}
\Text(290,-15)[l]{$-\ \lambda$}
\Text(315,-40)[lb]{\footnotesize{$(\gamma)$}}

\CArc(120,50)(5,0,360)
\Line(118,52)(122,48)
\Line(122,52)(118,48)
\CArc(150,50)(25,0,360)
\CArc(180,50)(5,0,360)
\Line(178,52)(182,48)
\Line(182,52)(178,48)
\Gluon(160,27)(160,73){2.5}{9}
\Photon(130,35)(158.5,73){-2}{9}
\Photon(129,36)(157.5,74){-2}{9}
\Vertex(160,73){2}
\Text(143,53.5)[c]{\rotatebox{65}{\bf{\big /}}}
\Text(143.4,53.9)[c]{\rotatebox{65}{\bf{\big /}}}

\CArc(220,50)(5,0,360)
\Line(218,52)(222,48)
\Line(222,52)(218,48)
\CArc(250,50)(25,0,360)
\CArc(280,50)(5,0,360)
\Line(278,52)(282,48)
\Line(282,52)(278,48)
\Gluon(260,27)(260,73){2.5}{9}
\PhotonArc(225,27)(23.7,356,90){2}{7.5}
\PhotonArc(225,27)(22.3,356,90){2}{7.5}
\Vertex(225,50){2}
\Text(245,41)[c]{\rotatebox{-25}{\bf{\big /}}}
\Text(244.6,41.4)[c]{\rotatebox{-25}{\bf{\big /}}}

\CArc(120,-15)(5,0,360)
\Line(118,-17)(122,-13)
\Line(122,-17)(118,-13)
\CArc(150,-15)(25,0,360)
\CArc(180,-15)(5,0,360)
\Line(178,-17)(182,-13)
\Line(182,-17)(178,-13)
\GlueArc(96,-15.7)(61.2,337,384){-2.5}{9}
\PhotonArc(205,-15)(61.2,157,203.8){2}{9}
\PhotonArc(205,-15)(60,157,203.8){2}{9}
\Vertex(150.5,10){2}
\Vertex(150.5,-40){2}

\CArc(220,-15)(5,0,360)
\Line(218,-17)(222,-13)
\Line(222,-17)(218,-13)
\CArc(250,-15)(25,0,360)
\CArc(280,-15)(5,0,360)
\Line(278,-17)(282,-13)
\Line(282,-17)(278,-13)
\Gluon(260,-38)(260,8){2.5}{9}
\PhotonArc(225,20)(35,270,340){-2}{7.5}
\PhotonArc(225,20)(36.4,270,340){-2}{7.5}
\Vertex(259.7,8){2}
\Vertex(225,-15){2}

\CArc(320,-15)(5,0,360)
\Line(318,-17)(322,-13)
\Line(322,-17)(318,-13)
\CArc(350,-15)(25,0,360)
\CArc(380,-15)(5,0,360)
\Line(378,-17)(382,-13)
\Line(382,-17)(378,-13)
\Gluon(360,-38)(360,8){2.5}{9} 
\PhotonArc(325,-38)(23.7,356,90){2}{7.5}
\PhotonArc(325,-38)(22.3,356,90){2}{7.5}
\Vertex(325,-15){2}
\Text(345,-24)[c]{\rotatebox{-25}{\bf{\big /}}}
\Text(344.6,-23.6)[c]{\rotatebox{-25}{\bf{\big /}}}

\end{picture}
\end{center}

Let us concentrate on the three topologies shown above. It is clear
that topology $(\alpha)$ can be generated {\it only} from diagram (b), and
so it must cancel against it. 
Topologies $(\beta)$ and $(\gamma)$ will be also
generated from diagrams (d) and (c);  
however these last two diagrams do not need to 
cancel in full against the one coming from (d) and (c), 
because, as we will see, topologically analogous contributions
will also appear from other diagrams.

As a second example, we consider the box diagram (b), which gives

\begin{center}
\begin{picture}(0,120)(190,-45)

\Text(-28,50)[l]{(b) $\to\ \lambda$}

\CArc(20,50)(5,0,360)
\Line(18,52)(22,48)
\Line(22,52)(18,48)
\CArc(50,50)(25,0,360)
\CArc(80,50)(5,0,360)
\Line(78,52)(82,48)
\Line(82,52)(78,48)
\Gluon(34.5,30.5)(64.5,70.5){-2.5}{9}
\Photon(65,30)(52,48){-2}{4}
\Photon(66,31)(53,49){-2}{4}
\Photon(47,55)(35,70){-2}{4}
\Photon(48,56)(36,71){-2}{4}
\Text(59,40)[c]{\rotatebox{-30}{\bf{\big /}}}
\Text(58.6,40.4)[c]{\rotatebox{-30}{\bf{\big /}}}
\Text(61,38)[c]{\rotatebox{-30}{\bf{\big /}}}
\Text(61.4,37.6)[c]{\rotatebox{-30}{\bf{\big /}}}

\Text(90,50)[l]{$=\  \lambda$}
\Text(190,50)[l]{$-\ \lambda$}
\Text(90,-15)[l]{$=\ \lambda$}
\Text(115,-40)[lb]{\footnotesize{$(\delta)$}}
\Text(190,-15)[l]{$-\ \lambda$}
\Text(215,-40)[lb]{\footnotesize{$(\varepsilon)$}}
\Text(290,-15)[l]{$-\ \lambda$}
\Text(315,-40)[lb]{\footnotesize{$(\xi)$}}

\CArc(120,50)(5,0,360)
\Line(118,52)(122,48)
\Line(122,52)(118,48)
\CArc(150,50)(25,0,360)
\CArc(180,50)(5,0,360)
\Line(178,52)(182,48)
\Line(182,52)(178,48)
\Gluon(135,30)(165,70){-2.5}{9}
\Photon(166.5,70)(166.5,30){-2}{7}
\Photon(165,70)(165,30){-2}{7}
\Vertex(165,70){2}
\Text(166,45)[c]{\rotatebox{-71}{\bf{\big /}}}
\Text(166,45.6)[c]{\rotatebox{-71}{\bf{\big /}}}

\CArc(220,50)(5,0,360)
\Line(218,52)(222,48)
\Line(222,52)(218,48)
\CArc(250,50)(25,0,360)
\CArc(280,50)(5,0,360)
\Line(278,52)(282,48)
\Line(282,52)(278,48)
\Gluon(235,30)(265,70){-2.5}{9}
\PhotonArc(225,15)(35.7,19,48){-2}{3}
\PhotonArc(225,15)(34.3,19,48){-2}{3}
\PhotonArc(225,15)(35.7,63,90){-2}{3}
\PhotonArc(225,15)(34.3,63,90){-2}{3}
\Vertex(225,50){2}
\Text(255,35)[c]{\rotatebox{-25}{\bf{\big /}}}
\Text(254.6,35.4)[c]{\rotatebox{-25}{\bf{\big /}}}

\CArc(120,-15)(5,0,360)
\Line(118,-17)(122,-13)
\Line(122,-17)(118,-13)
\CArc(150,-15)(25,0,360)
\CArc(180,-15)(5,0,360)
\Line(178,-17)(182,-13)
\Line(182,-17)(178,-13)
\PhotonArc(175,20)(35,200,270){2}{7.5}
\PhotonArc(175,20)(36.4,200,270){2}{7.5}
\Gluon(140,-38)(140,9){-2.5}{9}
\Vertex(140.3,8){2}
\Vertex(175,-15){2}

\CArc(220,-15)(5,0,360)
\Line(218,-17)(222,-13)
\Line(222,-17)(218,-13)
\CArc(250,-15)(25,0,360)
\CArc(280,-15)(5,0,360)
\Line(278,-17)(282,-13)
\Line(282,-17)(278,-13)
\GlueArc(305,-15)(61.2,157,204){-2.5}{9}
\PhotonArc(195,-15)(60,337,383.8){-2}{9}
\PhotonArc(195,-15)(61.2,337,383.8){-2}{9}
\Vertex(250.5,10){2}
\Vertex(250.5,-40){2}

\CArc(320,-15)(5,0,360)
\Line(318,-17)(322,-13)
\Line(322,-17)(318,-13)
\CArc(350,-15)(25,0,360)
\CArc(380,-15)(5,0,360)
\Line(378,-17)(382,-13)
\Line(382,-17)(378,-13)
\Gluon(335,-35)(365,5){-2.5}{9}
\PhotonArc(325,-50)(35.7,19,48){-2}{3}
\PhotonArc(325,-50)(34.3,19,48){-2}{3}
\PhotonArc(325,-50)(35.7,63,90){-2}{3}
\PhotonArc(325,-50)(34.3,63,90){-2}{3}
\Vertex(325,-15){2}
\Text(355,-30)[c]{\rotatebox{-25}{\bf{\big /}}}
\Text(355.4,-30.4)[c]{\rotatebox{-25}{\bf{\big /}}}

\end{picture}
\end{center}

As expected, topology $(\varepsilon)$ cancels against the diagram $(\alpha)$ of
the previous equation (after exchanging the order of the internal lines and
relabelling the internal momenta), but we have generated also two new
topologies: $(\delta)$ which will be generated as well by diagram (e), and
$(\zeta)$ which will be generated by (g). Letting untouched the vertical
propagator in diagrams (c),$\dots$,(h), we then arrive at the following equation

\begin{center}
\begin{picture}(0,100)(370,5)

\Text(385,95)[c]{(a)+(b)+(c)+(d)+(e)+(f)+(g)+(h) $\to$}

\Text(190,50)[l]{$\lambda$}
\Text(275,50)[l]{$-\ \lambda$}
\Text(375,50)[l]{$+\ \lambda$}
\Text(475,50)[l]{$-\ \lambda$}

\Text(200,25)[lb]{\footnotesize{$(\eta)$}}
\Text(300,25)[lb]{\footnotesize{$(\theta)$}}
\Text(400,25)[lb]{\footnotesize{$(\mu)$}}
\Text(500,25)[lb]{\footnotesize{$(\nu)$}}

\CArc(205,50)(5,0,360)
\Line(203,52)(207,48)
\Line(207,52)(203,48)
\CArc(235,50)(25,0,360)
\CArc(265,50)(5,0,360)
\Line(263,52)(267,48)
\Line(267,52)(263,48)
\Gluon(235.2,25)(235.2,75){2.5}{9}
\Photon(210,50.6)(232,50.6){2}{3.5}
\Photon(210,49.4)(232,49.4){2}{3.5}
\Photon(239,50.6)(260,50.6){-2}{3.5}
\Photon(239,49.4)(260,49.4){-2}{3.5}
\Vertex(210,50){2}
\Vertex(260,50){2}

\CArc(305,50)(5,0,360)
\Line(303,52)(307,48)
\Line(307,52)(303,48)
\CArc(335,50)(25,0,360)
\CArc(365,50)(5,0,360)
\Line(363,52)(367,48)
\Line(367,52)(363,48)
\Gluon(345,27.7)(345,72.3){2.5}{9}
\PhotonArc(320,50)(6.8,90,450){2}{9}
\PhotonArc(320,50)(8,90,450){2}{9}
\Vertex(310,50){2}

\CArc(405,50)(5,0,360)
\Line(403,52)(407,48)
\Line(407,52)(403,48)
\CArc(435,50)(25,0,360)
\CArc(465,50)(5,0,360)
\Line(463,52)(467,48)
\Line(467,52)(463,48)
\PhotonArc(460,85)(35,200,270){2}{7.5}
\PhotonArc(460,85)(36.4,200,270){2}{7.5}
\Gluon(425,27)(425,74){-2.5}{9}
\Vertex(425.3,73){2}
\Vertex(460,50){2}

\CArc(505,50)(5,0,360)
\Line(503,52)(507,48)
\Line(507,52)(503,48)
\CArc(535,50)(25,0,360)
\CArc(565,50)(5,0,360)
\Line(563,52)(567,48)
\Line(567,52)(563,48)
\Gluon(545,27)(545,73){2.5}{9}
\PhotonArc(510,85)(35,270,340){-2}{7.5}
\PhotonArc(510,85)(36.4,270,340){-2}{7.5}
\Vertex(544.7,73){2}
\Vertex(510,50){2}

\end{picture}
\end{center}

Actually the last two diagrams add up to zero. To see this,
we can use their $\mu\leftrightarrow\nu$ symmetry, to observe that
(the integral measures are suppressed)
\begin{equation}
(\mu)\sim\lambda{\rm Tr}\,\left[\gamma^\mu\frac1{\ka-m}\gamma^\rho
\frac1{\ka+\erre+\elle-m}\gamma^\nu\frac1{\ka+\erre-\qu-m}
\gamma^\sigma\frac1{\ka-\qu-m}\right]\Delta_{\rho\sigma}(r^2)
\frac1{\ell^4},
\end{equation}
while
\begin{eqnarray}
(\nu) &\sim& 
-\lambda{\rm Tr}\,\left[\gamma^\nu\frac1{\ka+\elle+\erre-m}
\gamma^\rho\frac1{\ka-m}\gamma^\mu\frac1{\ka+\elle-\qu-m}
\gamma^\sigma\frac1{\ka+\elle+\erre-\qu-m}\right]
\Delta_{\rho\sigma}(r^2)\frac1{\ell^4} \nonumber \\
& \stackrel{k+\ell+r\to k}{=} &
-\lambda{\rm Tr}\,\left[\gamma^\nu\frac1{\ka-m}
\gamma^\rho\frac1{\ka-\elle-\erre-m}\gamma^\mu
\frac1{\ka-\erre-\qu-m}\gamma^\sigma\frac1{\ka-\qu-m}\right]
\Delta_{\rho\sigma}(r^2)\frac1{\ell^4} \nonumber \\
& \stackrel{\stackrel{\stackrel{\scriptstyle{\ell\to -\ell}}{\scriptstyle{{r\to
-r}}}}{\mu\leftrightarrow\nu}}{=} &
-\lambda{\rm Tr}\,\left[\gamma^\mu\frac1{\ka-m}
\gamma^\rho\frac1{\ka+\erre+\elle-m}\gamma^\nu
\frac1{\ka+\erre-\qu-m}\gamma^\sigma\frac1{\ka-\qu-m}\right]
\Delta_{\rho\sigma}(r^2)\frac1{\ell^4}.
\end{eqnarray}

Moreover,
from the remaining diagrams we get 

\begin{center}
\begin{picture}(0,70)(120,20)

\Text(-60,50)[l]{(l)+(n)+(p)+(i)+(m)+(o) $\to\ 2\lambda$}
\Text(100,25)[lb]{\footnotesize{$(\xi)$}}
\Text(200,25)[lb]{\footnotesize{$(\rho)$}}

\CArc(105,50)(5,0,360)
\Line(103,52)(107,48)
\Line(107,52)(103,48)
\CArc(135,50)(25,0,360)
\CArc(165,50)(5,0,360)
\Line(163,52)(167,48)
\Line(167,52)(163,48)
\Photon(110,49.4)(160,49.4){2}{9}
\Photon(110,50.6)(160,50.6){2}{9}
\Vertex(110,50){2}
\Vertex(160,50){2}
\GlueArc(135,75)(17.2,201,339){2.5}{7.5}

\Text(175,50)[l]{$-\ 2\lambda$}

\CArc(205,50)(5,0,360)
\Line(203,52)(207,48)
\Line(207,52)(203,48)
\CArc(235,50)(25,0,360)
\CArc(265,50)(5,0,360)
\Line(263,52)(267,48)
\Line(267,52)(263,48)
\GlueArc(234,70)(13,6,184){2.5}{7.5}
\PhotonArc(260,105)(56.2,221,270){2}{9.5}
\PhotonArc(260,105)(55,221,270){2}{9.5}
\Vertex(260,50){2}
\Vertex(219,69.2){2}

\end{picture}
\end{center}
After having identified these cancellations of mixed order
( ${\cal O}(\lambda)$ and ${\cal O}(\lambda^2)$) we next 
find it convenient to pursue the remaining cancellations
treating separately the ${\cal O}(\lambda)$ and
${\cal O}(\lambda^2)$ GFP-dependent amplitudes.

\subsubsection{The ${\cal O}(\lambda^2)$ cancellation.}

For this case one has to replace the propagator $\begin{picture}(10,5)(0,0)
\SetScale{0.8}
\Gluon(5,3)(40,3){2}{6}
\end{picture}\qquad \ \ $ appearing in the diagrams $(\eta),\dots,(\rho)$ 
with the propagator $\begin{picture}(10,5)(0,0)
\SetScale{0.8}
\Photon(5,4)(40,4){2}{6}
\Photon(5,2.5)(40,2.5){2}{6}
\Text(19,3.25)[c]{\rotatebox{19}{\bf{\big /}}}
\Text(19.4,3.25)[c]{\rotatebox{19}{\bf{\big /}}}
\Text(16,3.25)[c]{\rotatebox{19}{\bf{\big /}}}
\Text(15.6,3.25)[c]{\rotatebox{19}{\bf{\big /}}}
\end{picture}\qquad \ $. It is then fairly easy to show that

\begin{center}
\begin{picture}(0,60)(160,20)

\Text(50,50)[l]{$(\xi)\to\ 2\lambda^2$}

\CArc(105,50)(5,0,360)
\Line(103,52)(107,48)
\Line(107,52)(103,48)
\CArc(135,50)(25,0,360)
\CArc(165,50)(5,0,360)
\Line(163,52)(167,48)
\Line(167,52)(163,48)
\PhotonArc(136,80)(40,230,307){2}{9}
\PhotonArc(136,80)(38.8,230,307){2}{9}
\PhotonArc(134,20)(40,410,487){2}{9}
\PhotonArc(134,20)(38.8,410,487){2}{9}
\Vertex(110,50){2}
\Vertex(160,50){2}

\Text(175,50)[l]{$-\  2\lambda^2$}

\CArc(210,50)(5,0,360)
\Line(208,52)(212,48)
\Line(212,52)(208,48)
\CArc(240,50)(25,0,360)
\CArc(270,50)(5,0,360)
\Line(268,52)(272,48)
\Line(272,52)(268,48)
\Photon(235,50.6)(265,50.6){2}{6}
\Photon(235,49.4)(265,49.4){2}{6}
\Photon(215,50.6)(228,50.6){2}{2.5}
\Photon(215,49.4)(228,49.4){2}{2.5}
\Vertex(215,50){2}
\Vertex(265,50){2}
\PhotonArc(225,50)(8,90,450){2}{9}
\PhotonArc(225,50)(6.8,90,450){2}{9}
\Vertex(215,50){2}

\end{picture}
\end{center}
which implies 
\begin{equation}
(\xi)+(\eta)=0.
\end{equation}
Moreover, considering diagram ($\theta$) we find the result 

\begin{center}
\begin{picture}(0,130)(215,-50)

\Text(48,50)[l]{$(\theta)\to\ 2\lambda^2$}
\Text(175,50)[l]{$-\ \lambda^2$}
\Text(275,50)[l]{$-\ \lambda^2$}

\CArc(105,50)(5,0,360)
\Line(103,52)(107,48)
\Line(107,52)(103,48)
\CArc(135,50)(25,0,360)
\CArc(165,50)(5,0,360)
\Line(163,52)(167,48)
\Line(167,52)(163,48)
\Photon(130,50.6)(160,50.6){2}{6}
\Photon(130,49.4)(160,49.4){2}{6}
\Photon(110,50.6)(123,50.6){2}{2.5}
\Photon(110,49.4)(123,49.4){2}{2.5}
\Vertex(110,50){2}
\Vertex(160,50){2}
\PhotonArc(120,50)(8,90,450){2}{9}
\PhotonArc(120,50)(6.8,90,450){2}{9}
\Vertex(110,50){2}

\CArc(205,50)(5,0,360)
\Line(203,52)(207,48)
\Line(207,52)(203,48)
\CArc(235,50)(25,0,360)
\CArc(265,50)(5,0,360)
\Line(263,52)(267,48)
\Line(267,52)(263,48)
\PhotonArc(220,50)(8,90,450){2}{9}
\PhotonArc(220,50)(6.8,90,450){2}{9}
\Vertex(210 ,50){2}
\PhotonArc(250,50)(8,-90,270){2}{9}
\PhotonArc(250,50)(6.8,-90,270){2}{9}
\Vertex(260,50){2}

\CArc(305,50)(5,0,360)
\Line(303,52)(307,48)
\Line(307,52)(303,48)
\CArc(335,50)(25,0,360)
\CArc(365,50)(5,0,360)
\Line(363,52)(367,48)
\Line(367,52)(363,48)
\PhotonArc(318.7,50)(5.5,90,450){2}{5}
\PhotonArc(318.7,50)(4.3,90,450){2}{5}
\PhotonArc(324,50)(12,90,450){2}{9}
\PhotonArc(324,50)(10.8,90,450){2}{9}
\Vertex(310,50){2}

\Text(67,-15)[l]{$=\ 2\lambda^2$}

\CArc(105,-15)(5,0,360)
\Line(103,-17)(107,-13)
\Line(107,-17)(103,-13)
\CArc(135,-15)(25,0,360)
\CArc(165,-15)(5,0,360)
\Line(163,-17)(167,-13)
\Line(167,-17)(163,-13)
\Photon(130,-14.4)(160,-14.4){2}{6}
\Photon(130,-15.6)(160,-15.6){2}{6}
\Photon(110,-14.4)(123,-14.4){2}{2.5}
\Photon(110,-15.6)(123,-15.6){2}{2.5}
\Vertex(110,-15){2}
\Vertex(160,-15){2}
\PhotonArc(120,-15)(8,90,450){2}{9}
\PhotonArc(120,-15)(6.8,90,450){2}{9}
\Vertex(110,-15){2}

\Text(173.5,-15)[l]{$-\ 2\lambda^2$}

\CArc(205,-15)(5,0,360)
\Line(203,-17)(207,-13)
\Line(207,-17)(203,-13)
\CArc(235,-15)(25,0,360)
\CArc(265,-15)(5,0,360)
\Line(263,-17)(267,-13)
\Line(267,-17)(263,-13)
\PhotonArc(220,-15)(8,90,450){2}{9}
\PhotonArc(220,-15)(6.8,90,450){2}{9}
\Vertex(210 ,-15){2}
\PhotonArc(250,-15)(8,-90,270){2}{9}
\PhotonArc(250,-15)(6.8,-90,270){2}{9}
\Vertex(260,-15){2}

\end{picture}
\end{center}
The fact that the last two diagrams in the first line of the previous equality
can be added up, reflects 
the freedom of moving at will  the photon (gluon) tadpole-like
loops in a given pinched diagram. This can be done because such loops
represent scalar quantities --defined in Eq.(\ref{SL}) --with no 
interactions left  at the vertex. Notice that this freedom does not 
interfere with our
notion of unphysical vertices, since both diagrams, written in either way, 
are equally unphysical. We will often use this property in what follows.     

Adding to the above result the contribution coming from diagram (q), we then 
get

\begin{center}
\begin{picture}(0,60)(160,20)

\Text(35,50)[l]{$(\theta)+$(q)$\, \to\  \lambda^2$}

\CArc(105,50)(5,0,360)
\Line(103,52)(107,48)
\Line(107,52)(103,48)
\CArc(135,50)(25,0,360)
\CArc(165,50)(5,0,360)
\Line(163,52)(167,48)
\Line(167,52)(163,48)
\Vertex(110,50){2}
\Vertex(160,50){2}
\PhotonArc(136,80)(40,230,307){2}{9}
\PhotonArc(136,80)(38.8,230,307){2}{9}
\PhotonArc(134,20)(40,410,487){2}{9}
\PhotonArc(134,20)(38.8,410,487){2}{9}

\Text(175,50)[l]{$-\ \lambda^2$}

\CArc(205,50)(5,0,360)
\Line(203,52)(207,48)
\Line(207,52)(203,48)
\CArc(235,50)(25,0,360)
\CArc(265,50)(5,0,360)
\Line(263,52)(267,48)
\Line(267,52)(263,48)
\PhotonArc(220,50)(8,90,450){2}{9}
\PhotonArc(220,50)(6.8,90,450){2}{9}
\Vertex(210,50){2}
\PhotonArc(250,50)(8,-90,270){2}{9}
\PhotonArc(250,50)(6.8,-90,270){2}{9}
\Vertex(260,50){2}

\end{picture}
\end{center}
a combination that should then cancel completely with the diagram $(\rho)$,
which at ${\cal O}(\lambda^2)$ reads
\begin{center}
\begin{equation}
\begin{picture}(0,70)(110,15)

\Text(48,50)[l]{$(\rho)\to\ 2\lambda^2$}
\Text(138.5,71.5)[l]{\footnotesize{$r$}}
\Text(130,50)[l]{\footnotesize{$\ell$}}
\Text(153,75)[l]{\footnotesize{$k+r+\ell$}}
\Text(105,65)[l]{\footnotesize{$k$}}

\CArc(105,50)(5,0,360)
\Line(103,52)(107,48)
\Line(107,52)(103,48)
\CArc(135,50)(25,0,360)
\CArc(165,50)(5,0,360)
\Line(163,52)(167,48)
\Line(167,52)(163,48)
\PhotonArc(160,105)(55,221,270){2}{10}
\PhotonArc(160,105)(53.8,221,270){2}{10}
\PhotonArc(120,15)(55,401,450){-2}{10}
\PhotonArc(120,15)(53.8,401,450){-2}{10}
\Vertex(160,50){2}
\Vertex(119.8,69.8){2}
\Text(143.5,64)[c]{\rotatebox{-7}{\bf{\big /}}}
\Text(143.9,63.6)[c]{\rotatebox{-7}{\bf{\big /}}}

\end{picture}

\label{poi}
\end{equation}
\end{center}

In order to make this cancellation manifest, we need to change the
given topology by means of pinching internal fermion propagators. 
Of course, the only available momentum $r$ in diagram $(\rho)$  
cannot pinch directly,
due to the obvious kinematic mismatch. 
However, one has
 
\begin{center}
\begin{picture}(0,130)(160,-50)

\Text(48,50)[l]{$(\rho)\ =\ \lambda^2$}

\CArc(105,50)(5,0,360)
\Line(103,52)(107,48)
\Line(107,52)(103,48)
\CArc(135,50)(25,0,360)
\CArc(165,50)(5,0,360)
\Line(163,52)(167,48)
\Line(167,52)(163,48)
\PhotonArc(160,105)(55,221,270){2}{10}
\PhotonArc(160,105)(53.8,221,270){2}{10}
\PhotonArc(120,15)(55,401,450){-2}{10}
\PhotonArc(120,15)(53.8,401,450){-2}{10}
\Vertex(160,50){2}
\Vertex(119.8,69.8){2}
\Text(143.5,64)[c]{\rotatebox{-7}{\bf{\big /}}}
\Text(143.9,63.6)[c]{\rotatebox{-7}{\bf{\big /}}}

\Text(175,50)[l]{$+\ \lambda^2$}

\CArc(205,50)(5,0,360)
\Line(203,52)(207,48)
\Line(207,52)(203,48)
\CArc(235,50)(25,0,360)
\CArc(265,50)(5,0,360)
\Line(263,52)(267,48)
\Line(267,52)(263,48)
\PhotonArc(260,105)(55,221,270){2}{10}
\PhotonArc(260,105)(53.8,221,270){2}{10}
\PhotonArc(220,15)(55,401,450){-2}{10}
\PhotonArc(220,15)(53.8,401,450){-2}{10}
\Vertex(260,50){2}
\Vertex(219.8,69.8){2}
\Text(238.5,55)[c]{\rotatebox{-7}{\bf{\big /}}}
\Text(238.9,54.6)[c]{\rotatebox{-7}{\bf{\big /}}}

\Text(68,-15)[l]{$=\ \lambda^2$}

\CArc(105,-15)(5,0,360)
\Line(103,-17)(107,-13)
\Line(107,-17)(103,-13)
\CArc(135,-15)(25,0,360)
\CArc(165,-15)(5,0,360)
\Line(163,-17)(167,-13)
\Line(167,-17)(163,-13)
\PhotonArc(120,-15)(8,90,450){2}{9}
\PhotonArc(120,-15)(6.8,90,450){2}{9}
\Vertex(110 ,-15){2}
\PhotonArc(150,-15)(8,-90,270){2}{9}
\PhotonArc(150,-15)(6.8,-90,270){2}{9}
\Vertex(160,-15){2}

\Text(175,-15)[l]{$-\ \lambda^2$}

\CArc(205,-15)(5,0,360)
\Line(203,-17)(207,-13)
\Line(207,-17)(203,-13)
\CArc(235,-15)(25,0,360)
\CArc(265,-15)(5,0,360)
\Line(263,-17)(267,-13)
\Line(267,-17)(263,-13)
\PhotonArc(236,15)(40,230,307){2}{9}
\PhotonArc(236,15)(38.8,230,307){2}{9}
\PhotonArc(234,-45)(40,410,487){2}{9}
\PhotonArc(234,-45)(38.8,410,487){2}{9}
\Vertex(210,-15){2}
\Vertex(260,-15){2}

\end{picture}
\end{center}
which implies finally
\begin{equation}
(\rho)+(\theta)+\rm{(q)}=0.
\end{equation}

\subsubsection{The ${\cal O}(\lambda)$ cancellation.}

For this case one has to replace the propagator $\begin{picture}(10,5)(0,0)
\SetScale{0.8}
\Gluon(5,3)(40,3){2}{6}
\end{picture}\mbox{}\qquad \ \ $
appearing in the diagrams $(\eta),\dots,(\rho)$ with
the Feynman propagator $\begin{picture}(10,5)(0,0)
\SetScale{0.8}
\Photon(5,3.25)(40,3.25){2}{6}
\end{picture}\qquad \ $.
However these diagrams represent only half of those contributing to
this order; clearly, the other half is  
obtained by inverting, with respect to the
previous calculation, the pinching and Feynman propagators. 
For example, for the
diagrams (c),$\dots$,(h) the new terms are obtained by pinching with the vertical
propagator, treating the other one as the
Feynman propagator.  
Now, notice that the topologies $(\eta)$ and $(\theta)$ can be
canceled only by these new contributions coming from the diagrams (a) and (b).
In fact one has 

\begin{center}
\begin{picture}(100,120)(120,-45)

\Text(35,50)[l]{(a)+(b) $\to\ \lambda$}

\CArc(105,50)(5,0,360)
\Line(103,52)(107,48)
\Line(107,52)(103,48)
\CArc(135,50)(25,0,360)
\CArc(165,50)(5,0,360)
\Line(163,52)(167,48)
\Line(167,52)(163,48)
\Photon(125,27)(125,73){-2}{9}
\Photon(145,27)(145,73){2}{9}
\Photon(143.6,27)(143.6,73){2}{9}
\Text(145,51.5)[c]{\rotatebox{-71}{\bf{\big /}}}
\Text(145,52.1)[c]{\rotatebox{-71}{\bf{\big /}}}
\Text(145,48.5)[c]{\rotatebox{-71}{\bf{\big /}}}
\Text(145,47.9)[c]{\rotatebox{-71}{\bf{\big /}}}

\Text(175,50)[l]{$+\ \lambda$}
\Text(175,-15)[l]{$+\ \lambda$}

\CArc(205,50)(5,0,360)
\Line(203,52)(207,48)
\Line(207,52)(203,48)
\CArc(235,50)(25,0,360)
\CArc(265,50)(5,0,360)
\Line(263,52)(267,48)
\Line(267,52)(263,48)
\Photon(220,30)(250,70){2}{9}
\Photon(219,31)(249,71){2}{9}
\Photon(250,30)(237,48){-2}{4}
\Photon(232,55)(220,70){-2}{4}
\Text(227,40)[c]{\rotatebox{65}{\bf{\big /}}}
\Text(227.4,40.4)[c]{\rotatebox{65}{\bf{\big /}}}
\Text(225,38)[c]{\rotatebox{65}{\bf{\big /}}}
\Text(224.6,37.6)[c]{\rotatebox{65}{\bf{\big /}}}

\Text(65,-15)[l]{$=\ - \lambda$}

\CArc(205,-15)(5,0,360)
\Line(203,-17)(207,-13)
\Line(207,-17)(203,-13)
\CArc(235,-15)(25,0,360)
\CArc(265,-15)(5,0,360)
\Line(263,-17)(267,-13)
\Line(267,-17)(263,-13)
\Photon(225,-38)(225,8){-2}{9}
\PhotonArc(250,-15)(8,-90,270){2}{9}
\PhotonArc(250,-15)(6.8,-90,270){2}{9}
\Vertex(260,-15){2}

\CArc(105,-15)(5,0,360)
\Line(103,-17)(107,-13)
\Line(107,-17)(103,-13)
\CArc(135,-15)(25,0,360)
\CArc(165,-15)(5,0,360)
\Line(163,-17)(167,-13)
\Line(167,-17)(163,-13)
\Photon(135,-40)(135,10){2}{9}
\Photon(110,-14.3)(132,-14.3){2}{3.5}
\Photon(110,-15.7)(132,-15.7){2}{3.5}
\Photon(139,-14.3)(160,-14.3){-2}{3.5}
\Photon(139,-15.7)(160,-15.7){-2}{3.5}
\Vertex(110,-15){2}
\Vertex(160,-15){2}

\end{picture}
\end{center}
which are exactly the topologies needed to cancel the $(\eta)$ and 
$(\theta)$ terms. 
In addition, the following new contributions are obtained from diagrams
(c),$\dots$,(n)

\begin{center}
\begin{picture}(100,100)(80,-50)

\Text(135,30)[c]{(c)+(f)+(g)+(i)+(m)+(o)+(d)+(e)+(h)+(p)+(l)+(n)$\to$}

\Text(-15,-15)[l]{$2\lambda$}

\CArc(5,-15)(5,0,360)
\Line(3,-17)(7,-13)
\Line(7,-17)(3,-13)
\CArc(35,-15)(25,0,360)
\CArc(65,-15)(5,0,360)
\Line(63,-17)(67,-13)
\Line(67,-17)(63,-13)
\PhotonArc(20,-15)(8,90,450){2}{9}
\PhotonArc(20,-15)(6.8,90,450){2}{9}
\Vertex(10,-15){2}
\PhotonArc(35,-40)(13,16.5,165){2}{7.5}

\Text(75,-15)[l]{$-\ 4\lambda$}

\CArc(105,-15)(5,0,360)
\Line(103,-17)(107,-13)
\Line(107,-17)(103,-13)
\CArc(135,-15)(25,0,360)
\CArc(165,-15)(5,0,360)
\Line(163,-17)(167,-13)
\Line(167,-17)(163,-13)
\Photon(110,-15.7)(165,-15.7){2}{9}
\Photon(110,-14.3)(165,-14.3){2}{9}
\Vertex(110,-15){2}
\Vertex(160,-15){2}
\PhotonArc(135,-40)(13,16.5,165){2}{7.5}

\Text(175,-15)[l]{$+\ 2\lambda$}

\CArc(205,-15)(5,0,360)
\Line(203,-17)(207,-13)
\Line(207,-17)(203,-13)
\CArc(235,-15)(25,0,360)
\CArc(265,-15)(5,0,360)
\Line(263,-17)(267,-13)
\Line(267,-17)(263,-13)
\PhotonArc(235,-35)(14.2,183,-4){2}{7.5}
\PhotonArc(260,-70)(55,90,138){-2}{9.5}
\PhotonArc(260,-70)(53.8,90,138){-2}{9.5}
\Vertex(220,-35){2}
\Vertex(260,-15){2}

\Text(0,-40)[lb]{\footnotesize{$(\sigma)$}}
\Text(100,-40)[lb]{\footnotesize{$(\tau)$}}
\Text(200,-40)[lb]{\footnotesize{$(\upsilon)$}}

\end{picture}
\end{center}

Finally, the two contributions coming from diagram (q) will add
up giving the final result
\begin{center}
\begin{picture}(100,60)(40,20)

\Text(-35,50)[l]{(q) $\to\ 2\lambda$}

\CArc(20,50)(5,0,360)
\Line(18,52)(22,48)
\Line(22,52)(18,48)
\CArc(50,50)(25,0,360)
\CArc(80,50)(5,0,360)
\Line(78,52)(82,48)
\Line(82,52)(78,48)
\Photon(25,50.7)(75,50.7){2}{9}
\Photon(25,49.3)(75,49.3){2}{9}
\Vertex(25,50){2}
\Vertex(75,50){2}
\PhotonArc(50,25)(13,16.5,165){2}{7.5}

\Text(90,50)[l]{$-\ 2\lambda$}

\CArc(120,50)(5,0,360)
\Line(118,52)(122,48)
\Line(122,52)(118,48)
\CArc(150,50)(25,0,360)
\CArc(180,50)(5,0,360)
\Line(178,52)(182,48)
\Line(182,52)(178,48)
\PhotonArc(135,50)(8,90,450){2}{9}
\PhotonArc(135,50)(6.8,90,450){2}{9}
\Vertex(125,50){2}
\PhotonArc(150,25)(13,16.5,165){2}{7.5}

\end{picture}
\end{center}
Adding all terms together, we finally find that
\begin{equation}
(\xi)+(\rho)+(\sigma)+(\tau)+(\upsilon)+{\rm (q)}=0,
\end{equation}
which completes the proof of the gauge cancellation in the Abelian case.

\subsection{\label{nac}Non-Abelian case.}

We next proceed to address the non-Abelian case. 
With respect to the Abelian case we have two main differences:
first, there are seven more diagrams 
(plus ghosts) to consider (see diagrams
(r),$\dots$,(y) in Fig.\ref{fig3}), all of which contain at least
one three-gluon vertex; second,
due to the non trivial color structure of the theory, 
the cancellations in the Abelian-like subset of graphs will 
not go through as before.

Let us first deal with this latter point. 
Taking into account  the factors of $i$ coming from the
Feynman rules, we have the following color prefactors:
\begin{eqnarray}
& & {\rm (a),\ (c),\ (d),\ (e),\ (f),\ (i),\ (l),\ (o),\ (p),\ (q)} 
\sim -C^2_f, \nonumber\\
& & {\rm (b),\ (g),\ (h),\ (m),\ (n)} \sim -C^2_f+\frac12C_AC_f,
\nonumber\\
& & {\rm (r),\ (s),\ (t),\ (u)}\sim\frac12 C_A C_f, \nonumber\\
& & {\rm (w),\ (x),\ (y)}\sim -\frac12 C_A C_f, 
\label{nadia} 
\end{eqnarray}
where $C_f$ and $C_A$ represent respectively the quadratic 
Casimir operators of the
fundamental and the adjoint representations of the $SU(N_c)$ group, {\it i.e.}
\begin{equation}
C^2_f=\frac{N^2_c-1}{2N_c}, \qquad C_A=N_c.
\end{equation}
It is then clear that, while for the
 parts of the Abelian-like diagrams
(a),$\dots$,(q) which are proportional to $C^2_f$
the cancellations proven in the previous
sections
will still go through,
the parts proportional to $C_A$
will survive, 
and will eventually cancel against contributions from the 
purely non-Abelian graphs.
Thus our first task is to determine the non-Abelian remainders of the 
diagrams appearing in the second line of Eq.(\ref{nadia}), 
at each order in $\lambda$.

At any order in $\lambda$ one has

\begin{equation}
\begin{picture}(0,100)(370,5)

\Text(385,95)[c]{(g)+(h)+(m)+(n) $\to$}

\CArc(205,50)(5,0,360)
\Line(203,52)(207,48)
\Line(207,52)(203,48)
\CArc(235,50)(25,0,360)
\CArc(265,50)(5,0,360)
\Line(263,52)(267,48)
\Line(267,52)(263,48)
\PhotonArc(260,25)(25,90,180){-2}{9}
\PhotonArc(260,25)(23.8,90,180){-2}{9}
\GlueArc(224,35)(13.8,155,320){2.5}{7.5}
\Vertex(260,50){2}
\Vertex(233.5,25){2}

\Text(175,50)[l]{$-\ 2\lambda$}
\Text(275,50)[l]{$-\ 2\lambda$}
\Text(375,50)[l]{$-\ 2\lambda$}
\Text(475,50)[l]{$+\ 2\lambda$}

\CArc(305,50)(5,0,360)
\Line(303,52)(307,48)
\Line(307,52)(303,48)
\CArc(335,50)(25,0,360)
\CArc(365,50)(5,0,360)
\Line(363,52)(367,48)
\Line(367,52)(363,48)
\GlueArc(333,30)(13,182,350){2.5}{7.5}
\PhotonArc(360,-5)(55,90,139){-2}{9.5}
\PhotonArc(360,-5)(56.2,90,139){-2}{9.5}
\Vertex(360,50){2}
\Vertex(319,31){2}

\CArc(405,50)(5,0,360)
\Line(403,52)(407,48)
\Line(407,52)(403,48)
\CArc(435,50)(25,0,360)
\CArc(465,50)(5,0,360)
\Line(463,52)(467,48)
\Line(467,52)(463,48)
\GlueArc(435,30)(13,182,354){2.5}{7.5}
\PhotonArc(435,35)(15,-15,200){2}{10}
\PhotonArc(435,35)(13.8,-15,200){2}{10}
\Vertex(449,29.5){2}
\Vertex(422,29.5){2}

\CArc(505,50)(5,0,360)
\Line(503,52)(507,48)
\Line(507,52)(503,48)
\CArc(535,50)(25,0,360)
\CArc(565,50)(5,0,360)
\Line(563,52)(567,48)
\Line(567,52)(563,48)
\PhotonArc(520,50)(8,90,450){2}{9}
\PhotonArc(520,50)(6.8,90,450){2}{9}
\Vertex(510,50){2}
\GlueArc(535,25)(14.2,16.5,165){2.5}{7.5}

\end{picture}
\label{one}
\end{equation}

Now, at ${\cal O}(\lambda^2)$ the above equation reads

\begin{center}
\begin{picture}(0,160)(340,-55)

\Text(340,95)[c]{(g)+(h)+(m)+(n) $\to$}

\CArc(310,50)(5,0,360)
\Line(308,52)(312,48)
\Line(312,52)(308,48)
\CArc(340,50)(25,0,360)
\CArc(370,50)(5,0,360)
\Line(368,52)(372,48)
\Line(372,52)(368,48)
\PhotonArc(365,-5)(55,90,139){2}{10}
\PhotonArc(365,-5)(53.8,90,139){2}{10}
\PhotonArc(324,86)(55,270,319){-2}{10}
\PhotonArc(324,86)(53.8,270,319){-2}{10}
\Vertex(365,50){2}
\Vertex(325.2,30.2){2}
\Text(348,36)[c]{\rotatebox{40}{\bf{\big /}}}
\Text(347.6,35.6)[c]{\rotatebox{40}{\bf{\big /}}}

\Text(180,50)[l]{$ 2\lambda^2$}
\Text(170,-15)[l]{$-\ 2\lambda^2$}
\Text(275,50)[l]{$+\ 2\lambda^2$}
\Text(380,50)[l]{$-\ 2\lambda^2$}
\Text(275,-15)[l]{$+\ 2\lambda^2$}

\Text(200,25)[lb]{\footnotesize{(i)}}
\Text(305,25)[lb]{\footnotesize{(ii)}}
\Text(200,-40)[lb]{\footnotesize{(iii)}}
\Text(305,-40)[lb]{\footnotesize{(iv)}}

\CArc(205,-15)(5,0,360)
\Line(203,-17)(207,-13)
\Line(207,-17)(203,-13)
\CArc(235,-15)(25,0,360)
\CArc(265,-15)(5,0,360)
\Line(263,-17)(267,-13)
\Line(267,-17)(263,-13)
\PhotonArc(220,-15)(8,90,450){2}{9}
\PhotonArc(220,-15)(6.8,90,450){2}{9}
\PhotonArc(250,-15)(8,-90,270){2}{9}
\PhotonArc(250,-15)(6.8,-90,270){2}{9}
\Vertex(210,-15){2}
\Vertex(260,-15){2}

\CArc(415,50)(5,0,360)
\Line(413,52)(417,48)
\Line(417,52)(413,48)
\CArc(445,50)(25,0,360)
\CArc(475,50)(5,0,360)
\Line(473,52)(477,48)
\Line(477,52)(473,48)
\PhotonArc(445,32.5)(14.2,196.5,345){2}{7.5}
\PhotonArc(445,32.5)(15.4,196.5,345){2}{7.5}
\PhotonArc(445,35)(15,-15,200){2}{10}
\PhotonArc(445,35)(13.8,-15,200){2}{10}
\Vertex(459,30){2}
\Vertex(431,29.5){2}
\Text(445.5,17)[c]{\rotatebox{19}{\bf{\big /}}}
\Text(448.5,17)[c]{\rotatebox{19}{\bf{\big /}}}
\Text(444.9,17)[c]{\rotatebox{19}{\bf{\big /}}}
\Text(449.1,17)[c]{\rotatebox{19}{\bf{\big /}}}

\CArc(205,50)(5,0,360)
\Line(203,52)(207,48)
\Line(207,52)(203,48)
\CArc(235,50)(25,0,360)
\CArc(265,50)(5,0,360)
\Line(263,52)(267,48)
\Line(267,52)(263,48)
\PhotonArc(210,25)(25,0,90){2}{9}
\PhotonArc(210,25)(23.8,0,90){2}{9}
\PhotonArc(260,25)(25,90,180){2}{9}
\PhotonArc(260,25)(23.8,90,180){2}{9}
\Vertex(210,50){2}
\Vertex(235,25){2}
\Vertex(260,50){2}
\Text(225,44)[c]{\rotatebox{155}{\bf{\big /}}}
\Text(225.45,43.6)[c]{\rotatebox{155}{\bf{\big /}}}

\CArc(310,-15)(5,0,360)
\Line(308,-17)(312,-13)
\Line(312,-17)(308,-13)
\CArc(340,-15)(25,0,360)
\CArc(370,-15)(5,0,360)
\Line(368,-17)(372,-13)
\Line(372,-17)(368,-13)
\Photon(335,-14.3)(365,-14.3){2}{6}
\Photon(335,-15.7)(365,-15.7){2}{6}
\Photon(315,-14.3)(328,-14.3){2}{2.5}
\Photon(315,-15.7)(328,-15.7){2}{2.5}
\PhotonArc(325,-15)(8,90,450){2}{9}
\PhotonArc(325,-15)(6.8,90,450){2}{9}
\Vertex(315,-15){2}
\Vertex(365,-15){2}

\end{picture}
\end{center}

On the other hand

\begin{center}
\begin{picture}(0,120)(340,-50)

\Text(145,50)[l]{(b) $\to\ 2\lambda^2$}

\Text(170,-15)[l]{$+\ 2\lambda^2$}
\Text(275,50)[l]{$-\ \lambda^2$}
\Text(380,50)[l]{$+\ \lambda^2$}
\Text(275,-15)[l]{$-\ 2\lambda^2$}
\Text(380,-15)[l]{$-\ \lambda^2$}

\Text(200,25)[lb]{\footnotesize{(v)}}
\Text(305,25)[lb]{\footnotesize{(vi)}}
\Text(410,25)[lb]{\footnotesize{(vii)}}
\Text(200,-40)[lb]{\footnotesize{(viii)}}
\Text(305,-40)[lb]{\footnotesize{(ix)}}

\CArc(205,50)(5,0,360)
\Line(203,52)(207,48)
\Line(207,52)(203,48)
\CArc(235,50)(25,0,360)
\CArc(265,50)(5,0,360)
\Line(263,52)(267,48)
\Line(267,52)(263,48)
\PhotonArc(260,105)(55,221,270){2}{10}
\PhotonArc(260,105)(53.8,221,270){2}{10}
\PhotonArc(220,15)(55,401,450){-2}{10}
\PhotonArc(220,15)(53.8,401,450){-2}{10}
\Vertex(260,50){2}
\Vertex(219.8,69.8){2}
\Text(243.5,64)[c]{\rotatebox{-7}{\bf{\big /}}}
\Text(243.9,63.6)[c]{\rotatebox{-7}{\bf{\big /}}}

\CArc(310,50)(5,0,360)
\Line(308,52)(312,48)
\Line(312,52)(308,48)
\CArc(340,50)(25,0,360)
\CArc(370,50)(5,0,360)
\Line(368,52)(372,48)
\Line(372,52)(368,48)
\PhotonArc(315,75)(25,270,0){-2}{9}
\PhotonArc(315,75)(23.8,270,0){-2}{9}
\PhotonArc(365,75)(25,180,270){-2}{9}
\PhotonArc(365,75)(23.8,180,270){-2}{9}
\Vertex(315,50){2}
\Vertex(365,50){2}
\Vertex(340,75){2}
\Text(330,56)[c]{\rotatebox{65}{\bf{\big /}}}
\Text(330.45,56.4)[c]{\rotatebox{65}{\bf{\big /}}}

\CArc(415,50)(5,0,360)
\Line(413,52)(417,48)
\Line(417,52)(413,48)
\CArc(445,50)(25,0,360)
\CArc(475,50)(5,0,360)
\Line(473,52)(477,48)
\Line(477,52)(473,48)
\PhotonArc(420,25)(25,0,90){-2}{9}
\PhotonArc(420,25)(23.8,0,90){-2}{9}
\PhotonArc(470,25)(25,90,180){-2}{9}
\PhotonArc(470,25)(23.8,90,180){-2}{9}
\Vertex(420,50){2}
\Vertex(470,50){2}
\Vertex(445,25){2}
\Text(455,44)[c]{\rotatebox{65}{\bf{\big /}}}
\Text(455.45,44.4)[c]{\rotatebox{65}{\bf{\big /}}}

\CArc(205,-15)(5,0,360)
\Line(203,-17)(207,-13)
\Line(207,-17)(203,-13)
\CArc(235,-15)(25,0,360)
\CArc(265,-15)(5,0,360)
\Line(263,-17)(267,-13)
\Line(267,-17)(263,-13)
\PhotonArc(236,15)(40,230,307){2}{9}
\PhotonArc(236,15)(38.8,230,307){2}{9}
\PhotonArc(234,-45)(40,410,487){2}{9}
\PhotonArc(234,-45)(38.8,410,487){2}{9}
\Vertex(210,-15){2}
\Vertex(260,-15){2}

\CArc(310,-15)(5,0,360)
\Line(308,-17)(312,-13)
\Line(312,-17)(308,-13)
\CArc(340,-15)(25,0,360)
\CArc(370,-15)(5,0,360)
\Line(368,-17)(372,-13)
\Line(372,-17)(368,-13)
\Photon(335,-15.7)(365,-15.7){2}{6}
\Photon(335,-14.3)(365,-14.3){2}{6}
\Photon(315,-15.7)(328,-15.7){2}{2.5}
\Photon(315,-14.3)(328,-14.3){2}{2.5}
\PhotonArc(325,-15)(8,90,450){2}{9}
\PhotonArc(325,-15)(6.8,90,450){2}{9}
\Vertex(315,-15){2}
\Vertex(365,-15){2}

\CArc(415,-15)(5,0,360)
\Line(413,-17)(417,-13)
\Line(417,-17)(413,-13)
\CArc(445,-15)(25,0,360)
\CArc(475,-15)(5,0,360)
\Line(473,-17)(477,-13)
\Line(477,-17)(473,-13)
\PhotonArc(391,-15.7)(60,337,24){-2}{9}
\PhotonArc(391,-15.7)(61.2,337,24){-2}{9}
\PhotonArc(500,-15)(61.2,157,203.8){2}{9}
\PhotonArc(500,-15)(60,157,203.8){2}{9}
\Vertex(445,10){2}
\Vertex(445.5,-40){2}
\Text(441,-16.5)[c]{\rotatebox{-71}{\bf{\big /}}}
\Text(441,-17.1)[c]{\rotatebox{-71}{\bf{\big /}}}
\Text(441,-13.5)[c]{\rotatebox{-71}{\bf{\big /}}}
\Text(441,-12.9)[c]{\rotatebox{-71}{\bf{\big /}}}

\end{picture}
\end{center}

It is straightforward to verify, using the procedure presented 
following Eq.(\ref{poi}), that 

\begin{equation}
{\rm (ii)+(iii)+(v)+(viii)} =0, \qquad
{\rm (i)+(vi)+(vii)}= 0. 
\end{equation}
Finally, (iv) and (ix) cancel directly. Thus, at ${\cal
O}(\lambda^2)$ we are left with the non-Abelian remainder

\begin{center}
\begin{equation}
\begin{picture}(0,70)(230,10)

\Text(55,50)[l]{(g)+(h)+(m)+(n)+(b) $\to\ -2\lambda^2$}

\CArc(205,50)(5,0,360)
\Line(203,52)(207,48)
\Line(207,52)(203,48)
\CArc(235,50)(25,0,360)
\CArc(265,50)(5,0,360)
\Line(263,52)(267,48)
\Line(267,52)(263,48) 
\PhotonArc(235,32.5)(14.2,196.5,345){2}{7.5}
\PhotonArc(235,32.5)(15.4,196.5,345){2}{7.5}
\PhotonArc(235,35)(15,-15,200){2}{10}
\PhotonArc(235,35)(13.8,-15,200){2}{10}
\Vertex(249,30){2}
\Vertex(221,29.5){2}
\Text(235.5,17)[c]{\rotatebox{19}{\bf{\big /}}}
\Text(238.5,17)[c]{\rotatebox{19}{\bf{\big /}}}
\Text(234.9,17)[c]{\rotatebox{19}{\bf{\big /}}}
\Text(239.1,17)[c]{\rotatebox{19}{\bf{\big /}}}

\Text(275,50)[l]{$-\ \,\lambda^2$}

\CArc(305,50)(5,0,360)
\Line(303,52)(307,48)
\Line(307,52)(303,48)
\CArc(335,50)(25,0,360)
\CArc(365,50)(5,0,360)
\Line(363,52)(367,48)
\Line(367,52)(363,48)
\PhotonArc(281,49.3)(60,337,24){-2}{9}
\PhotonArc(281,49.3)(61.2,337,24){-2}{9}
\PhotonArc(390,50)(61.2,157,203.8){2}{9}
\PhotonArc(390,50)(60,157,203.8){2}{9}
\Vertex(335,75){2}
\Vertex(335.5,25){2}
\Text(331,51.5)[c]{\rotatebox{-71}{\bf{\big /}}}
\Text(331,52.1)[c]{\rotatebox{-71}{\bf{\big /}}}
\Text(331,48.5)[c]{\rotatebox{-71}{\bf{\big /}}}
\Text(331,47.9)[c]{\rotatebox{-71}{\bf{\big /}}}

\end{picture}
\label{AR}
\end{equation}
\end{center}
which will be canceled later.

In the ${\cal O}(\lambda)$ case, we have to substitute the propagator
$\begin{picture}(10,5)(0,0)
\SetScale{0.8}
\Gluon(5,4)(40,4){2}{6}
\end{picture} \qquad \ \ $ appearing in (\ref{one}) with
the Feynman propagator $\begin{picture}(10,5)(0,0)
\SetScale{0.8}
\Photon(5,3.25)(40,3.25){2}{6}
\end{picture}\qquad \ $.
Moreover,
as already mentioned, we have still to compute 
the GFP-dependent contributions (linear in $\lambda$)
of diagrams
(g), (h), (m), (n), where the Feynman and pinching propagators 
are reversed with respect to that presented in Eq.(\ref{one}).
Thus, we obtain:

\begin{center}
\begin{picture}(0,170)(370,-60)

\Text(275,50)[l]{$+\ 2\lambda$}
\Text(375,50)[l]{$-\ 2\lambda$}
\Text(175,-15)[l]{$-\ 2\lambda$}
\Text(275,-15)[l]{$-\ 2\lambda$}
\Text(375,-15)[l]{$+\ 2\lambda$}
\Text(475,50)[l]{$+\ 2\lambda$}
\Text(185,50)[l]{$2\lambda$}
\Text(475,-15)[l]{$+\ 2\lambda$}

\Text(385,95)[c]{(g)+(h)+(m)+(n) $\to$}

\CArc(205,50)(5,0,360)
\Line(203,52)(207,48)
\Line(207,52)(203,48)
\CArc(235,50)(25,0,360)
\CArc(265,50)(5,0,360)
\Line(263,52)(267,48)
\Line(267,52)(263,48)
\Photon(245,27)(245,73){2}{9}
\PhotonArc(220,50)(8,90,450){2}{9}
\PhotonArc(220,50)(6.8,90,450){2}{9}
\Vertex(210,50){2}

\CArc(305,50)(5,0,360)
\Line(303,52)(307,48)
\Line(307,52)(303,48)
\CArc(335,50)(25,0,360)
\CArc(365,50)(5,0,360)
\Line(363,52)(367,48)
\Line(367,52)(363,48)
\Photon(335,25)(335,75){2}{9}
\Photon(310,50.7)(332,50.7){2}{3.5}
\Photon(310,49.3)(332,49.3){2}{3.5}
\Photon(339,50.7)(360,50.7){-2}{3.5}
\Photon(339,49.3)(360,49.3){-2}{3.5}
\Vertex(310,50){2}
\Vertex(360,50){2}

\CArc(405,50)(5,0,360)
\Line(403,52)(407,48)
\Line(407,52)(403,48)
\CArc(435,50)(25,0,360)
\CArc(465,50)(5,0,360)
\Line(463,52)(467,48)
\Line(467,52)(463,48)
\PhotonArc(435,35)(15,-15,200){2}{10}
\PhotonArc(435,35)(13.8,-15,200){2}{10}
\Vertex(449.5,29.5){2}
\Vertex(420,29.5){2}
\PhotonArc(435,32.5)(15.4,196.5,345){2}{7.5}

\CArc(205,-15)(5,0,360)
\Line(203,-17)(207,-13)
\Line(207,-17)(203,-13)
\CArc(235,-15)(25,0,360)
\CArc(265,-15)(5,0,360)
\Line(263,-17)(267,-13)
\Line(267,-17)(263,-13)
\PhotonArc(260,-40)(25,90,180){-2}{7}
\PhotonArc(260,-40)(23.8,90,180){-2}{7}
\Photon(235,-40)(235,10){2}{9}
\Vertex(235,-40){2}
\Vertex(260,-15){2}

\CArc(305,-15)(5,0,360)
\Line(303,-17)(307,-13)
\Line(307,-17)(303,-13)
\CArc(335,-15)(25,0,360)
\CArc(365,-15)(5,0,360)
\Line(363,-17)(367,-13)
\Line(367,-17)(363,-13)
\PhotonArc(310,-40)(25,0,90){-2}{7}
\PhotonArc(310,-40)(23.8,0,90){-2}{7}
\Photon(335,-40)(335,10){2}{9}
\Vertex(335,-40){2}
\Vertex(310,-15){2}

\CArc(405,-15)(5,0,360)
\Line(403,-17)(407,-13)
\Line(407,-17)(403,-13)
\CArc(435,-15)(25,0,360)
\CArc(465,-15)(5,0,360)
\Line(463,-17)(467,-13)
\Line(467,-17)(463,-13)
\PhotonArc(420,-15)(8,90,450){2}{9}
\PhotonArc(420,-15)(6.8,90,450){2}{9}
\Vertex(410,-15){2}
\PhotonArc(435,-40)(14.2,16.5,165){2}{7.5}

\CArc(505,50)(5,0,360)
\Line(503,52)(507,48)
\Line(507,52)(503,48)
\CArc(535,50)(25,0,360)
\CArc(565,50)(5,0,360)
\Line(563,52)(567,48)
\Line(567,52)(563,48)
\PhotonArc(560,25)(25,90,180){-2}{9}
\PhotonArc(560,25)(23.8,90,180){-2}{9}
\PhotonArc(525,35)(13.8,162,305){-2}{6}
\Vertex(560,50){2}
\Vertex(533.5,25){2}

\CArc(505,-15)(5,0,360)
\Line(503,-17)(507,-13)
\Line(507,-17)(503,-13)
\CArc(535,-15)(25,0,360)
\CArc(565,-15)(5,0,360)
\Line(563,-17)(567,-13)
\Line(567,-17)(563,-13)
\PhotonArc(535,-35)(14.2,183,-4){2}{7.5}
\PhotonArc(560,-70)(55,90,139){2}{9.5}
\PhotonArc(560,-70)(53.8,90,139){2}{9.5}
\Vertex(560,-15){2}
\Vertex(519.8,-35.2){2}

\end{picture}
\end{center}

Finally, from (b) we obtain: 

\begin{center}
\begin{picture}(0,70)(370,10)

\Text(155,50)[l]{(b)$\to\ 2\lambda$}
\Text(275,50)[l]{$+\ 2\lambda$}
\Text(375,50)[l]{$-\ 2\lambda$}
\Text(475,50)[l]{$-\ 2\lambda$}

\CArc(205,50)(5,0,360)
\Line(203,52)(207,48)
\Line(207,52)(203,48)
\CArc(235,50)(25,0,360)
\CArc(265,50)(5,0,360)
\Line(263,52)(267,48)
\Line(267,52)(263,48)
\PhotonArc(260,25)(25,90,180){-2}{7}
\PhotonArc(260,25)(23.8,90,180){-2}{7}
\Photon(235,25)(235,75){2}{9}
\Vertex(235,25){2}
\Vertex(260,50){2}

\CArc(305,50)(5,0,360)
\Line(303,52)(307,48)
\Line(307,52)(303,48)
\CArc(335,50)(25,0,360)
\CArc(365,50)(5,0,360)
\Line(363,52)(367,48)
\Line(367,52)(363,48)
\PhotonArc(310,25)(25,0,90){-2}{7}
\PhotonArc(310,25)(23.8,0,90){-2}{7}
\Photon(335,25)(335,75){2}{9}
\Vertex(335,25){2}
\Vertex(310,50){2}

\CArc(405,50)(5,0,360)
\Line(403,52)(407,48)
\Line(407,52)(403,48)
\CArc(435,50)(25,0,360)
\CArc(465,50)(5,0,360)
\Line(463,52)(467,48)
\Line(467,52)(463,48)
\Photon(435,25)(435,75){2}{9}
\Photon(410,50.7)(432,50.7){2}{3.5}
\Photon(410,49.3)(432,49.3){2}{3.5}
\Photon(439,50.7)(460,50.7){-2}{3.5}
\Photon(439,49.3)(460,49.3){-2}{3.5}
\Vertex(410,50){2}
\Vertex(460,50){2}

\CArc(505,50)(5,0,360)
\Line(503,52)(507,48)
\Line(507,52)(503,48)
\CArc(535,50)(25,0,360)
\CArc(565,50)(5,0,360)
\Line(563,52)(567,48)
\Line(567,52)(563,48)
\PhotonArc(481,49.3)(60,337,24){-2}{9}
\PhotonArc(590,50)(61.2,157,203.8){2}{9}
\PhotonArc(590,50)(60,157,203.8){2}{9}
\Vertex(535,75){2}
\Vertex(535,25){2}

\end{picture}
\end{center}

Thus, putting all terms 
together, we arrive at the ${\cal O}(\lambda)$ non-Abelian remainder:

\begin{center}
\begin{equation}
\begin{picture}(0,100)(380,5)

\Text(385,95)[c]{(g)+(h)+(m)+(n)+(b) $\to$}
\Text(185,50)[l]{$2\lambda$}
\Text(275,50)[l]{$-\ 2\lambda$}
\Text(375,50)[l]{$+\ 4\lambda$}
\Text(475,50)[l]{$-\ 4\lambda$}

\CArc(205,50)(5,0,360)
\Line(203,52)(207,48)
\Line(207,52)(203,48)
\CArc(235,50)(25,0,360)
\CArc(265,50)(5,0,360)
\Line(263,52)(267,48)
\Line(267,52)(263,48)
\Photon(245,27)(245,73){2}{9}
\PhotonArc(220,50)(8,90,450){2}{9}
\PhotonArc(220,50)(6.8,90,450){2}{9}
\Vertex(210,50){2}

\CArc(305,50)(5,0,360)
\Line(303,52)(307,48)
\Line(307,52)(303,48)
\CArc(335,50)(25,0,360)
\CArc(365,50)(5,0,360)
\Line(363,52)(367,48)
\Line(367,52)(363,48)
\PhotonArc(281,49.3)(60,337,24){-2}{9}
\PhotonArc(390,50)(61.2,157,203.8){2}{9}
\PhotonArc(390,50)(60,157,203.8){2}{9}
\Vertex(335,75){2}
\Vertex(335,25){2}

\CArc(405,50)(5,0,360)
\Line(403,52)(407,48)
\Line(407,52)(403,48)
\CArc(435,50)(25,0,360)
\CArc(465,50)(5,0,360)
\Line(463,52)(467,48)
\Line(467,52)(463,48)
\PhotonArc(420,50)(8,90,450){2}{9}
\PhotonArc(420,50)(6.8,90,450){2}{9}
\Vertex(410,50){2}
\PhotonArc(435,25)(14.2,16.5,165){2}{7.5}

\CArc(505,50)(5,0,360)
\Line(503,52)(507,48)
\Line(507,52)(503,48)
\CArc(535,50)(25,0,360)
\CArc(565,50)(5,0,360)
\Line(563,52)(567,48)
\Line(567,52)(563,48)
\PhotonArc(535,35)(13.8,-15,200){2}{10}
\PhotonArc(535,35)(15,-15,200){2}{10}
\Vertex(549.5,29.5){2}
\Vertex(520,29.5){2}
\PhotonArc(535,32.5)(15.4,196.5,345){2}{7.5}

\end{picture}
\label{NAR}
\end{equation}
\end{center}

We next concentrate on the purely non-Abelian diagrams  (q),$\dots$,(y).
In this case we will split the calculation from the beginning 
into different orders
in $\lambda$. 
Notice that the ${\cal O}(\lambda^3)$
cancellation is automatically accomplished due to the  elementary WI
\begin{equation} 
k_1^\mu k_2^\nu k_3^\rho\Gamma_{\mu\nu\rho}(k_1,k_2,k_3)=0,
\end{equation}
satisfied by the three gluon vertex. Therefore we only have to collect
${\cal O}(\lambda^2)$ and ${\cal O}(\lambda)$ contributions.
 
\subsubsection{The ${\cal O}(\lambda^2)$ cancellation.}

In dealing with the non-Abelian diagrams, 
we have found it more economical to carry out 
all possible cancellations before letting the longitudinal momenta 
act on the three gluon vertex. 
Consider, for example the diagrams (r) and (s): each one 
gives rise to three possible ${\cal O}(\lambda^2)$
diagrams, as shown below

\begin{center}
\begin{picture}(0,120)(160,-40)

\Text(35,25)[lb]{\footnotesize{$(\alpha)$}}
\Text(135,25)[lb]{\footnotesize{$(\beta)$}}
\Text(235,25)[lb]{\footnotesize{$(\gamma)$}}

\Text(35,-40)[lb]{\footnotesize{$(\delta)$}}
\Text(135,-40)[lb]{\footnotesize{$(\varepsilon)$}}
\Text(235,-40)[lb]{\footnotesize{$(\zeta)$}}

\Text(0,50)[l]{${\rm (r)}\to$}
\Text(122.5,50)[c]{$+$}
\Text(222.5,50)[c]{$+$}

\Text(0,-15)[l]{${\rm (s)}\to$}
\Text(122.5,-15)[c]{$+$}
\Text(222.5,-15)[c]{$+$}

\CArc(40,50)(5,0,360)
\Line(38,52)(42,48)
\Line(42,52)(38,48)
\CArc(70,50)(25,0,360)
\CArc(100,50)(5,0,360)
\Line(98,52)(102,48)
\Line(102,52)(98,48)
\Photon(50,65)(70,50){2}{5}
\Photon(51,66)(71,51){2}{5}
\Photon(71.5,50)(90,65){2}{5}
\Photon(70.5,51)(89,66){2}{5}
\Photon(70,25)(70,50){2}{6}
\GCirc(70,50){1}{1}
\Text(82.5,60)[c]{\rotatebox{65}{\bf{\big /}}}
\Text(82.1,59.6)[c]{\rotatebox{65}{\bf{\big /}}}
\Text(84.5,62)[c]{\rotatebox{65}{\bf{\big /}}}
\Text(84.1,61.6)[c]{\rotatebox{65}{\bf{\big /}}}
\Text(57.5,60)[c]{\rotatebox{155}{\bf{\big /}}}
\Text(57.9,59.6)[c]{\rotatebox{155}{\bf{\big /}}}
\Text(55.5,62)[c]{\rotatebox{155}{\bf{\big /}}}
\Text(55.9,61.6)[c]{\rotatebox{155}{\bf{\big /}}}

\CArc(140,50)(5,0,360)
\Line(138,52)(142,48)
\Line(142,52)(138,48)
\CArc(170,50)(25,0,360)
\CArc(200,50)(5,0,360)
\Line(198,52)(202,48)
\Line(202,52)(198,48)
\Photon(170.7,25)(170.7,50){2}{6}
\Photon(169.3,25)(169.3,50){2}{6}
\Photon(150,65)(170,50){2}{5}
\Photon(151,66)(171,51){2}{5}
\Photon(170,50)(189,66){2}{5}
\GCirc(170,50){1}{1}
\Text(170.5,39)[c]{\rotatebox{-71}{\bf{\big /}}}
\Text(170.5,39.6)[c]{\rotatebox{-71}{\bf{\big /}}}
\Text(170.5,36)[c]{\rotatebox{-71}{\bf{\big /}}}
\Text(170.5,35.4)[c]{\rotatebox{-71}{\bf{\big /}}}
\Text(157.5,60)[c]{\rotatebox{155}{\bf{\big /}}}
\Text(157.9,59.6)[c]{\rotatebox{155}{\bf{\big /}}}
\Text(155.5,62)[c]{\rotatebox{155}{\bf{\big /}}}
\Text(155.9,61.6)[c]{\rotatebox{155}{\bf{\big /}}}

\CArc(240,50)(5,0,360)
\Line(238,52)(242,48)
\Line(242,52)(238,48)
\CArc(270,50)(25,0,360)
\CArc(300,50)(5,0,360)
\Line(298,52)(302,48)
\Line(302,52)(298,48)
\Photon(270.7,25)(270.7,50){2}{6}
\Photon(269.3,25)(269.3,50){2}{6}
\Photon(250,65)(270,50){2}{5}
\Photon(271.5,50)(290,65){2}{5}
\Photon(270.5,51)(289,66){2}{5}
\GCirc(270,50){1}{1}
\Text(271,39)[c]{\rotatebox{-71}{\bf{\big /}}}
\Text(271,39.6)[c]{\rotatebox{-71}{\bf{\big /}}}
\Text(271,36)[c]{\rotatebox{-71}{\bf{\big /}}}
\Text(271,35.4)[c]{\rotatebox{-71}{\bf{\big /}}}
\Text(282.5,60)[c]{\rotatebox{65}{\bf{\big /}}}
\Text(282.1,59.6)[c]{\rotatebox{65}{\bf{\big /}}}
\Text(284.5,62)[c]{\rotatebox{65}{\bf{\big /}}}
\Text(284.1,61.6)[c]{\rotatebox{65}{\bf{\big /}}}

\CArc(40,-15)(5,0,360)
\Line(38,-17)(42,-13)
\Line(42,-17)(38,-13)
\CArc(70,-15)(25,0,360)
\CArc(100,-15)(5,0,360)
\Line(98,-17)(102,-13)
\Line(102,-17)(98,-13)
\Photon(70,10)(70,-15){2}{5}
\Photon(50,-30)(70,-15){2}{5}
\Photon(51,-31)(71,-16){2}{5}
\Photon(71.5,-15)(90,-30){2}{5}
\Photon(70.5,-16)(89,-31){2}{5}
\GCirc(70,-15){1}{1}
\Text(57.5,-25)[c]{\rotatebox{65}{\bf{\big /}}}
\Text(57.1,-25.4)[c]{\rotatebox{65}{\bf{\big /}}}
\Text(59.5,-23)[c]{\rotatebox{65}{\bf{\big /}}}
\Text(59.1,-23.4)[c]{\rotatebox{65}{\bf{\big /}}}
\Text(82.5,-25)[c]{\rotatebox{155}{\bf{\big /}}}
\Text(82.9,-25.4)[c]{\rotatebox{155}{\bf{\big /}}}
\Text(80.5,-23)[c]{\rotatebox{155}{\bf{\big /}}}
\Text(80.9,-23.4)[c]{\rotatebox{155}{\bf{\big /}}}

\CArc(140,-15)(5,0,360)
\Line(138,-17)(142,-13)
\Line(142,-17)(138,-13)
\CArc(170,-15)(25,0,360)
\CArc(200,-15)(5,0,360)
\Line(198,-17)(202,-13)
\Line(202,-17)(198,-13)
\Photon(170.7,10)(170.7,-15){2}{5}
\Photon(169.3,10)(169.3,-15){2}{5}
\Photon(150,-30)(170,-15){2}{5}
\Photon(151,-31)(171,-16){2}{5}
\Photon(170,-15)(189,-31){2}{5}
\GCirc(170,-15){1}{1}
\Text(157.5,-25)[c]{\rotatebox{65}{\bf{\big /}}}
\Text(157.1,-25.4)[c]{\rotatebox{65}{\bf{\big /}}}
\Text(159.5,-23)[c]{\rotatebox{65}{\bf{\big /}}}
\Text(159.1,-23.4)[c]{\rotatebox{65}{\bf{\big /}}}
\Text(170.5,-1)[c]{\rotatebox{-71}{\bf{\big /}}}
\Text(170.5,-0.4)[c]{\rotatebox{-71}{\bf{\big /}}}
\Text(170.5,-4)[c]{\rotatebox{-71}{\bf{\big /}}}
\Text(170.5,-4.6)[c]{\rotatebox{-71}{\bf{\big /}}}

\CArc(240,-15)(5,0,360)
\Line(238,-17)(242,-13)
\Line(242,-17)(238,-13)
\CArc(270,-15)(25,0,360)
\CArc(300,-15)(5,0,360)
\Line(298,-17)(302,-13)
\Line(302,-17)(298,-13)
\Photon(270.7,10)(270.7,-15){2}{5}
\Photon(269.3,10)(269.3,-15){2}{5}
\Photon(271.5,-15)(290,-30){2}{5}
\Photon(270.5,-16)(289,-31){2}{5}
\Photon(250,-30)(270,-15){2}{5}
\GCirc(270,-15){1}{1}
\Text(282.5,-25)[c]{\rotatebox{155}{\bf{\big /}}}
\Text(282.9,-25.4)[c]{\rotatebox{155}{\bf{\big /}}}
\Text(280.5,-23)[c]{\rotatebox{155}{\bf{\big /}}}
\Text(280.9,-23.4)[c]{\rotatebox{155}{\bf{\big /}}}
\Text(271,-1)[c]{\rotatebox{-71}{\bf{\big /}}}
\Text(271,-0.4)[c]{\rotatebox{-71}{\bf{\big /}}}
\Text(271,-4)[c]{\rotatebox{-71}{\bf{\big /}}}
\Text(271,-4.6)[c]{\rotatebox{-71}{\bf{\big /}}}

\end{picture}
\end{center}
Similarly, the basic topologies obtained from 
diagrams (t) and (u) can be easily worked out.
After some elementary manipulations involving 
further pinching in order to allow 
the combination/cancellation of seemingly different
topologies,
(but without acting on the three-gluon vertex), 
one finally arrives at 

\begin{center}
\begin{picture}(0,130)(277,-50)

\Text(80,50)[l]{$(\beta)+(\gamma)+(\delta)$+(t) 
$\to\ 2\lambda^2$}
\Text(70,-15)[l]{$(\alpha)+(\varepsilon)+(\zeta)$+(u) $\to\ 
-2\lambda^2$}
 
\Text(275,50)[l]{$-\ \lambda^2$}
\Text(375,50)[l]{$-\ \lambda^2$}
\Text(275,-15)[l]{$-\ \lambda^2$}
\Text(375,-15)[l]{$+\ \lambda^2$}

\CArc(205,50)(5,0,360)
\Line(203,52)(207,48)
\Line(207,52)(203,48)
\CArc(235,50)(25,0,360)
\CArc(265,50)(5,0,360)
\Line(263,52)(267,48)
\Line(267,52)(263,48)
\PhotonArc(243,60)(10,210,570){2}{11}
\PhotonArc(243,60)(8.8,210,570){2}{11}
\Vertex(250,69.5){2}
\PhotonArc(234,67)(14,165,270){-2}{5}
\GCirc(234.5,53.5){1}{1}
\Text(237,65)[c]{\rotatebox{65}{\bf{\big /}}}
\Text(236.6,64.6)[c]{\rotatebox{65}{\bf{\big /}}}
\Text(239,67)[c]{\rotatebox{65}{\bf{\big /}}}
\Text(239.4,67.4)[c]{\rotatebox{65}{\bf{\big /}}}
\Text(251,54)[c]{\rotatebox{65}{\bf{\big /}}}
\Text(251.4,54.4)[c]{\rotatebox{65}{\bf{\big /}}}

\CArc(305,50)(5,0,360)
\Line(303,52)(307,48)
\Line(307,52)(303,48)
\CArc(335,50)(25,0,360)
\CArc(365,50)(5,0,360)
\Line(363,52)(367,48)
\Line(367,52)(363,48)
\PhotonArc(335,37)(10,0,360){2}{11}
\PhotonArc(335,37)(8.8,0,360){2}{11}
\Photon(335,75)(335,47){2}{6}
\Vertex(335,25){2}
\GCirc(335,46){1}{1}
\Text(327,38.5)[c]{\rotatebox{-71}{\bf{\big /}}}
\Text(327,39.1)[c]{\rotatebox{-71}{\bf{\big /}}}
\Text(327,35.5)[c]{\rotatebox{-71}{\bf{\big /}}}
\Text(327,34.9)[c]{\rotatebox{-71}{\bf{\big /}}}
\Text(345.6,36.7)[c]{\rotatebox{-71}{\bf{\big /}}}
\Text(345.6,37.3)[c]{\rotatebox{-71}{\bf{\big /}}}

\CArc(405,50)(5,0,360)
\Line(403,52)(407,48)
\Line(407,52)(403,48)
\CArc(435,50)(25,0,360)
\CArc(465,50)(5,0,360)
\Line(463,52)(467,48)
\Line(467,52)(463,48)
\PhotonArc(448,50)(10,90,450){2}{11}
\PhotonArc(448,50)(8.8,90,450){2}{11}
\Vertex(460,50){2}
\PhotonArc(434,67)(14,166,280){-2}{5}
\GCirc(437.5,53){1}{1}
\Text(449.7,60)[c]{\rotatebox{19}{\bf{\big /}}}
\Text(450.3,60)[c]{\rotatebox{19}{\bf{\big /}}}
\Text(449.7,40)[c]{\rotatebox{19}{\bf{\big /}}}
\Text(450.3,40)[c]{\rotatebox{19}{\bf{\big /}}}

\CArc(205,-15)(5,0,360)
\Line(203,-17)(207,-13)
\Line(207,-17)(203,-13)
\CArc(235,-15)(25,0,360)
\CArc(265,-15)(5,0,360)
\Line(263,-17)(267,-13)
\Line(267,-17)(263,-13)
\PhotonArc(243,-25)(10,200,560){2}{11}
\PhotonArc(243,-25)(8.8,200,560){2}{11}
\Vertex(250,-35){2}
\PhotonArc(230,-31)(14,76,180){2}{5}
\GCirc(234.5,-18.5){1}{1}
\Text(237,-30)[c]{\rotatebox{155}{\bf{\big /}}}
\Text(236.6,-29.6)[c]{\rotatebox{155}{\bf{\big /}}}
\Text(239,-32)[c]{\rotatebox{155}{\bf{\big /}}}
\Text(239.4,-32.4)[c]{\rotatebox{155}{\bf{\big /}}}
\Text(251,-19)[c]{\rotatebox{155}{\bf{\big /}}}
\Text(250.6,-18.6)[c]{\rotatebox{155}{\bf{\big /}}}

\CArc(305,-15)(5,0,360)
\Line(303,-17)(307,-13)
\Line(307,-17)(303,-13)
\CArc(335,-15)(25,0,360)
\CArc(365,-15)(5,0,360)
\Line(363,-17)(367,-13)
\Line(367,-17)(363,-13)
\PhotonArc(335,-2)(10,180,540){2}{11}
\PhotonArc(335,-2)(8.8,180,540){2}{11}
\Photon(335,-40)(335,-12){2}{6}
\Vertex(335,10){2}
\GCirc(335,-11){1}{1}
\Text(327,-1.5)[c]{\rotatebox{-71}{\bf{\big /}}}
\Text(327,-0.9)[c]{\rotatebox{-71}{\bf{\big /}}}
\Text(327,-4.5)[c]{\rotatebox{-71}{\bf{\big /}}}
\Text(327,-5.1)[c]{\rotatebox{-71}{\bf{\big /}}}
\Text(345.6,-3.3)[c]{\rotatebox{-71}{\bf{\big /}}}
\Text(345.6,-2.7)[c]{\rotatebox{-71}{\bf{\big /}}}

\CArc(405,-15)(5,0,360)
\Line(403,-17)(407,-13)
\Line(407,-17)(403,-13)
\CArc(435,-15)(25,0,360)
\CArc(465,-15)(5,0,360)
\Line(463,-17)(467,-13)
\Line(467,-17)(463,-13)
\PhotonArc(448,-15)(10,90,450){2}{11}
\PhotonArc(448,-15)(8.8,90,450){2}{11}
\Vertex(460,-15){2}
\PhotonArc(433,-32)(14,436,550){2}{5}
\GCirc(437,-18){1}{1}
\Text(449.7,-5)[c]{\rotatebox{19}{\bf{\big /}}}
\Text(450.3,-5)[c]{\rotatebox{19}{\bf{\big /}}}
\Text(449.7,-25)[c]{\rotatebox{19}{\bf{\big /}}}
\Text(450.3,-25)[c]{\rotatebox{19}{\bf{\big /}}}

\end{picture}
\end{center}

We next consider  diagrams (w),(x) and (y), and the
corresponding ghost and fermion diagrams (z$_1$),$\dots$, (z$_6$). 
Let us introduce the one-loop gluon self-energy
$$
\begin{picture}(100,30)(30,10)

\Text(-45,25)[l]{$\Pi_{\mu\nu}(q,\lambda)\equiv\ \frac12$}

\Photon(20,25)(40,25){2}{4}
\GlueArc(52,25)(12,0,180){2}{6}
\GlueArc(52,25)(12,180,360){2}{6}
\Photon(64,25)(84,25){2}{4}
\GCirc(39,25){1}{1}
\GCirc(65,25){1}{1}

\Text(89.5,25)[c]{$+$}
\Text(168.5,25)[c]{$+$}

\Photon(95,25)(115,25){2}{4}
\DashCArc(129,25)(14,0,360){1}
\Photon(143,25)(163,25){2}{4}

\Photon(174,25)(194,25){2}{4}
\CArc(208,25)(14,0,360)
\Photon(222,25)(242,25){2}{4}

\end{picture}
$$
which, due to its transversality, satisfies
\begin{equation}
q^\mu\Pi_{\mu\nu}(q,\lambda)=0.
\end{equation}
This fact will then imply that as far as the ghost and fermion diagrams 
(z$_1$),$\dots$,(z$_6$) are concerned, one can effectively
fix the Feynman gauge $\lambda=0$ right from
the start, while for the diagrams (w), (x), and (y) the above transversality 
condition has the consequence of putting the external propagators 
({\it i.e.}
those touching the fermion loop) in the Feynman gauge.
Thus in these latter
graphs the pinching momenta can act on the three gluon vertex {\it only}, 
triggering the elementary WI
\begin{equation}
k_1^\mu k_2^\nu\Gamma_{\mu\nu\rho}(k_1,k_2,\ell)=
-\frac12\ell^2\left(k_1-k_2\right)_{\rho}+\frac12\ell\cdot\left(
k_1-k_2\right)\ell_{\rho}.
\label{due}
\end{equation}
The first term represents an inverse propagator times a momentum which in
general cannot pinch, whereas the second term represents  
an effective three gluon vertex times a pinching momentum. Thus, for example,

\begin{center}
\begin{picture}(0,60)(340,20)

\Text(170,50)[l]{(x) $\to$}
\Text(275,50)[l]{$=\ \lambda^2$}
\Text(375,50)[l]{$-\ \lambda^2$}

\CArc(205,50)(5,0,360)
\Line(203,52)(207,48)
\Line(207,52)(203,48)
\CArc(235,50)(25,0,360)
\CArc(265,50)(5,0,360)
\Line(263,52)(267,48)
\Line(267,52)(263,48)
\PhotonArc(235,46.5)(10,-1,359){2}{11}
\PhotonArc(235,46.5)(8.8,-1,359){2}{11}
\PhotonArc(235,33.5)(18,130,182){2}{4}
\PhotonArc(235,33.5)(18,-1,52){-2}{4}
\GCirc(224,48){1}{1}\GCirc(246,48){1}{1}
\Text(236.5,55.9)[c]{\rotatebox{19}{\bf{\big /}}}
\Text(237.1,55.9)[c]{\rotatebox{19}{\bf{\big /}}}
\Text(234.5,55.9)[c]{\rotatebox{19}{\bf{\big /}}}
\Text(233.9,55.9)[c]{\rotatebox{19}{\bf{\big /}}}
\Text(236.5,37.1)[c]{\rotatebox{19}{\bf{\big /}}}
\Text(237.1,37.1)[c]{\rotatebox{19}{\bf{\big /}}}
\Text(234.5,37.1)[c]{\rotatebox{19}{\bf{\big /}}}
\Text(233.9,37.1)[c]{\rotatebox{19}{\bf{\big /}}}

\CArc(305,50)(5,0,360)
\Line(303,52)(307,48)
\Line(307,52)(303,48)
\CArc(335,50)(25,0,360)
\CArc(365,50)(5,0,360)
\Line(363,52)(367,48)
\Line(367,52)(363,48)
\PhotonArc(327,40)(10,-5,355){2}{11}
\PhotonArc(327,40)(8.8,-5,355){2}{11}
\PhotonArc(336,33.5)(14,-13,90){-2}{5}
\GCirc(336,46.5){1}{1}
\Vertex(319.5,30.5){2}
\Text(334,33)[c]{\rotatebox{65}{\bf{\big /}}}
\Text(333.6,32.6)[c]{\rotatebox{65}{\bf{\big /}}}
\Text(336,35)[c]{\rotatebox{65}{\bf{\big /}}}
\Text(336.4,35.4)[c]{\rotatebox{65}{\bf{\big /}}}
\Text(322,46)[c]{\rotatebox{65}{\bf{\big /}}}
\Text(321.6,45.6)[c]{\rotatebox{65}{\bf{\big /}}}

\CArc(405,50)(5,0,360)
\Line(403,52)(407,48)
\Line(407,52)(403,48)
\CArc(435,50)(25,0,360)
\CArc(465,50)(5,0,360)
\Line(463,52)(467,48)
\Line(467,52)(463,48)
\PhotonArc(435,50)(10,0,360){2}{11}
\PhotonArc(435,50)(8.8,0,360){2}{11}
\Photon(410,50)(421,50){2}{3}
\PhotonArc(435,25)(22,26,62){-2}{4}
\GCirc(445,45.3){1}{1}
\put(426,50){\rotatebox{45}{\GBoxc(0,0)(5,5){0}}}
\Vertex(410,50){2}
\Text(436.7,59.4)[c]{\rotatebox{19}{\bf{\big /}}}
\Text(437.3,59.4)[c]{\rotatebox{19}{\bf{\big /}}}
\Text(436.7,40.6)[c]{\rotatebox{19}{\bf{\big /}}}
\Text(437.3,40.6)[c]{\rotatebox{19}{\bf{\big /}}}

\end{picture}
\end{center}
where we did not draw a tadpole-like diagram since it is zero upon integration 
in the internal momenta, and the black box represents the unphysical 
effective vertex defined in (\ref{due}), {\it i.e.}

\begin{center}
\begin{equation}
\begin{picture}(0,40)(25,10)

\Photon(0,25)(20.4,25){2}{4}
\Gluon(25,27.5)(40,40){2}{5}
\Gluon(25,22.5)(40,10){-2}{5}
\put(23.8,25){\rotatebox{45}{\GBoxc(0,0)(5,5){0}}}

\Text(42,40)[l]{\footnotesize{$k_1$}}
\Text(42,10)[l]{\footnotesize{$k_2$}}
\Text(0,32)[l]{\footnotesize{$\ell$}}

\Text(50,25)[l]{$\equiv i\frac12\ell\cdot\left(
k_1-k_2\right)$}

\label{UV1}

\end{picture}
\end{equation}
\end{center}

Proceeding in this way we find (recall that there is a relative minus sign
between diagrams (r),$\dots$,(u) and (w),$\dots$,(y))

\begin{center}
\begin{picture}(0,150)(330,-45)

\Text(335,95)[c]{{\rm (r)+(s)+(t)+(u)+(w)+(x)+(y)}$\to$}

\Text(180,50)[l]{$2 \lambda^2$}
\Text(275,50)[l]{$-\ \lambda^2$}
\Text(375,50)[l]{$-\ 2\lambda^2$}
\Text(165,-15)[l]{$=\ 2 \lambda^2$}
\Text(275,-15)[l]{$-\ \lambda^2$}

\Text(195,-40)[lb]{\footnotesize{$(\eta)$}}
\Text(295,-40)[lb]{\footnotesize{$(\theta)$}}

\CArc(205,50)(5,0,360)
\Line(203,52)(207,48)
\Line(207,52)(203,48)
\CArc(235,50)(25,0,360)
\CArc(265,50)(5,0,360)
\Line(263,52)(267,48)
\Line(267,52)(263,48)
\PhotonArc(243,60)(10,210,570){2}{11}
\PhotonArc(243,60)(8.8,210,570){2}{11}
\Vertex(250,69.5){2}
\PhotonArc(234,67)(14,165,270){-2}{5}
\GCirc(234.5,53.5){1}{1}
\Text(237,65)[c]{\rotatebox{65}{\bf{\big /}}}
\Text(236.6,64.6)[c]{\rotatebox{65}{\bf{\big /}}}
\Text(239,67)[c]{\rotatebox{65}{\bf{\big /}}}
\Text(239.4,67.4)[c]{\rotatebox{65}{\bf{\big /}}}
\Text(251,54)[c]{\rotatebox{65}{\bf{\big /}}}
\Text(251.4,54.4)[c]{\rotatebox{65}{\bf{\big /}}}

\CArc(305,50)(5,0,360)
\Line(303,52)(307,48)
\Line(307,52)(303,48)
\CArc(335,50)(25,0,360)
\CArc(365,50)(5,0,360)
\Line(363,52)(367,48)
\Line(367,52)(363,48)
\PhotonArc(335,37)(10,0,360){2}{11}
\PhotonArc(335,37)(8.8,0,360){2}{11}
\Photon(335,75)(335,47){2}{6}
\Vertex(335,25){2}
\GCirc(335,46){1}{1}
\Text(327,38.5)[c]{\rotatebox{-71}{\bf{\big /}}}
\Text(327,39.1)[c]{\rotatebox{-71}{\bf{\big /}}}
\Text(327,35.5)[c]{\rotatebox{-71}{\bf{\big /}}}
\Text(327,34.9)[c]{\rotatebox{-71}{\bf{\big /}}}
\Text(345.6,36.7)[c]{\rotatebox{-71}{\bf{\big /}}}
\Text(345.6,37.3)[c]{\rotatebox{-71}{\bf{\big /}}}

\CArc(410,50)(5,0,360)
\Line(408,52)(412,48)
\Line(412,52)(408,48)
\CArc(440,50)(25,0,360)
\CArc(470,50)(5,0,360)
\Line(468,52)(472,48)
\Line(472,52)(468,48)
\PhotonArc(453,50)(10,90,450){2}{11}
\PhotonArc(453,50)(8.8,90,450){2}{11}
\Vertex(465,50){2}
\PhotonArc(439,67)(14,166,280){-2}{5}
\GCirc(442.5,53){1}{1}
\Text(454.7,60)[c]{\rotatebox{19}{\bf{\big /}}}
\Text(455.3,60)[c]{\rotatebox{19}{\bf{\big /}}}
\Text(454.7,40)[c]{\rotatebox{19}{\bf{\big /}}}
\Text(455.3,40)[c]{\rotatebox{19}{\bf{\big /}}}

\CArc(205,-15)(5,0,360)
\Line(203,-17)(207,-13)
\Line(207,-17)(203,-13)
\CArc(235,-15)(25,0,360)
\CArc(265,-15)(5,0,360)
\Line(263,-17)(267,-13)
\Line(267,-17)(263,-13)
\PhotonArc(243,-5)(10,210,570){2}{11}
\PhotonArc(243,-5)(8.8,210,570){2}{11}
\Vertex(250,4.5){2}
\PhotonArc(234,2)(14,165,270){-2}{5}
\GCirc(234.5,-11.5){1}{1}
\Text(237,0)[c]{\rotatebox{65}{\bf{\big /}}}
\Text(236.6,-0.4)[c]{\rotatebox{65}{\bf{\big /}}}
\Text(239,2)[c]{\rotatebox{65}{\bf{\big /}}}
\Text(239.4,2.4)[c]{\rotatebox{65}{\bf{\big /}}}
\Text(251,-11)[c]{\rotatebox{65}{\bf{\big /}}}
\Text(251.4,-10.6)[c]{\rotatebox{65}{\bf{\big /}}}

\CArc(305,-15)(5,0,360)
\Line(303,-17)(307,-13)
\Line(307,-17)(303,-13)
\CArc(335,-15)(25,0,360)
\CArc(365,-15)(5,0,360)
\Line(363,-17)(367,-13)
\Line(367,-17)(363,-13)
\PhotonArc(335,-28)(10,0,360){2}{11}
\PhotonArc(335,-28)(8.8,0,360){2}{11}
\Photon(335,10)(335,-18){2}{6}
\Vertex(335,-40){2}
\GCirc(335,-19){1}{1}
\Text(345.6,-26.5)[c]{\rotatebox{-71}{\bf{\big /}}}
\Text(345.6,-25.9)[c]{\rotatebox{-71}{\bf{\big /}}}
\Text(345.6,-29.5)[c]{\rotatebox{-71}{\bf{\big /}}}
\Text(345.6,-30.1)[c]{\rotatebox{-71}{\bf{\big /}}}
\Text(327,-28.3)[c]{\rotatebox{-71}{\bf{\big /}}}
\Text(327,-27.7)[c]{\rotatebox{-71}{\bf{\big /}}}

\end{picture}
\end{center}
where the last step is achieved by allowing 
the second diagram in the
first line to pinch further. 
The coefficient multiplying this equation is $C_fC_A/2$.
Notice that the effective vertex introduced in Eq.(\ref{UV1})
does not appear at this point.

We can finally act on the three gluon vertex with the remaining pinching
momenta, to obtain

\begin{center}
\begin{picture}(0,135)(320,-45)

\Text(150,50)[l]{$(\eta) =\ 2\lambda^2$}

\Text(200,25)[lb]{\footnotesize{(i)}}
\Text(305,25)[lb]{\footnotesize{(ii)}}
\Text(200,-40)[lb]{\footnotesize{(iii)}}
\Text(300,-40)[lb]{\footnotesize{(vi)}}
\Text(400,-40)[lb]{\footnotesize{(v)}}

\Text(275,50)[l]{$+\ 2\lambda^2$}
\Text(155,-15)[l]{$(\theta) =\ \lambda^2$}
\Text(275,-15)[l]{$+\ \lambda^2$}
\Text(375,-15)[l]{$-\ \lambda^2$}

\CArc(205,50)(5,0,360)
\Line(203,52)(207,48)
\Line(207,52)(203,48)
\CArc(235,50)(25,0,360)
\CArc(265,50)(5,0,360)
\Line(263,52)(267,48)
\Line(267,52)(263,48)
\PhotonArc(235,65)(15,165,380){2}{10}
\PhotonArc(235,65)(13.8,165,380){2}{10}
\PhotonArc(235,67.5)(15.4,16.5,165){2}{7.5}
\PhotonArc(235,67.5)(14.2,16.5,165){2}{7.5}
\Text(233.5,82.3)[c]{\rotatebox{19}{\bf{\big /}}}
\Text(232.9,82.3)[c]{\rotatebox{19}{\bf{\big /}}}
\Text(236.5,82.3)[c]{\rotatebox{19}{\bf{\big /}}}
\Text(237.1,82.3)[c]{\rotatebox{19}{\bf{\big /}}}
\Vertex(249.5,70.5){2}
\Vertex(220,70.5){2}

\CArc(310,50)(5,0,360)
\Line(308,52)(312,48)
\Line(312,52)(308,48)
\CArc(340,50)(25,0,360)
\CArc(370,50)(5,0,360)
\Line(368,52)(372,48)
\Line(372,52)(368,48)
\PhotonArc(348,60)(10,210,570){2}{11}
\PhotonArc(348,60)(8.8,210,570){2}{11}
\PhotonArc(315,85)(35,270,310){-2}{5}
\Vertex(355,69.5){2}
\put(338.5,57.7){\rotatebox{45}{\GBoxc(0,0)(5,5){0}}}
\Vertex(315,50){2}
\Text(343,66)[c]{\rotatebox{65}{\bf{\big /}}}
\Text(343.4,66.4)[c]{\rotatebox{65}{\bf{\big /}}}

\CArc(205,-15)(5,0,360)
\Line(203,-17)(207,-13)
\Line(207,-17)(203,-13)
\CArc(235,-15)(25,0,360)
\CArc(265,-15)(5,0,360)
\Line(263,-17)(267,-13)
\Line(267,-17)(263,-13)
\PhotonArc(181,-15.7)(60,337,24){-2}{9}
\PhotonArc(181,-15.7)(61.2,337,24){-2}{9}
\PhotonArc(290,-15)(61.2,157,203.8){2}{9}
\PhotonArc(290,-15)(60,157,203.8){2}{9}
\Vertex(235,10){2}
\Vertex(235,-40){2}
\Text(229.8,-13.5)[c]{\rotatebox{-71}{\bf{\big /}}}
\Text(229.8,-12.9)[c]{\rotatebox{-71}{\bf{\big /}}}
\Text(229.8,-16.5)[c]{\rotatebox{-71}{\bf{\big /}}}
\Text(229.8,-17.1)[c]{\rotatebox{-71}{\bf{\big /}}}

\CArc(305,-15)(5,0,360)
\Line(303,-17)(307,-13)
\Line(307,-17)(303,-13)
\CArc(335,-15)(25,0,360)
\CArc(365,-15)(5,0,360)
\Line(363,-17)(367,-13)
\Line(367,-17)(363,-13)
\PhotonArc(343,-25)(10,204,564){2}{11}
\PhotonArc(343,-25)(8.8,204,564){2}{11}
\PhotonArc(310,-50)(35,50,90){-2}{5}
\Vertex(350.5,-34){2}
\put(335,-23.3){\rotatebox{45}{\GBoxc(0,0)(5,5){0}}}
\Vertex(310,-15){2}
\Text(350.8,-17.8)[c]{\rotatebox{155}{\bf{\big /}}}
\Text(350.4,-17.4)[c]{\rotatebox{155}{\bf{\big /}}}

\CArc(405,-15)(5,0,360)
\Line(403,-17)(407,-13)
\Line(407,-17)(403,-13)
\CArc(435,-15)(25,0,360)
\CArc(465,-15)(5,0,360)
\Line(463,-17)(467,-13)
\Line(467,-17)(463,-13)
\PhotonArc(427,-25)(10,220,580){2}{11}
\PhotonArc(427,-25)(8.8,220,580){2}{11}
\PhotonArc(460,-50)(35,90,130){-2}{5}
\Vertex(419.5,-34){2}
\Vertex(460,-15){2}
\put(439.5,-23){\rotatebox{45}{\GBoxc(0,0)(5,5){0}}}
\Text(420.8,-20.6)[c]{\rotatebox{65}{\bf{\big /}}}
\Text(420.4,-21)[c]{\rotatebox{65}{\bf{\big /}}}

\end{picture}
\end{center}
It is then clear how the final steps proceed:
the combination ${\rm (i)+(iii)}$ cancels the ${\cal O}(\lambda ^2)$ 
non-Abelian remainder of Eq.(\ref{AR}), while, as can be easily shown 
\begin{equation}
{\rm (ii)+(vi)+(v)}=0.
\end{equation}
This completes the proof of the cancellation of the 
${\cal O}(\lambda^2)$ terms.

\subsubsection{The ${\cal O}(\lambda)$ cancellation.}

As in the previous case, the strategy will be to achieve
the widest possible
cancellation between diagrams, avoiding 
to act on the three 
gluon vertex. First of all,
each one of the diagrams (s),$\dots$,(u)
will again generate three contributions, which are obtained from the ${\cal
O}(\lambda^2)$ ones by trading one of the propagators 
$\begin{picture}(10,5)(0,0)
\SetScale{0.8}
\Photon(5,4)(40,4){2}{6}
\Photon(5,2.5)(40,2.5){2}{6}
\Text(19,3.25)[c]{\rotatebox{19}{\bf{\big /}}}
\Text(19.4,3.25)[c]{\rotatebox{19}{\bf{\big /}}}
\Text(16,3.25)[c]{\rotatebox{19}{\bf{\big /}}}
\Text(15.6,3.25)[c]{\rotatebox{19}{\bf{\big /}}}
\end{picture}\qquad \ $
for a Feynman propagator.
Then, taking all these diagrams into account, we arrive at the equation  

\begin{center}
\begin{picture}(0,50)(360,30)

\Text(230,50)[l]{(r)+(s)+(t)+(u) $\to\ 4\lambda$}
\Text(415,50)[l]{$+\ 8\lambda$}

\CArc(345,50)(5,0,360)
\Line(343,52)(347,48)
\Line(347,52)(343,48)
\CArc(375,50)(25,0,360)
\CArc(405,50)(5,0,360)
\Line(403,52)(407,48)
\Line(407,52)(403,48)
\PhotonArc(375,63)(8.8,450,270){2}{5.5}
\PhotonArc(375,63)(10,450,270){2}{5.5}
\PhotonArc(375,63)(10,270,450){-2}{6.5}
\Photon(375,25)(375,53){2}{6}
\Vertex(373.5,75){2}
\GCirc(375,54){1}{1}
\Text(366.6,63.3)[c]{\rotatebox{-71}{\bf{\big /}}}
\Text(366.6,62.7)[c]{\rotatebox{-71}{\bf{\big /}}}

\CArc(445,50)(5,0,360)
\Line(443,52)(447,48)
\Line(447,52)(443,48)
\CArc(475,50)(25,0,360)
\CArc(505,50)(5,0,360)
\Line(503,52)(507,48)
\Line(507,52)(503,48)
\PhotonArc(467,60)(10,125,315){2}{5.5}
\PhotonArc(467,60)(8.8,125,315){2}{5.5}
\PhotonArc(467,60)(10,315,125){-2}{6.5}
\Vertex(459,69){2}
\PhotonArc(476,67)(14,262,371){2}{5}
\GCirc(474,54.5){1}{1}
\Text(463,53)[c]{\rotatebox{155}{\bf{\big /}}}
\Text(462.6,53.4)[c]{\rotatebox{155}{\bf{\big /}}}

\end{picture}
\end{center}

We are now left with the topologies (w),$\dots$,(z$_3$). As in the ${\cal
O}(\lambda^2)$ case, for the external propagators 
$\lambda=0$. However, contrary
to the previous case where each one gave a single contribution, at ${\cal
O}(\lambda)$ each of the topologies (w), (x) and (y)   
gives rise to two equal contributions (hence the factor of 2 in 
Eq.(\ref{yyy}), below). 
Moreover, for these diagrams 
the pinching momenta can only act on the three gluon vertex,
triggering the elementary WI
\begin{equation}
k_1^\mu\Gamma_{\mu\nu\rho}(k_1,k_2,\ell)=(\ell^2g_{\nu\rho}-
\ell_{\nu}\ell_{\rho})-
(k_2^2g_{\nu\rho}-k_{2\nu}k_{2\rho}).
\label{wi3}
\end{equation}
The first and third term of the above expression
represent two inverse propagators in the Feynman gauge; 
the second and fourth terms contain instead two
longitudinal momenta each, one acting on the external fermion loop, 
and the other
one on the remaining three gluon vertex.
For example, considering diagram (w), we find

\begin{center}
\begin{equation}
\begin{picture}(0,120)(300,-40)

\Text(150,50)[l]{(w) $\to\ 2 \lambda$}

\Text(200,-40)[lb]{\footnotesize{$(\mu)$}}
\Text(300,-40)[lb]{\footnotesize{$(\nu)$}}
\Text(400,-40)[lb]{\footnotesize{$(\xi)$}}

\Text(170,-15)[l]{$=\ 2\lambda$}
\Text(275,-15)[l]{$+\ 2\lambda$}
\Text(375,-15)[l]{$+\ 2\lambda$}

\CArc(205,50)(5,0,360)
\Line(203,52)(207,48)
\Line(207,52)(203,48)
\CArc(235,50)(25,0,360)
\CArc(265,50)(5,0,360)
\Line(263,52)(267,48)
\Line(267,52)(263,48)
\PhotonArc(235,46.5)(10,-1,179){2}{5.5}
\PhotonArc(235,46.5)(8.8,-1,179){2}{5.5}
\PhotonArc(235,46.5)(10,179,-1){-2}{6.5}
\PhotonArc(235,33.5)(18,130,182){2}{4}
\PhotonArc(235,33.5)(18,-1,52){-2}{4}
\GCirc(224,48){1}{1}
\GCirc(246,48){1}{1}
\Text(237.5,55.9)[c]{\rotatebox{19}{\bf{\big /}}}
\Text(238.1,55.9)[c]{\rotatebox{19}{\bf{\big /}}}
\Text(234.5,55.9)[c]{\rotatebox{19}{\bf{\big /}}}
\Text(233.9,55.9)[c]{\rotatebox{19}{\bf{\big /}}}

\CArc(205,-15)(5,0,360)
\Line(203,-17)(207,-13)
\Line(207,-17)(203,-13)
\CArc(235,-15)(25,0,360)
\CArc(265,-15)(5,0,360)
\Line(263,-17)(267,-13)
\Line(267,-17)(263,-13)
\PhotonArc(227,-25)(10,35,225){2}{5.5}
\PhotonArc(227,-25)(8.8,35,225){2}{5.5}
\PhotonArc(227,-25)(10,225,35){-2}{6.5}
\PhotonArc(236,-31.5)(14,-13,90){-2}{5}
\GCirc(235.5,-18.5){1}{1}
\Vertex(218,-32.5){2}
\Text(222,-19)[c]{\rotatebox{65}{\bf{\big /}}}
\Text(221.6,-19.4)[c]{\rotatebox{65}{\bf{\big /}}}

\CArc(305,-15)(5,0,360)
\Line(303,-17)(307,-13)
\Line(307,-17)(303,-13)
\CArc(335,-15)(25,0,360)
\CArc(365,-15)(5,0,360)
\Line(363,-17)(367,-13)
\Line(367,-17)(363,-13)
\PhotonArc(335,-15)(10,-20,180){2}{5.5}
\PhotonArc(335,-15)(8.8,-20,180){2}{5.5}
\PhotonArc(335,-15)(10,180,-20){-2}{6.5}
\Photon(310,-15)(321,-15){2}{3}
\PhotonArc(335,-40)(22,26,65){-2}{4}
\GCirc(344.2,-18.7){1}{1}
\put(325.5,-15){\rotatebox{45}{\GBoxc(0,0)(5,5){1}}}
\Vertex(310,-15){2}
\Text(318.1,-15)[c]{\rotatebox{19}{\bf{\big /}}}
\Text(317.5,-15)[c]{\rotatebox{19}{\bf{\big /}}}
\Text(337.3,-5.6)[c]{\rotatebox{19}{\bf{\big /}}}
\Text(336.7,-5.6)[c]{\rotatebox{19}{\bf{\big /}}}

\CArc(405,-15)(5,0,360)
\Line(403,-17)(407,-13)
\Line(407,-17)(403,-13)
\CArc(435,-15)(25,0,360)
\CArc(465,-15)(5,0,360)
\Line(463,-17)(467,-13)
\Line(467,-17)(463,-13)
\PhotonArc(435,-18.5)(10,-1,359){2}{11}
\PhotonArc(435,-18.5)(8.8,-1,179){2}{5.5}
\PhotonArc(435,-31.5)(18,130,182){2}{4}
\PhotonArc(435,-31.5)(18,-1,52){-2}{4}
\put(427,-17){\rotatebox{45}{\GBoxc(0,0)(5,5){1}}}
\GCirc(446,-17){1}{1}
\Text(436.3,-9.1)[c]{\rotatebox{19}{\bf{\big /}}}
\Text(436.9,-9.1)[c]{\rotatebox{19}{\bf{\big /}}}
\Text(435.1,-27.9)[c]{\rotatebox{19}{\bf{\big /}}}
\Text(434.5,-27.9)[c]{\rotatebox{19}{\bf{\big /}}}
\Text(438.1,-27.9)[c]{\rotatebox{19}{\bf{\big /}}}
\Text(438.7,-27.9)[c]{\rotatebox{19}{\bf{\big /}}}

\label{yyy}

\end{picture}
\end{equation}
\end{center}

The diagram $(\mu)$ comes from the first term in 
Eq.(\ref{wi3}), while the third
term in the WI produces in this case a tadpole-like diagram 
which is odd in the integrated momentum and vanishes. The terms 
$(\nu)$ and $(\xi)$ are generated from the second 
and fourth
term in Eq.(\ref{wi3}), respectively. Notice that
a new unphysical effective three gluon vertex
\begin{center}
\begin{picture}(0,40)(25,10)

\Photon(0,25)(20,25){2}{4}
\put(23.8,25){\rotatebox{45}{\GBoxc(0,0)(5,5){1}}}
\Gluon(25,27.5)(40,40){2}{5}
\Gluon(25,22.5)(40,10){-2}{5}

\Text(42,40)[l]{\footnotesize{$k_1$}}
\Text(42,10)[l]{\footnotesize{$k_2$}}
\Text(0,32)[l]{\footnotesize{$\ell$}}
\Text(50,25)[l]{$\equiv i$}

\end{picture}
\end{center}
has appeared.
After acting on the remaining three gluon
vertex of diagram $(\xi)$ we get 

\begin{center}
\begin{picture}(0,50)(290,25)

\Text(155,50)[l]{$(\xi)=\ 2\lambda$}
\Text(275,50)[l]{$+\ 2\lambda$}

\CArc(205,50)(5,0,360)
\Line(203,52)(207,48)
\Line(207,52)(203,48)
\CArc(235,50)(25,0,360)
\CArc(265,50)(5,0,360)
\Line(263,52)(267,48)
\Line(267,52)(263,48) 
\PhotonArc(243,40)(10,150,305){-2}{6.5}
\PhotonArc(243,40)(10,305,150){2}{5.5}
\PhotonArc(243,40)(8.8,305,150){2}{5.5}
\Vertex(251,31.5){2}
\PhotonArc(230,34)(14,76,180){2}{5}
\put(235,47.5){\rotatebox{45}{\GBoxc(0,0)(5,5){1}}}
\Text(237,35)[c]{\rotatebox{155}{\bf{\big /}}}
\Text(236.6,35.4)[c]{\rotatebox{155}{\bf{\big /}}}
\Text(239,33)[c]{\rotatebox{155}{\bf{\big /}}}
\Text(239.4,32.6)[c]{\rotatebox{155}{\bf{\big /}}}

\CArc(305,50)(5,0,360)
\Line(303,52)(307,48)
\Line(307,52)(303,48)
\CArc(335,50)(25,0,360)
\CArc(365,50)(5,0,360)
\Line(363,52)(367,48)
\Line(367,52)(363,48)
\PhotonArc(335,50)(10,10,200){2}{5.5}
\PhotonArc(335,50)(8.8,10,200){2}{5.5}
\PhotonArc(335,50)(10,200,10){-2}{6.5}
\Photon(360,50)(348.5,50){2}{3}
\PhotonArc(335,25)(22,115,154){2}{4}
\put(326.5,45){\rotatebox{45}{\GBoxc(0,0)(5,5){1}}}
\put(346.5,50.5){\rotatebox{45}{\GBoxc(0,0)(5,5){0}}}
\Text(336.9,40.6)[c]{\rotatebox{19}{\bf{\big /}}}
\Text(336.3,40.6)[c]{\rotatebox{19}{\bf{\big /}}}

\end{picture}
\end{center}
Proceeding in this way we arrive at the result

\begin{center}
\begin{picture}(0,100)(510,5)

\Text(525,95)[c]{(r)+(s)+(t)+(u)+(w)+(x)+(y) $\to$}
\Text(415,50)[l]{$+\ 4\lambda$}
\Text(515,50)[l]{$-\ 2\lambda$}
\Text(615,50)[l]{$-\ 4\lambda$}
\Text(325,50)[l]{$2\lambda$}

\Text(340,25)[lb]{\footnotesize{$(\pi)$}}
\Text(440,25)[lb]{\footnotesize{$(\rho)$}}
\Text(540,25)[lb]{\footnotesize{$(\sigma)$}}
\Text(640,25)[lb]{\footnotesize{$(\tau)$}}

\CArc(345,50)(5,0,360)
\Line(343,52)(347,48)
\Line(347,52)(343,48)
\CArc(375,50)(25,0,360)
\CArc(405,50)(5,0,360)
\Line(403,52)(407,48)
\Line(407,52)(403,48)
\PhotonArc(375,63)(8.8,450,270){2}{5.5}
\PhotonArc(375,63)(10,450,270){2}{5.5}
\PhotonArc(375,63)(10,270,450){-2}{6.5}
\Photon(375,25)(375,53){2}{6}
\Vertex(374,75){2}
\GCirc(375,54){1}{1}
\Text(367,62.8)[c]{\rotatebox{-71}{\bf{\big /}}}
\Text(367,63.4)[c]{\rotatebox{-71}{\bf{\big /}}}

\CArc(445,50)(5,0,360)
\Line(443,52)(447,48)
\Line(447,52)(443,48)
\CArc(475,50)(25,0,360)
\CArc(505,50)(5,0,360)
\Line(503,52)(507,48)
\Line(507,52)(503,48)
\PhotonArc(467,60)(10,125,315){2}{5.5}
\PhotonArc(467,60)(8.8,125,315){2}{5.5}
\PhotonArc(467,60)(10,315,125){-2}{6.5}
\Vertex(460,69.5){2}
\PhotonArc(476,67)(14,262,371){2}{5}
\GCirc(474,54.5){1}{1}
\Text(462,55)[c]{\rotatebox{155}{\bf{\big /}}}
\Text(462.4,54.6)[c]{\rotatebox{155}{\bf{\big /}}}

\CArc(545,50)(5,0,360)
\Line(543,52)(547,48)
\Line(547,52)(543,48)
\CArc(575,50)(25,0,360)
\CArc(605,50)(5,0,360)
\Line(603,52)(607,48)
\Line(607,52)(603,48)
\PhotonArc(575,37)(10,90,270){-2}{6.5}
\PhotonArc(575,37)(10,270,90){2}{5.5}
\PhotonArc(575,37)(8.8,270,90){2}{5.5}
\Photon(575,75)(575,48.5){2}{6}
\Vertex(576.5,25){2}
\put(577,48){\rotatebox{45}{\GBoxc(0,0)(5,5){1}}}
\Text(586.4,38.5)[c]{\rotatebox{-71}{\bf{\big /}}}
\Text(586.4,39.1)[c]{\rotatebox{-71}{\bf{\big /}}}
\Text(586.4,35.5)[c]{\rotatebox{-71}{\bf{\big /}}}
\Text(586.4,34.9)[c]{\rotatebox{-71}{\bf{\big /}}}

\CArc(645,50)(5,0,360)
\Line(643,52)(647,48)
\Line(647,52)(643,48)
\CArc(675,50)(25,0,360)
\CArc(705,50)(5,0,360)
\Line(703,52)(707,48)
\Line(707,52)(703,48)
\PhotonArc(683,40)(10,150,305){-2}{6.5}
\PhotonArc(683,40)(10,305,150){2}{5.5}
\PhotonArc(683,40)(8.8,305,150){2}{5.5}
\Vertex(691,31.5){2}
\PhotonArc(670,34)(14,76,180){2}{5}
\put(675.5,46.5){\rotatebox{45}{\GBoxc(0,0)(5,5){1}}}
\Text(677,35)[c]{\rotatebox{155}{\bf{\big /}}}
\Text(676.6,35.4)[c]{\rotatebox{155}{\bf{\big /}}}
\Text(679,33)[c]{\rotatebox{155}{\bf{\big /}}}
\Text(679.4,32.6)[c]{\rotatebox{155}{\bf{\big /}}}

\end{picture}
\end{center}
where the group-theoretical coefficient of this equation is $C_AC_f/2$.

The final step of the proof is achieved 
by acting on the three gluon vertex in $(\rho)$ and $(\pi)$
with
the remaining pinching momenta.
One has, keeping in mind Eq.(\ref{wi3}),

\begin{center}
\begin{picture}(0,130)(370,-45)

\Text(150,50)[l]{$(\pi)=\ 2\lambda$}
\Text(150,-15)[l]{$(\rho)=\ 4\lambda$}

\Text(275,50)[l]{$-\ 4\lambda$}
\Text(375,50)[l]{$+\ 2\lambda$}
\Text(475,50)[l]{$-\ 2\lambda$}

\Text(275,-15)[l]{$+\ 4\lambda$}
\Text(375,-15)[l]{$-\ 4\lambda$}
\Text(475,-15)[l]{$+\ 4\lambda$}

\Text(200,25)[lb]{\footnotesize{(i)}}
\Text(300,25)[lb]{\footnotesize{(ii)}}
\Text(400,25)[lb]{\footnotesize{(iii)}}
\Text(500,25)[lb]{\footnotesize{(iv)}}
\Text(200,-40)[lb]{\footnotesize{(v)}}
\Text(300,-40)[lb]{\footnotesize{(vi)}}
\Text(400,-40)[lb]{\footnotesize{(vii)}}
\Text(500,-40)[lb]{\footnotesize{(viii)}}

\CArc(205,50)(5,0,360)
\Line(203,52)(207,48)
\Line(207,52)(203,48)
\CArc(235,50)(25,0,360)
\CArc(265,50)(5,0,360)
\Line(263,52)(267,48)
\Line(267,52)(263,48)
\PhotonArc(181,49.3)(60,337,24){-2}{9}
\PhotonArc(290,50)(61.2,157,203.8){2}{9}
\PhotonArc(290,50)(60,157,203.8){2}{9}
\Vertex(235,75){2}
\Vertex(235,25){2}

\CArc(305,50)(5,0,360)
\Line(303,52)(307,48)
\Line(307,52)(303,48)
\CArc(335,50)(25,0,360)
\CArc(365,50)(5,0,360)
\Line(363,52)(367,48)
\Line(367,52)(363,48)
\PhotonArc(327,60)(10,120,360){2}{6.5}
\PhotonArc(327,60)(8.8,120,360){2}{6.5}
\PhotonArc(327,60)(10,360,120){-2}{4.5}
\PhotonArc(360,85)(35,230,270){-2}{5}
\Vertex(319.5,69.5){2}
\Vertex(360,50){2}
\put(337.5,58.5){\rotatebox{45}{\GBoxc(0,0)(5,5){1}}}
\Text(350.9,51)[c]{\rotatebox{-7}{\bf{\big /}}}
\Text(350.4,51.1)[c]{\rotatebox{-7}{\bf{\big /}}}

\CArc(405,50)(5,0,360)
\Line(403,52)(407,48)
\Line(407,52)(403,48)
\CArc(435,50)(25,0,360)
\CArc(465,50)(5,0,360)
\Line(463,52)(467,48)
\Line(467,52)(463,48)
\PhotonArc(435,63)(8.8,90,270){2}{5.5}
\PhotonArc(435,63)(10,90,270){2}{5.5}
\PhotonArc(435,63)(10,270,90){-2}{6.5}
\Photon(435,25)(435,53){2}{6}
\Vertex(434,75){2}
\put(436.5,53.7){\rotatebox{45}{\GBoxc(0,0)(5,5){1}}}
\Text(446.4,64.5)[c]{\rotatebox{-71}{\bf{\big /}}}
\Text(446.4,65.1)[c]{\rotatebox{-71}{\bf{\big /}}}
\Text(446.4,61.5)[c]{\rotatebox{-71}{\bf{\big /}}}
\Text(446.4,60.9)[c]{\rotatebox{-71}{\bf{\big /}}}

\CArc(505,50)(5,0,360)
\Line(503,52)(507,48)
\Line(507,52)(503,48)
\CArc(535,50)(25,0,360)
\CArc(565,50)(5,0,360)
\Line(563,52)(567,48)
\Line(567,52)(563,48)
\Photon(545,27)(545,73){2}{9}
\PhotonArc(520,50)(8,90,450){2}{9}
\PhotonArc(520,50)(6.8,90,450){2}{9}
\Vertex(510,50){2}

\CArc(205,-15)(5,0,360)
\Line(203,-17)(207,-13)
\Line(207,-17)(203,-13)
\CArc(235,-15)(25,0,360)
\CArc(265,-15)(5,0,360)
\Line(263,-17)(267,-13)
\Line(267,-17)(263,-13)
\PhotonArc(235,0)(15,165,380){2}{10}
\PhotonArc(235,0)(13.8,165,380){2}{10}
\Vertex(249.5,5.5){2}
\Vertex(220,5.5){2}
\PhotonArc(235,2.5)(15.4,16.5,165){2}{7.5}

\CArc(305,-15)(5,0,360)
\Line(303,-17)(307,-13)
\Line(307,-17)(303,-13)
\CArc(335,-15)(25,0,360)
\CArc(365,-15)(5,0,360)
\Line(363,-17)(367,-13)
\Line(367,-17)(363,-13)
\PhotonArc(360,20)(35,230,270){-2}{5}
\PhotonArc(327,-5)(10,120,360){2}{6.5}
\PhotonArc(327,-5)(8.8,120,360){2}{6.5}
\PhotonArc(327,-5)(10,360,120){-2}{4.5}
\Vertex(319.5,4.5){2}
\Vertex(360,-15){2}
\put(337.5,-6.6){\rotatebox{45}{\GBoxc(0,0)(5,5){1}}}
\Text(350.9,-14)[c]{\rotatebox{-7}{\bf{\big /}}}
\Text(350.4,-13.9)[c]{\rotatebox{-7}{\bf{\big /}}}

\CArc(405,-15)(5,0,360)
\Line(403,-17)(407,-13)
\Line(407,-17)(403,-13)
\CArc(435,-15)(25,0,360)
\CArc(465,-15)(5,0,360)
\Line(463,-17)(467,-13)
\Line(467,-17)(463,-13)
\PhotonArc(420,-15)(8,90,450){2}{9}
\PhotonArc(420,-15)(6.8,90,450){2}{9}
\Vertex(410,-15){2}
\PhotonArc(435,-40)(14.2,16.5,165){2}{7.5}

\CArc(505,-15)(5,0,360)
\Line(503,-17)(507,-13)
\Line(507,-17)(503,-13)
\CArc(535,-15)(25,0,360)
\CArc(565,-15)(5,0,360)
\Line(563,-17)(567,-13)
\Line(567,-17)(563,-13)
\Vertex(519,4){2}
\PhotonArc(536,2)(14,270,371){2}{5}
\PhotonArc(527,-5)(10,125,315){2}{5.5}
\PhotonArc(527,-5)(8.8,125,315){2}{5.5}
\PhotonArc(527,-5)(10,315,125){-2}{6.5}
\put(537.5,-11.5){\rotatebox{45}{\GBoxc(0,0)(5,5){1}}}
\Text(535,1)[c]{\rotatebox{155}{\bf{\big /}}}
\Text(535.4,0.6)[c]{\rotatebox{155}{\bf{\big /}}}
\Text(533,3)[c]{\rotatebox{155}{\bf{\big /}}}
\Text(532.6,3.4)[c]{\rotatebox{155}{\bf{\big /}}}

\end{picture}
\end{center}
Then the sum (i)+(iv)+(v)+(vii) cancels against 
the ${\cal O}(\lambda)$ non-Abelian
remainder (\ref{NAR}), (ii) and (vi) cancel directly, while finally
\begin{equation}
(\tau)+{\rm (viii)}=(\sigma)+{\rm (iii)}=0.
\end{equation}
This complete the proof of the non-Abelian gauge cancellation.

We see that, as happened in the one-loop case,
the GFP-dependent contributions coming from the 
original graphs of Fig.\ref{fig3} defining
$\Sigma^{(2)}(p,\xi)$, 
cancel exactly against
equal but opposite {\it propagator-like} contributions coming from
vertex-like and box-like graphs. 
Thus,
one is left with the ``pure'' GFP-independent
one-loop fermion \mbox{self-energy}, $\widehat{\Sigma}^{(2)}(p)$, which 
again 
concides with the $\Sigma^{(2)}_F (p) \equiv \Sigma^{(2)}(p,1)$, 
{\it i.e.}
\begin{equation}
\widehat{\Sigma}^{(2)}(p) = \Sigma^{(2)}_F (p),
\label{GFPII}
\end{equation}
which constitutes the central result of this paper. 

\section{The absorptive construction: The one-loop case}\label{absorptive}

In the next two sections we  will show in detail how one may construct
the  two-loop PT  effective  fermion self-energy  using unitarity  and
analyticity  arguments \cite{Papavassiliou:1996zn}.  The  general idea
is  the following:  The imaginary  parts  of the  two-loop PT  fermion
self-energies $\widehat{\Sigma}^{(2)}$  of QED and QCD  are related by
the optical  theorem to precisely identifiable and  very special parts
of four different cross-sections.  In particular, for the case of QED,
the two-particle Cutkosky cuts of $\widehat{\Sigma}^{(2)}$ are related
to  the   the  ``genuine''  $s$-channel  part  of   the  the  one-loop
cross-section for the  process $\gamma Q \to \gamma  Q$, while, at the
same time, the  three-particle Cutkosky cuts of the  same quantity are
related  to  the  ``genuine''  $s$-channel  parts  of  the  tree-level
cross-sections for the processes $\gamma  Q \to \gamma Q $, $\gamma Q
\to Q \bar Q Q$, and  $\gamma Q \to  \gamma \gamma Q$.   The corresponding
processes  for the QCD  can be  obtained by  replacing the  photons by
gluons ($\gamma \leftrightarrow G$) in the final states.  The key word
in  the above  description  is the  word  ``genuine'': By  ``genuine''
$s$-channel part we mean the $s$-channel part obtained {\it after} the
longitudinal terms of the polarization vectors involved have triggered
the WIs of the various  amplitudes.  These WIs implement themselves in
ways that do not respect  the original $s-t$ channel separation of the
amplitude, as given  by the Feynman graphs; instead,  various $s$- and
$t$-channel contributions are non-trivially  mixed, in such ways as to
finally  result in fundamental  cancellations. It  turns out  that all
such contributions  can again be  pictorially represented by  means of
unphysical elementary vertices, a fact which facilitates significantly
their identification.

In this section we will set up the formalism, adopted 
to the fermion self-energy,
and discuss in detail the one-loop case; 
the two-loop generalization will be presented
in the next section.

\subsection{QED}

The optical theorem for the case of forward scattering assumes the form 
\begin{equation}
\Im m 
\langle a | T | a \rangle = \frac{1}{2}
\sum_{i} (2\pi)^{4} \delta^{(4)}(p_{a}-p_{i})
{\langle i | T | a \rangle}^{*}
\langle i | T | a \rangle \, ,
\label{ab1}
\end{equation}
where the sum $\sum_{i}$ should be understood to be over the entire phase space
of all allowed on-shell intermediate states $i$.
After expanding the $T$ matrix in powers of $g$, {\it i.e.}
$ T = \sum_{n=2} T^{[n]} $, we have that
\begin{equation}
\Im m 
\langle a | T^{[n]} | a \rangle = \frac{1}{2}
\sum_{i} (2\pi)^{4} \delta^{(4)}(p_{a}-p_{i})
\sum_{k} 
{\langle i | T^{[k]} | a \rangle}^{*}
\langle i | T^{[n-k]} | a \rangle \, .
\label{ab2}
\end{equation}

In the particular case of QED, if in the initial states we have a $ \gamma Q$  
{\it i.e.} $ |a\rangle = |\gamma Q \rangle$,
we have for the first non-trivial order, $n=4$, in the
$T$:

\begin{equation}
\Im m 
\langle \gamma Q | T^{[4]} | \gamma Q \rangle =
\frac{1}{2} \int (dPS)
{\langle \gamma Q | T^{[2]} | \gamma Q \rangle}^{*}
\langle \gamma Q | T^{[2]} |\gamma Q \rangle \, ,
\label{ab3}
\end{equation}
where $(dP\!S)$ denotes the (two-body) phase-space integration.
Next we introduce the short-hand notation 
\begin{eqnarray}
{\cal A}^{[n]} &\equiv &
\Im m \langle \gamma Q | T^{[n]} 
| \gamma Q \rangle \, ,
\nonumber\\
{\cal T}^{[k]} &\equiv & 
\langle \gamma Q | T^{[k]} |\gamma Q \rangle.  
\label{ab5}
\end{eqnarray}
To avoid notational clutter we will suppress the Lorentz index
corresponding to the external photon.
Using the above notation, 
and suppressing the phase-space integrations, we have 
\begin{equation}
{\cal A}^{[4]} =\frac{1}{2} 
 {\cal T}^{[2]}_{\mu}\, 
P^{\mu\mu'}(p_1)\, {\cal T}^{[2]*}_{\mu'}\,
\, ,
\label{ab8}
\end{equation}
where 
$P_{\mu\nu}$ is the
polarization tensor for photons or gluons,
\begin{equation}
P_{\mu\nu}(p,n,\eta )\ =\ -g_{\mu\nu}+ \frac{n_{\mu} p_{\nu}
+n_{\nu} p_{\mu} }{n\cdot p} -
\eta \frac{p_{\mu}p_{\nu}}{{(n\cdot p)}^2}\, ,
\label{PhotPol}
\end{equation}
with $n_{\mu}$ being  
an arbitrary four-vector, and $\eta$ a gauge parameter.

The amplitude 
${\cal T}^{[2]}_{\mu}$ consists of 
$s$ - channel and $t$ - channel contributions, {\it i.e.}
\begin{equation}
{\cal T}^{[2]}_{\mu} = 
{\cal T}^{[2]}_{s\, \mu} + 
{\cal T}^{[2]}_{t\, \mu},
\end{equation}
(see Fig.\ref{figqed}).
From the gauge symmetry we know that 
\begin{equation}
p_1^{\mu} {\cal T}_{\mu} = 0,
\label{WIBAS}
\end{equation}
to all orders. Clearly, by virtue of 
Eq.(\ref{WIBAS})
all reference to the  unphysical quantities $n_{\mu}$ and $\eta$ disappears. We
emphasize however that the action of the  momentum $p_1^{\mu}$ does not respect
the $s-t$ separation given by the initial set of Feynman diagrams. Instead, the
action of $p_1^{\mu}$ gives rise to cancellations between the two sets.  In
particular we have that

\begin{center}

\begin{picture}(100,50)(0,15)

\Text(-20,50)[l]{$p_1^{\mu} {\cal T}^{[2]}_{s\,\mu } = 
{\cal R}$}
\Text(-20,30)[l]{$p_1^{\mu} {\cal T}^{[2]}_{t\,\mu } = 
- {\cal R}$}
\Text(100,40)[l]{${\cal R}\ =$}

\Line(120,17.5)(130,40)
\Photon(120,62.5)(130,40){2}{6}
\Line(130,40)(165,40)
\Photon(130,40)(150,62.5){2}{7}
\Vertex(130,40){2}
\end{picture}

\end{center}
Notice that the term ${\cal R}$ contains always an unphysical vertex.  

Therefore, 
\begin{eqnarray}
{\cal A}^{[4]} &=& \frac{1}{2}\,\, {\cal T}^{[2]} 
\otimes {\cal T}^{[2]*} \nonumber\\ 
&=& \frac{1}{2}\,\,  
\left({\cal T}^{[2]}_{s}+ {\cal T}^{[2]}_{t}\right)
\otimes 
\left({\cal T}^{[2]*}_{s}+ {\cal T}^{[2]*}_{t}\right)
\nonumber\\
&=& {\cal A}^{[4]}_{ss} + {\cal A}^{[4]}_{st} + {\cal A}^{[4]}_{tt}
\, ,
\label{asd}
\end{eqnarray}
with
\begin{eqnarray}
{\cal A}^{[4]}_{ss} &=& \frac{1}{2}\,  
{\cal T}^{[2]}_{s} \otimes {\cal T}^{[2]*}_{s},
\nonumber\\ 
{\cal A}^{[4]}_{st} &=& 
\frac{1}{2} \, \left( 
{\cal T}^{[2]}_{s}\otimes {\cal T}^{[2]*}_{t}
+ {\cal T}^{[2]}_{t}\otimes {\cal T}^{[2]*}_{s} \right),
 \nonumber\\ 
{\cal A}^{[4]}_{tt} &=& \frac{1}{2}\,  
{\cal T}^{[2]}_{t}\otimes {\cal T}^{[2]*}_{t}.
\label{asd1}
\end{eqnarray}
Next we will focus on ${\cal A}^{[4]}_{ss}$, which is the
``genuine'' $s$-channel part, {\it i.e.} the $s$-channel contribution
after the longitudinal parts of  $P^{\mu\mu'}(p_1)$ have been
eliminated. We 
will cast ${\cal A}^{[4]}_{ss}$ in the form
\begin{equation}
{\cal A}^{[4]}_{ss} = (e\gamma_{\alpha})\,
S_0 (p)\, A^{[2]}_{ss}(p)\, S_0 (p) \,
(e\gamma^{\alpha}),
\end{equation}
and then identify 
\begin{equation}
\Im m {\widehat{\Sigma}}^{(1)} (p) = A^{[2]}_{ss}(p),
\label{trew}
\end{equation}
where $\widehat{\Sigma}^{(1)} (p)$ is the one-loop fermion
self-energy under construction.
 
At this point it is straighforward  to verify that 
\begin{equation}
A^{[2]}_{ss}(p) = 
{\cal C}_{2}\left\{\widehat{\Sigma}^{(1)}(p)\right\}
=
{\cal C}_{2}\left\{  \Sigma_F^{(1)}(p)\right\} \, ,
\end{equation}
where ${\cal C}_{n}\{$\dots$\}$ is the operator which carries out the
the n-particle 
Cutkosky cuts to the quantity appearing inside the curly brackets. In this
case the two-particle cut involves a (massless) $\gamma$ and a $Q$
of mass $m$. 
The real part of $\widehat{\Sigma}^{(1)} (p)$
can be obtained directly from 
$A^{[2]}_{ss}(p)$ by means of a (twice-subtracted) dispersion relation.
In particular
\begin{equation}
\Re e \widehat{\Sigma}^{(1)}(p) =
\int_{t_2}^{\infty} dt\, \frac{A^{[2]}_{ss}(t)}{t-p^2} \, , 
\label{DR}
\end{equation}
where $t_2 = m^2$ is the two-body threshold. 
After subtracting twice ``on-shell'' one obtains the 
corresponding renormalized quantity
\begin{equation}
\Re e \widehat{\Sigma}^{(1)}_{\scriptscriptstyle R}(p) = (p^2 -m^2)^2
\int_{t_2}^{\infty} dt \, \frac{A^{[2]}_{ss}(t)}
{{(t-p^2)}(t-m^2)^2}.
\label{SDR}
\end{equation}

\subsection{QCD}

The one-loop QCD case can be directly derived from the QED 
analysis presented above. 
 In particular, when applying the optical theorem
one must consider a quark ($Q$) and a gluon 
($G$) as an intermediate state, {\it i.e.}
\begin{equation}
\Im m 
\langle \gamma Q | T^{[4]} | \gamma Q \rangle =
\frac{1}{2} \int (dP\!S)
{\langle G Q | T^{[2]} | \gamma Q \rangle}^{*}
\langle G Q | T^{[2]} |\gamma Q \rangle \, ,
\label{ab4}
\end{equation}
and define the corresponding quantities (we suppress color)
\begin{eqnarray}
{\cal A}^{[n]} &\equiv &
\Im m \langle \gamma Q | T^{[n]} | \gamma Q \rangle \, ,
\nonumber\\
{\cal T}^{[k]} &\equiv & 
\langle G Q | T^{[k]}|\gamma Q \rangle.  
\label{ab5b}
\end{eqnarray}
From this point on 
the analysis is exactly analogous to that presented for QED. 
The fact that the QED and QCD constructions coincide
is special to the one-loop case, 
and, as we will see in the next section, 
is not true in higher orders.

\section{The two-loop absorptive construction.}

As mentioned at the beginning of the previous section, in the two-loop
case we have  two distinct types of contributions:  (i) those that are
the one-loop  corrections to the two-to-two particle  process $\gamma Q
\to  \gamma  Q$ ($\gamma  Q  \to  G Q$  in  the  case  of QCD),  whose
tree-level analysis was considered in the previous section; (ii) those
that come from tree-level two-to-three particle processes.

There is  one additional fact we  will use in the  analysis below: The
one-loop contributions to $\gamma Q \to  \gamma Q$ \, ($\gamma Q \to G
Q$ in the case of QCD) considered in (i) can be effectively brought in
the Feynman  gauge, starting from  any other gauge, using  a procedure
exactly  analogous to  that used  in section  I. In  particular, using
nothing but elementary  WIs, the reader should be able  to see how all
longitudinal contributions inside the Feynman diagrams of Fig.\ref{fig5}
are equivalent to unphysical vertices, which cancel algebraically.

Before  we can  proceed with  the details  of the  two-loop absorptive
construction, some additional comments  are in order.  In the previous
section we  have distinguished between the  tree-level $s$-channel and
$t$-channel contributions, shown in Fig.4, using the obvious criterion
of whether a  diagram depends on the Mandelstam  variable $s$ (Fig.4a)
or $t$  (Fig.4b). Notice however that, 
in addition to the $t$-variable, the $t-channel$ fermion propagator 
in Fig.4b dependes explicitly on the mass of the incoming (test) 
fermion.
A similar  distinction between $s$-channel and 
$t$-channel contributions
needs be establisheed 
in  this  section; however,
additional  care is  needed when  classifying the various diagrams.  
Clearly,
diagrams that  are one-loop corrections to  the tree-level $t$-channel
graph  of Fig.4b,  such  as  those shown  in  Fig.5e, Fig.5f,  Fig.5g,
Fig.5h, Fig.5m,  Fig.5n, Fig.5p, and  Fig.5q, 
will be  characterized as
$t$-channel graphs.  In addition,  those graphs that arise as one-loop
vertex or wave-function corrections to the {\it incoming}
particles  of  Fig.4a,  such  as  Fig.5i  and  Fig.5o,  will  also  be
classified as $t$-channel graphs.   Finally, graphs as those 
shown in Fig.5a
-- Fig.5d, which are one-loop  corrections to either the $s$-dependent
off-shell propagator  or vertex  and wave-function corrections  to the
{\it outgoing} particles of the tree-level $s$-channel of Fig.4a, will
be  characterized   as  $s$-channel   graphs.   At  first   sight  the
characterization  of the graphs  in Fig.5i  and Fig.5o  as $t$-channel
graphs may seem unusual, since  there is no explicit $t$-dependence in
them;  indeed, both graphs  depend on  $s$, but,  in addition,  on the
masses of  the incoming particles.   Thus, in general, if  such graphs
were to  be considered as parts  of the two-loop  self-energy which is
being constructed absorptively, they would introduce in it an explicit
process-dependence.   This  would clearly  be  a  drawback, since  the
off-shell gauge-invariant fermion 
self-energies one  wants to define should be
universal, i.e.   process-independent.  To appreciate  this point, let
us  imagine  that  instead  of  the  flavour-conserving  processes  we
consider here (in which case the mass of the incoming on-shell fermion
is the same as that of  the off-shell one, and the 
external photons are mass-less), we were instead studying a
process containing a flavour-nonconserving interaction, such as $W^{+}
b  \to  t  \gamma$,  or $W^{+}  b  \to  t  Z$  attempting to  define
absorptively  the part  of the  off-shell top-quark  ($t$) self-energy
that contains a $t$  and a $\gamma$ or a $t$ and  a $Z$.  In that case
graphs  such as  those in  Fig.5i and  Fig.5o, together  with  the $W$
wave-function graphs  (not shown), would introduce into  the $t \gamma$
and  $t Z$  ``widths'' an  unphysical dependence  on $m_b$  and $M_W$.
Thus, according  to this definition, the $s$-channel  graphs are those
graphs which do not  contain information about the kinematical details
of the incoming test-particles.

\subsection{QED}

There are three different thresholds, to be denoted by
${\frak a} \equiv \gamma Q $,  ${\frak b} \equiv \gamma \gamma Q$, 
and ${\frak c} \equiv  Q \bar{Q} Q $. Thus,
\begin{eqnarray}
\Im m 
\langle q\bar{q}| T^{[6]} | q\bar{q} \rangle &=&
\frac{1}{2} \int (dP\!S)_{\frak a} \,\,
2 \Re e \bigg[ {\langle \gamma Q 
| T^{[4]}| \gamma Q \rangle}^{*}
\langle \gamma Q  | T^{[2]} |\gamma Q\rangle \bigg]
\nonumber\\
&+&
\frac{1}{2} \int (dP\!S)_{\frak b }
{\langle \gamma \gamma Q | T^{[3]}
| \gamma Q \rangle}^{*}
\langle \gamma \gamma Q| T^{[3]} 
| \gamma Q \rangle \, 
\nonumber\\
&+&
\frac{1}{2} \int (dP\!S)_{\frak c }
{\langle Q \bar{Q} Q | T^{[3]}
| \gamma Q \rangle}^{*}
\langle Q \bar{Q} Q| T^{[3]} 
| \gamma Q \rangle \, .
\nonumber\\
\label{ewab4}
\end{eqnarray}
Then we have, suppressing the phase-space integrations,
and using the previously introduced notation
\begin{eqnarray} 
{\cal A}^{[6]} &=&
\Re e \bigg(
{\cal T}^{[4]}_{{\frak a}\, \mu}\,\, 
P^{\mu\mu'}(p_1)\,\, {\cal T}^{[2]*}_{{\frak a} \,\mu'}\bigg)
+ \frac{1}{2} \, {\cal T}^{[3]}_{{\frak b}\, \mu\nu}\,
P^{\mu\mu'}(p_1) P^{\nu\nu'}(p_2)
{\cal T}^{[3]*}_{{\frak b}\, \mu'\nu'}
+ \frac{1}{2} \, {\cal T}^{[3]}_{\frak c }
{\cal T}^{[3]*}_{\frak c }
\nonumber\\
&\equiv&
{\cal A}^{[6]}_{\frak a } + {\cal A}^{[6]}_{\frak b} 
+ {\cal A}^{[6]}_{\frak c}
\, .
\label{qwer}
\end{eqnarray}

From the gauge symmetry we know that 
\begin{eqnarray}
p_1^{\mu} \, {\cal T}_{{\frak b}\,\mu\nu} &=& 0,
\nonumber\\
p_2^{\nu} \, {\cal T}_{{\frak b}\,\mu\nu} &=& 0,
\label{WIBAS2}
\end{eqnarray}
to all orders. Again, the same situation 
explained in
the one-loop case
is true now, 
namely the fact that the WI mixes contributions
between the $s$- and $t$- channels, which all contain unphysical
vertices. In particular, at order~$e^6$

\begin{center}
\begin{equation}
\begin{picture}(100,35)(0,30)

\Text(-50,50)[l]{$p_1^{\mu} \, ({\cal T}^{[3]}_{{\frak b}\,s})_{\mu}^{\nu} = 
{\cal R}^{[3]\,\nu}_{{\frak b}}$}
\Text(-50,30)[l]{$p_1^{\mu} \, ({\cal T}^{[3]}_{{\frak b}\,t})_{\mu}^{\nu} = 
- {\cal R}^{[3]\,\nu}_{{\frak b}}$}
\Text(85,40)[l]{${\cal R}^{[3]\,\nu}_{{\frak b}}\ =$}
\Text(170,60)[l]{\footnotesize{$\nu$}}

\Line(120,17.5)(130,40)
\Photon(120,62.5)(130,40){2}{6}
\Line(130,40)(180,40)
\Photon(130,40)(150,62.5){2}{7}
\Photon(165,40)(165,62.5){2}{6}
\Vertex(130,40){2}

\end{picture}
\label{SPQED}
\end{equation}
\end{center}
and an identical equation holds when contracting with $p_{2\nu}$.
Notice that again the term ${\cal R}^{[3]\,\nu}$ contains an unphysical vertex.

Finally,   
\begin{equation}
{\cal A}^{[6]}_{\frak i } =  
{\cal A}^{[6]}_{{\frak i } \, ss} + 
{\cal A}^{[6]}_{{\frak i }\, st} + {\cal A}^{[6]}_{{\frak i } \, tt}\,, 
\,\,\,\,\, \frak i = {\frak a }, {\frak b }, {\frak c }
\label{cvbn}
\end{equation}
with
\begin{eqnarray}
{\cal A}^{[6]}_{{\frak a }\, ss}  &=& 
\Re e \left[{\cal T}^{[4]}_{{\frak a }\, s} 
\otimes {\cal T}^{[2]*}_{{\frak a }\, s}\right], 
\nonumber\\
{\cal A}^{[6]}_{{\frak a }\, st} &=&
\Re e \left[{\cal T}^{[4]}_{{\frak a }\, s}
\otimes {\cal T}^{[2]*}_{{\frak a }\, t}
+ {\cal T}^{[4]}_{{\frak a }\, t}\otimes 
{\cal T}^{[2]*}_{{\frak a }\, s}\right],
\nonumber\\
{\cal A}^{[6]}_{{\frak a }\, tt} &=& 
\Re e \left[{\cal T}^{[4]}_{{\frak a }\, t} 
\otimes {\cal T}^{[2]*}_{{\frak a }\, t}\right], 
\label{Aa}
\end{eqnarray}
and
\begin{eqnarray}
{\cal A}^{[6]}_{{\frak m }\, ss} &=& \frac{1}{2}\,  
{\cal T}^{[3]}_{{\frak m }\, s} 
\otimes {\cal T}^{[3]*}_{{\frak m }\, s},
\nonumber\\ 
{\cal A}^{[6]}_{{\frak m }\, st} &=& 
\frac{1}{2} \, \left( 
{\cal T}^{[3]}_{{\frak m }\, s}
\otimes {\cal T}^{[3]*}_{{\frak m }\, t}
+ {\cal T}^{[3]}_{{\frak m }\, t}
\otimes {\cal T}^{[3]*}_{{\frak m }\, s} \right),
 \nonumber\\ 
{\cal A}^{[6]}_{{\frak m }\, tt} &=& \frac{1}{2}\,  
{\cal T}^{[3]}_{{\frak m }\, t}
\otimes {\cal T}^{[3]*}_{{\frak m }\, t}\,,
\,\,\,\,\,\, \frak m = {\frak b }, {\frak c }.
\label{2Lasd1}
\end{eqnarray}
Let us next consider the $ss$ parts, $A^{[6]}_{{\frak i}\, ss}(p)$. 
Unlike the one-loop case, where in Eq.(\ref{trew}) the entire 
$ss$ part $A^{[2]}_{ss}(p)$ was identified with 
$\Im m {\widehat{\Sigma}}^{(1)} (p)$, 
now we must  
identify $A^{[6]}_{{\frak i}\, ss}(p)$ 
with the imaginary parts of both 
the two-loop {\it one-particle irreducible} fermion self-energy
$\widehat{\Sigma}^{(2)}(p)$ and the {\it one-particle reducible}
string of two $\widehat{\Sigma}^{(1)}(p)$ ; of course the latter
contributions are known from the one-loop construction of the previous
section. 
Thus,
\begin{eqnarray}
\Im m {\widehat{\Sigma}}^{(2)}(p) &=& 
A^{[6]}_{{\frak i}\, ss}(p) - 
2\, \Im m {\widehat{\Sigma}}^{(1)}(p) \, \Re e {\widehat{\Sigma}}^{(1)}(p)
\nonumber\\
&\equiv& A^{[6]\, {\rm 1PI}}_{{\frak i}\, ss}(p)
\end{eqnarray}
where the superscript ``1PI'' stand for  
``one-particle irreducible''. One can verify at this point that 
\begin{equation}
A^{[6]\, {\rm 1PI}}_{{\frak i}\, ss}(p) = \, 
{\cal C}_{{\frak i} }\left\{\widehat{\Sigma}^{(2)}(p)\right\} \,
= {\cal C}_{{\frak i} }\left\{\widehat{\Sigma}^{(2)}_F (p)\right\}\,,
\,\,\,\,\,\,\frak i = {\frak a }, {\frak b }, {\frak c }.
\end{equation}
Clearly, 
the two-particle cut involves a $\gamma$ and a $Q$, whereas
the  three-particle cuts involves two $\gamma$'s and a $Q$,
and three $Q$'s, respectively.
Of course, for massless photons the two cuts coincide.
The real part of $\widehat{\Sigma}^{(2)}(p)$ 
can be obtained directly from the three quantities
$A^{[6]}_{{\frak i} \, ss}(p)$ 
by means of an appropriate dispersion relation.
In particular
\begin{equation}
\Re e \widehat{\Sigma}^{(2)}(p) = \sum_{\frak i} 
\int_{t_{\frak i}}^{\infty} dt\, 
\frac{A^{[6]\, {\rm 1PI}}_{{\frak i} \, ss}(t)}{t-p^2},
\end{equation}
with $t_{\frak a} = t_{\frak b} = m^2$ and $t_{\frak c} = 9 m^2$.
Again, after subtracting twice ``on-shell'', one obtains the 
corresponding renormalized quantity 
$\Re e \widehat{\Sigma}^{(2)}_{\scriptscriptstyle R}(p)$.

\subsection{QCD}

There are three different thresholds, to be denoted by
${\frak a} \equiv G Q $,  ${\frak b} \equiv G G Q$, 
and ${\frak c} \equiv  Q \bar{Q} Q $. So, at order
$e^2 g^4_s$ we have

\begin{eqnarray}
\Im m 
\langle q\bar{q}| T^{[6]} | q\bar{q} \rangle &=&
\frac{1}{2} \int (dP\!S)_{\frak a} \,\,
2 \Re e \bigg[ {\langle G Q 
| T^{[4]}| \gamma Q \rangle}^{*}
\langle G Q  | T^{[2]} |\gamma Q\rangle \bigg]
\nonumber\\
&+&
\frac{1}{2} \int (dP\!S)_{\frak b }
{\langle G G Q | T^{[3]}
| \gamma Q \rangle}^{*}
\langle G G Q| T^{[3]} 
| \gamma Q \rangle \, 
\nonumber\\
&+&
\frac{1}{2} \int (dP\!S)_{\frak c }
{\langle Q \bar{Q} Q | T^{[3]}
| \gamma Q \rangle}^{*}
\langle Q \bar{Q} Q| T^{[3]} 
| \gamma Q \rangle \, .
\nonumber\\
\label{ewab4b}
\end{eqnarray}

We next turn to the tree-level WIs satisfied 
by the QCD amplitudes appearing above, when contacted by the momenta
originating from the polarization tensor(s) of the final state gluon(s). 
To begin with, Eq.(\ref{qwer}) holds exactly as in the QED case. 
It is worthwhile studying how this  tree-level WI 
is realized at the diagrammatic level; 
a non-trivial interplay of terms containing
unphysical vertices takes place, 
allowing contributions originating from different 
kinematic channels to cancel against each-other. 
The diagrams contributing to the process $\gamma Q \to GQ$ 
at one loop  
are shown
in Fig.\ref{fig5}. For brevity, we will illustrate the point 
by focusing only on the non-Abelian diagrams (d),
(h) and (q) of Fig.\ref{fig5}. 
Using the elementary WI (\ref{wi3}), we find the
following equality

\begin{center}
\begin{picture}(100,125)(70,-45)

\Text(-35,50)[l]{$p_{1}^{\mu}$(d) $=\ \lambda\ $}
\Text(-6.5,-10)[l]{$=\ \lambda$}
\Text(15,-38)[r]{\footnotesize{($\alpha$)}}
\Text(115,-38)[r]{\footnotesize{($\beta$)}}

\Line(20,27.5)(30,50)
\Photon(20,72.5)(30,50){2}{6}
\Line(30,50)(90,50)
\Photon(30,50)(60,60){2}{7}
\Photon(60,60)(75,50){-2}{4}
\Photon(60,60)(70,72.5){2}{4}
\put(60,60){\rotatebox{45}{\GBoxc(0,0)(5,5){1}}}
\Vertex(30,50){2}
\Text(52,57.6)[c]{\rotatebox{35}{\bf{\big /}}}
\Text(52.5,57.85)[c]{\rotatebox{35}{\bf{\big /}}}

\Text(95,50)[l]{$-\ \lambda$}

\Line(120,27.5)(130,50)
\Photon(120,72.5)(130,50){2}{6}
\Line(130,50)(190,50)
\PhotonArc(160,59)(7,180,540){2}{9}
\Photon(165.5,65)(180,72.5){-2}{4}
\Vertex(160,50){2}
\put(165.5,65){\rotatebox{45}{\GBoxc(0,0)(5,5){1}}}
\Text(154,59.6)[c]{\rotatebox{-71}{\bf{\big /}}}
\Text(154,60.2)[c]{\rotatebox{-71}{\bf{\big /}}}

\Text(195,50)[l]{$-\ \lambda$}

\Line(220,27.5)(230,50)
\Photon(220,72.5)(230,50){2}{6}
\Line(230,50)(290,50)
\PhotonArc(260,59)(7,180,540){2}{9}
\Photon(265.5,65)(280,72.5){-2}{4}
\Vertex(260,50){2}
\put(265.5,65){\rotatebox{45}{\GBoxc(0,0)(5,5){1}}}
\Text(267,59.6)[c]{\rotatebox{-71}{\bf{\big /}}}
\Text(267,60.2)[c]{\rotatebox{-71}{\bf{\big /}}}

\Text(295,50)[l]{$+\ \cdots$}

\Photon(20,12.5)(30,-10){2}{6}
\Line(20,-32.5)(30,-10)
\Line(30,-10)(90,-10)
\Photon(30,-10)(60,0){2}{7}
\Photon(60,0)(75,-10){-2}{4}
\Photon(60,0)(70,12.5){2}{4}
\put(60,0){\rotatebox{45}{\GBoxc(0,0)(5,5){1}}}
\Vertex(30,-10){2}
\Text(52,-2.4)[c]{\rotatebox{35}{\bf{\big /}}}
\Text(52.5,-2.15)[c]{\rotatebox{35}{\bf{\big /}}}

\Text(95,-10)[l]{$-\ \lambda$}

\Photon(120,12.5)(130,-10){2}{6}
\Line(120,-32.5)(130,-10)
\Line(130,-10)(141,-10)
\Line(150,-10)(190,-10)
\PhotonArc(139,-10)(7,90,450){2}{9}
\Photon(145.3,-4.4)(159.8,1.8){-2}{4}
\put(145.5,-4.7){\rotatebox{45}{\GBoxc(0,0)(5,5){1}}}
\Vertex(130,-10){2}

\Text(195,-10)[l]{$+\ \cdots$}

\end{picture}
\end{center}
where the ellipses stand for diagrams that will cancel against contributions 
left over from the Abelian-like diagrams. Moreover one has

\begin{center}
\begin{picture}(100,195)(70,-115)

\Text(-25,50)[l]{$p_{1}^{\mu}$(h)$=\ -\ \lambda$}
\Text(110,50)[l]{$-\ \lambda$}
\Text(210,50)[l]{$+\ \cdots$}

\Text(-15,-50)[l]{$p_{1}^{\mu}$(q)$=\ \lambda$}
\Text(110,-50)[l]{$+\ \lambda$}
\Text(230,-50)[l]{$+\ \cdots$}

\Text(50,-10)[r]{\footnotesize{($\gamma$)}}
\Text(150,-10)[r]{\footnotesize{($\delta$)}}
\Text(50,-110)[r]{\footnotesize{($\zeta$)}}
\Text(145,-80)[r]{\footnotesize{($\varepsilon$)}}

\put(10,80){\rotatebox{-90}{
\Line(30,50)(30,90)
\Line(30,50)(90,50)
\Photon(30,50)(60,60){2}{7}
\Photon(60,60)(75,50){-2}{4}
\Photon(60,60)(70,72.5){2}{4}
\put(60,60){\rotatebox{45}{\GBoxc(0,0)(5,5){1}}}
\Vertex(30,50){2}
\Text(52,57.6)[c]{\rotatebox{35}{\bf{\big /}}}
\Text(52.5,57.85)[c]{\rotatebox{35}{\bf{\big /}}}}}

\put(110,180){\rotatebox{-90}{
\Line(130,50)(130,90)
\Line(130,50)(141,50)
\Line(150,50)(190,50)
\PhotonArc(139,50)(7,90,450){2}{9}
\Photon(145.3,55.7)(155.8,75.8){-2}{5}
\put(145.5,55.2){\rotatebox{45}{\GBoxc(0,0)(5,5){1}}}
\Vertex(130,50){2}}}

\put(10,-20){\rotatebox{-90}{
\Line(30,50)(30,90)
\Line(30,50)(90,50)
\Photon(30,50)(60,60){2}{7}
\Photon(60,60)(75,50){-2}{4}
\Photon(60,60)(70,72.5){2}{4}
\put(60,60){\rotatebox{45}{\GBoxc(0,0)(5,5){1}}}
\Vertex(30,50){2}
\Text(52,57.6)[c]{\rotatebox{35}{\bf{\big /}}}
\Text(52.5,57.85)[c]{\rotatebox{35}{\bf{\big /}}}}}

\put(130,-100){\Line(20,27.5)(30,50)
\Photon(20,72.5)(30,50){2}{6}
\Line(30,50)(90,50)
\Photon(30,50)(60,60){2}{7}
\Photon(60,60)(75,50){-2}{4}
\Photon(60,60)(70,72.5){2}{4}
\put(60,60){\rotatebox{45}{\GBoxc(0,0)(5,5){1}}}
\Vertex(30,50){2}
\Text(52,57.6)[c]{\rotatebox{35}{\bf{\big /}}}
\Text(52.5,57.85)[c]{\rotatebox{35}{\bf{\big /}}}}

\put(30,0){\Photon(20,72.5)(30,50){2}{6}}
\put(130,0){\Photon(20,72.5)(30,50){2}{6}}
\put(30,-100){\Photon(20,72.5)(30,50){2}{6}}

\end{picture}
\end{center}
so that taking into account that diagrams (h) and (q) carry a (group
theoretical) relative minus sign with respect to diagram (d), 
we find the cancellations
\begin{equation}
(\alpha)+(\zeta)=0, \qquad (\beta)+(\delta)=0, \qquad (\gamma)+(\varepsilon)=0.
\end{equation}

Thus, the analysis regarding the sector 
${\frak a}$ and ${\frak c}$ is exactly analogous to that of QED.
The only difference is related to the  sector ${\frak b}$,
and originates from the fact that 
Eq.(\ref{WIBAS2}) and Eq.(\ref{SPQED}) are altered, due to the
appearance of ghost-related contributions.
In particular,
\begin{eqnarray}
p_1^{\mu} \, ({\cal T}^{[3]}_{{\frak b}\,s })_{\mu}^{\nu} &=& 
{\cal S}^{[3]}_{{\frak b} \,s}\, p_{2}^{\nu} 
+ {\cal R}^{[3]\,\nu}_{{\frak b}}, \nonumber\\
p_1^{\mu} \, 
({\cal T}^{[3]}_{{\frak b}\,t})_{\mu}^{\nu} &=& 
{\cal S}^{[3]}_{{\frak b} \,t} \, p_{2}^{\nu}
- {\cal R}^{[3]\,\nu}_{{\frak b}}, 
\label{SPQCD}
\end{eqnarray}
and by Bose symmetry the same equation is valid
when $p_1 \leftrightarrow p_2$, and
$\mu \leftrightarrow \nu$. In addition, acting with $p_2^{\nu}$ on both sides
of Eq.(\ref{SPQCD}), and using 
the on-shell conditions $p_1^2 = p_2^2 = 0$ we have 
\begin{equation}
p_1^{\mu} p_2^{\nu} {\cal T}^{[3]}_{{\frak b}\,\mu\nu} = 0 .  
\end{equation}

To see how this WI is enforced diagrammatically, we
next turn to the set of 
diagrams contributing to the tree-level two-to-three  
process, listed in Fig.\ref{fig6}.
Contracting with the external momenta
$k_{1}^{\mu}$, taking into account the different color
structure of the diagrams, and recalling  
the elementary WI (\ref{WIzero}), we obtain

\begin{center}

\begin{picture}(0,150)(90,-60)

\Text(-50,50)[l]{$p_{1}^{\mu}$[(a)+(b)]$\ =\ $}

\Text(65,70)[r]{\scriptsize{$\nu$}}
\Text(145,70)[r]{\scriptsize{$\nu$}}
\Text(62,-2.5)[r]{\scriptsize{$\nu$}}
\Text(135,-13)[r]{\scriptsize{$\nu$}}
\Text(227,3)[r]{\scriptsize{$\nu$}}

\Line(20,27.5)(30,50)
\Photon(20,72.5)(30,50){2}{6}
\Line(30,50)(70,50)
\Gluon(50,50)(70,70){2}{6}
\Gluon(30,50)(50,70){2}{6}
\Vertex(30,50){1.8}

\Text(80,50)[l]{$\ +$}

\Line(100,27.5)(110,50)
\Photon(100,72.5)(110,50){2}{6}
\Line(110,50)(150,50)
\Gluon(130,50)(150,70){-2}{6}
\Gluon(130,50)(110,70){2}{6}
\Vertex(130,50){2}

\Text(-35,-20)[l]{$p_{1}^{\mu}$(c)$\ =\ -$}

\Photon(20,7.5)(30,-20){2}{6}
\Line(20,-42.5)(30,-20)
\Line(30,-20)(70,-20)
\Gluon(50,-20)(70,0){-2}{6}
\Gluon(50,-20)(30,0){2}{6}
\Vertex(50,-20){2}

\Text(80,-20)[l]{$\ +$}

\Photon(100,7.5)(110,-20){2}{6}
\Line(100,-42.5)(110,-20)
\Line(110,-20)(150,-20)
\GlueArc(128,-20)(18,90,180){2}{6}
\Gluon(128,-2)(148,-15){-2}{5}
\Gluon(128,-2)(148,11){2}{5}
\put(128,-2){\rotatebox{45}{\GBoxc(0,0)(5,5){1}}}
\Vertex(110,-20){2}

\Text(160,-20)[l]{$\ -$}

\Photon(180,7.5)(190,-20){2}{6}
\Line(180,-42.5)(190,-20)
\Line(190,-20)(248,-20)
\GlueArc(228,-20)(18,90,180){2}{6}
\DashArrowLine(228,-2)(248,-15){1}
\DashArrowLine(248,11)(228,-2){1}

\Text(15,20)[r]{\footnotesize{($\alpha$)}}
\Text(95,20)[r]{\footnotesize{($\beta$)}}
\Text(15,-50)[r]{\footnotesize{($\gamma$)}}
\Text(95,-50)[r]{\footnotesize{($\delta$)}}
\Text(175,-50)[r]{\footnotesize{($\varepsilon$)}}

\end{picture}
\end{center}

The cancellation between diagrams  ($\beta$) and ($\gamma$) corresponds to the
standard BRS-enforced $s$-$t$-channel cancellation \cite{ChengLi} taking place in
the tree-level amplitude $QQ \to GG$, which now appears embedded as a sub-process
in the amplitude $\gamma Q  \to GG Q$ that we consider here. As happens in the case
of the $QQ \to GG$ example, the  diagram ($\varepsilon$)  gives rise to the correct
ghost structure. As for diagrams  ($\alpha$) and ($\delta$), which contain 
external unphysical vertices, will cancel against similar contributions originating
from the $t$-channel graphs. Specifically,

\begin{center}

\begin{picture}(0,150)(90,-60)

\Text(-80,50)[l]{$p_{1}^{\mu}$[(d)+(e)+(f)]$\ =\ - $}

\Text(70,42)[r]{\scriptsize{$\nu$}}
\Text(135,19)[r]{\scriptsize{$\nu$}}
\Text(208,39)[r]{\scriptsize{$\nu$}}
\Text(70,-29)[r]{\scriptsize{$\nu$}}
\Text(139,-2.5)[l]{\scriptsize{$\nu$}}

\Line(30,20)(30,50)
\Photon(20,72.5)(30,50){2}{6}
\Line(30,50)(70,50)
\Gluon(30,50)(50,70){2}{6}
\Gluon(30,35)(70,35){2}{8}
\Vertex(30,50){2}

\Text(80,50)[l]{$\ -$}

\Line(110,20)(110,50)
\Photon(100,72.5)(110,50){2}{6}
\Line(110,50)(148,50)
\GlueArc(128,50)(18,180,270){2}{6}
\Gluon(128,32)(148,45){2}{5}
\Gluon(128,32)(148,19){-2}{5}
\put(128,32){\rotatebox{45}{\GBoxc(0,0)(5,5){1}}}
\Vertex(110,50){2}

\Text(160,50)[l]{$\ -$}

\Line(190,20)(190,50)
\Photon(180,72.5)(190,50){2}{6}
\Line(190,50)(228,50)
\Gluon(190,32)(208,32){2}{4}
\DashArrowLine(228,45)(208,32){1}
\DashArrowLine(208,32)(228,19){1}

\Text(-55,-20)[l]{$p_{1}^{\mu}$[(g)+(h)]$\ =$}

\Line(30,-20)(30,-50)
\Photon(20,2.5)(30,-20){2}{6}
\Line(30,-20)(70,-20)
\Gluon(30,-20)(50,0){2}{6}
\Gluon(30,-35)(70,-35){2}{8}
\Vertex(30,-20){2}

\Text(80,-20)[l]{$\ -$}

\Line(110,-20)(110,-50)
\Photon(100,2.5)(110,-20){2}{6}
\Line(110,-20)(150,-20)
\Gluon(130,-20)(150,0){2}{6}
\Gluon(110,-20)(130,0){2}{6}
\Vertex(110,-20){2}

\Text(25,15)[r]{\footnotesize{($\zeta$)}}
\Text(105,15)[r]{\footnotesize{($\eta$)}}
\Text(185,15)[r]{\footnotesize{($\theta$)}}
\Text(25,-55)[r]{\footnotesize{($\xi$)}}
\Text(105,-55)[r]{\footnotesize{($\pi$)}}

\end{picture}
\end{center}
We can then identify the cancellations
\begin{equation}
(\alpha)+(\pi)=0, \qquad (\delta)+(\eta)=0, \qquad (\zeta)+(\xi)=0,
\end{equation}
so that we are left only with the correct ghost structures, {\it i.e.}
diagrams ($\varepsilon$) and ($\theta$).

Thus, whereas Eq.(\ref{Aa}) and the part of 
Eq.(\ref{2Lasd1}) with ${\frak i} = {\frak c}$ remain unchanged, the
part of Eq.(\ref{2Lasd1}) with ${\frak i} = {\frak b}$
gets modified as follows:
\begin{eqnarray}
{\cal A}^{[6]}_{{\frak b }\, ss} &=& \frac{1}{2}\,  
 \bigg( {\cal T}^{[3]}_{{\frak b }\, s} 
\otimes {\cal T}^{[3]*}_{{\frak b }\, s} 
-2 \, {\cal S}^{[3]}_{{\frak b} \,s}\otimes
{\cal S}^{[3]*}_{{\frak b} \,s}
\bigg)\,,
\nonumber\\ 
{\cal A}^{[6]}_{{\frak b }\, st} &=& 
\frac{1}{2} \, \bigg[\bigg( 
{\cal T}^{[3]}_{{\frak b }\, s}
\otimes {\cal T}^{[3]*}_{{\frak m }\, t}
- 2 \, {\cal S}^{[3]}_{{\frak b} \,s}\otimes
{\cal S}^{[3]*}_{{\frak b} \,t}\bigg)
+ \bigg( {\cal T}^{[3]}_{{\frak b }\, t}
\otimes {\cal T}^{[3]*}_{{\frak b }\, s} 
- 2 \, {\cal S}^{[3]}_{{\frak b} \,t}\otimes
{\cal S}^{[3]*}_{{\frak b} \,s}
\bigg)\bigg]\,,
 \nonumber\\ 
{\cal A}^{[6]}_{{\frak b }\, tt} &=& \frac{1}{2}\,  
 \bigg( {\cal T}^{[3]}_{{\frak b }\, t}
\otimes {\cal T}^{[3]*}_{{\frak b }\, t}
-2\, {\cal S}^{[3]}_{{\frak b} \,t}\otimes
{\cal S}^{[3]*}_{{\frak b} \,t}\bigg)
\,.
\label{2Lasd1qcd}
\end{eqnarray}
Beyond this point is easy to see that the analysis 
following Eq.(\ref{2Lasd1}) 
of the QED case applies unchanged to the QCD case as well. 

\section{Discussion and Conclusions}

In  this paper  we  have  shown {\it  explicitly}  that the  off-shell
two-loop  fermion self-energy  constructed  by means  of  the PT  {\it
coincides} with the conventional fermion self-energy calculated in the
covariant (renormalizable)  Feynman gauge. This  has been demonstrated
by systematically  tracking down the  action of all  terms originating
from the longitudinal parts  of the tree-level gauge boson propagators
(photons or gluons)  appearing inside  the Feynman  diagrams
contributing  to manifestly gauge-invariant  amplitudes. It  turns out
that all  such terms  give rise to  unphysical vertices,  which cancel
diagrammatically in  the entire physical  amplitude, without affecting
the kinematical structure  of the various sub-amplitudes (propagators,
vertices,  boxes).   We have  followed  two  different but  physically
equivalent approaches.   First we have shown the  cancellations at the
level  of  the  full  two-loop  amplitude.  Then  we  have  shown  the
cancellations for  the two-and-three body  cross-sections which appear
on the right-hand side of the optical theorem.

It is worth commenting on the relation of the results 
established here and those appearing in 
\cite{Watson:1999vv}. 
That  work was an  early  attempt to  define  what the  pinch
technique should  be beyond one loop.   At the time  
it was written
the central issue
had been how to deal  with the internal three-gluon vertices appearing
inside Feynman  diagrams, i.e. three-gluon vertices all  three legs of
which are associated with gauge field propagating inside the loop.  In
particular, one  needed to  establish a well-defined  criterion, which
would allow one  to unambiguously decide whether and  how the internal
vertices should be split into pinching and a non-pinching parts.  What
was  proposed  in  \cite{Watson:1999vv}  was  to  split  the  internal
vertices  following  as a  guiding  principle  some  type of  skeleton
expansion  of  the  quark  two-loop self-energy.  In  particular,  the
starting point for the  construction has been the general diagrammatic
representation  of   the  two-loop  quark  self-energy   in  terms  of
renormalized   one-loop  two-point   and  three-point   functions  and
tree-level    Bethe-Salpeter-type   quark-gluon    scattering   kernel
insertions  in  the one-loop  quark  self-energy.   According to  this
construction, even if one were to  start out in the Feynman gauge, 
pinching momenta stemming from  the internal three-gluon vertex in the
diagram of Fig.3t  (or Fig.3u) would give rise  to pinch contributions
which  should  be  removed  from  the effective  PT  propagator  under
construction.  Thus, the final answer differs from that of the Feynman
gauge; in  fact, it does  not coincide with  any of the  known gauges.
However, this procedure was in contradiction with the already existing
absorptive PT constructions 
\cite{Papavassiliou:1995fq,Papavassiliou:1996fn}, according to 
which the imaginary  parts of the PT Green's  functions are related by
means  of  the optical  theorem  to  precisely  identifiable parts  of
physical cross-sections.  These latter  parts are constructed by using
again  the  PT  rearrangement, but  this  time  not  at the  level  of
amplitudes ($S$-matrix  elements) but at the  level of cross-sections.
This fundamental  PT property relating  real and imaginary parts  is a
non-trivial realization  of the optical  theorem at the level  of {\it
individual} Green's functions.  As has been explained in detail in
\cite{Papavassiliou:2000az}, any attempt to rearrange  
the internal vertices leads to a violation of
the aforementioned property; therefore  
internal three-gluon vertices should remain unchanged,
and only external ones need be modified. Therefore, in the case of
the fermion propagator the only pinching momenta originate from the
longitudinal parts of the internal gluon propagators.

If one accepts that the cancellation mechanism presented here persists
to all orders, then so does the main result of this paper, namely that
the PT fermion  self-energy coincides with that in  the Feynman gauge.
This is so  because, as already mentioned earlier, 
in the  case of the fermion
self-energy   all   three-boson   vertices  are   ``internal'',
{\it i.e.}  there are no  further pinching  contributions stemming  from the
usual PT  rearrangement of three-boson  vertices.  Should that  be the
case,  it would  be  interesting to  study  the form  of  the SD  this
self-energy satisfies.
For example,
the  SD equation for the electron  propagator   
$S$ will be given by (see Fig.\ref{fig7})  
\begin{equation}
S^{-1}(p)=  {S_{0}^{-1}}(p)  - e^2 \int  \frac{d^dk}{(2\pi)^d}
\gamma_{\mu}  S(p+k) \Gamma_{\nu}(p,k) \Delta^{\mu\nu}(k)  
\label{SD1}
\end{equation}
 where $\Gamma_{\nu}(p,k)$ is the full photon-electron-electron
vertex,    
and the full photon propagator $\Delta_{\mu\nu}$ 
now assumes the form
\begin{equation}
\Delta_{\mu\nu}(q) = 
(g_{\mu\nu} - q_{\mu}q_{\nu}/q^2)\, [q^2-\Pi(q)]^{-1}
+ q_{\mu}q_{\nu}/q^4
\label{vacpol}
\end{equation}
where the scalar quantity $\Pi(q)$
is related to the full  vacuum polarization
$\Pi_{\mu\nu}$  
by 
\begin{equation}
\Pi_{\mu\nu} = (g_{\mu\nu} - q_{\mu}q_{\nu}/q^2) \,\Pi(q)
\end{equation}

It will be interesting to study this form of SD equation
and its implication for fermion mass generation,
particularly in the cases
QED$_3$ and QED$_4$. Notice that the presence of 
$\Gamma_{\nu}(p,k)$ forces one to solve a system of coupled 
SD equations, or, alternatively, resort to 
a gauge-technique inspired Ansatz for $\Gamma_{\nu}(p,k)$ \,
\cite{Salam:1963sa}.

It is well known that one can construct a gauge-invariant operator
out of a gauge-variant one 
by means
of a path-order exponential containing the gauge field $A$ \cite{Wilson:1974jj}.
In the case of the fermion propagator 
\mbox{$S(x,y) = \langle 0|\psi(x) \bar{\psi}(y)|0\rangle$}
the corresponding
gauge-invariant propagator $S_{PO}$ reads 
(``PO'' stands for 
``path-ordered'')
\begin{equation}
S_{PO} (x,y) = 
\langle 0|\psi(x) P \exp\left(i \int_{x}^{y} dz \cdot A(z)  
\right)\bar{\psi}(y)|0\rangle 
\end{equation}
It would be interesting to explore possible connections between
the ``path-ordered'' propagator and the PT propagator constructed
in this paper, together with various other related formalisms,
which have appeared in the literature
\cite{Catani:1993fe,Bagan:1997su}.
  
Finally, it would be important to extend the methodology and results
of this paper to the case of the electroweak sector of the Standard Model,
especially given the phenomenological relevance of (resonant) top-quark
production.

\acknowledgements
The work of D.B. is supported by the Ministerio of Educaci\'on, Cultura y
Deporte, Spain, under Grant DGICYT-PB97-1227, and 
the research of J.P. is supported by CICYT, Spain, under Grant AEN-99/0692. 

\newpage

\newpage

\newpage

\begin{figure}

\begin{picture}(0,100)(-120,-20)

\Text(20,18)[c]{\footnotesize{(a)}}
\Text(135,18)[c]{\footnotesize{(b)}}

\Text(55,70)[l]{\scriptsize{$p_1$}}
\Text(40,70)[l]{\scriptsize{$\mu$}}
\Text(185,22)[r]{\scriptsize{$p_1$}}
\Text(185,38)[r]{\scriptsize{$\mu$}}

\Line(20,27.5)(30,50)
\Photon(20,72.5)(30,50){2}{6}
\Line(30,50)(70,50)
\Gluon(50,72.5)(50,50){-2}{6}

\Line(145,10)(145,50)
\Photon(135,72.5)(145,50){2}{6}
\Line(145,50)(185,50)
\Gluon(145,30)(185,30){-2}{10}

\end{picture}

\caption{The tree level one particle phase space appearing in the one 
loop QED absorptive
PT construction.  Diagram (a) defines the $s$-channel amplitude 
${\cal T}^{[2]}_{s}$,   while diagram (b) defines the $t$-channel amplitude  
${\cal T}^{[2]}_{t}$.}

\label{figqed}

\end{figure}
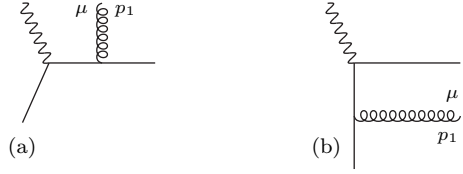

\begin{figure}

\begin{picture}(100,100)(0,0)

\Text(20,18)[c]{\footnotesize{(a)}}
\Text(135,18)[c]{\footnotesize{(b)}}
\Text(250,18)[c]{\footnotesize{(c)}}
\Text(365,18)[c]{\footnotesize{(d)}}

\Text(75,70)[l]{\scriptsize{$p_1$}}
\Text(60,70)[l]{\scriptsize{$\mu$}}
\Text(180,70)[l]{\scriptsize{$p_1$}}
\Text(165,70)[l]{\scriptsize{$\mu$}}
\Text(285,70)[l]{\scriptsize{$p_1$}}
\Text(270,70)[l]{\scriptsize{$\mu$}}
\Text(412,80)[l]{\scriptsize{$p_1$}}
\Text(395,80)[l]{\scriptsize{$\mu$}}

\Line(20,27.5)(30,50)
\Photon(20,72.5)(30,50){2}{6}
\Line(30,50)(90,50)
\GlueArc(50,50)(10,0,180){2}{8}
\Gluon(70,72.5)(70,50){-2}{6}

\Line(135,27.5)(145,50)
\Photon(135,72.5)(145,50){2}{6}
\Line(145,50)(205,50)
\GlueArc(175,50)(10,180,360){2}{8}
\Gluon(175,72.5)(175,50){-2}{6}

\Line(250,27.5)(260,50)
\Photon(250,72.5)(260,50){2}{6}
\Line(260,50)(320,50)
\GOval(300,50)(2,4)(0){0}
\Gluon(280,72.5)(280,50){-2}{6}

\Line(365,27.5)(375,50)
\Photon(365,72.5)(375,50){2}{6}
\Line(375,50)(435,50)
\GlueArc(405,50)(10,0,180){2}{8}
\Gluon(405,82.5)(405,62){-2}{6}
\GCirc(405,62){1}{1}

\end{picture}

\begin{picture}(100,100)(0,-25)

\Text(20,-18)[c]{\footnotesize{(e)}}
\Text(135,-18)[c]{\footnotesize{(f)}}
\Text(250,-18)[c]{\footnotesize{(g)}}
\Text(365,-18)[c]{\footnotesize{(h)}}

\Text(70,2)[r]{\scriptsize{$p_1$}}
\Text(70,18)[r]{\scriptsize{$\mu$}}
\Text(185,12)[r]{\scriptsize{$p_1$}}
\Text(185,28)[r]{\scriptsize{$\mu$}}
\Text(300,22)[r]{\scriptsize{$p_1$}}
\Text(300,38)[r]{\scriptsize{$\mu$}}
\Text(425,12)[r]{\scriptsize{$p_1$}}
\Text(425,28)[r]{\scriptsize{$\mu$}}

\Line(30,-10)(30,50)
\Photon(20,72.5)(30,50){2}{6}
\Line(30,50)(70,50)
\GlueArc(30,30)(10,270,450){2}{8}
\Gluon(30,10)(70,10){-2}{10}

\Line(145,-10)(145,50)
\Photon(135,72.5)(145,50){2}{6}
\Line(145,50)(185,50)
\GlueArc(145,20)(10,90,270){2}{8}
\Gluon(145,20)(185,20){-2}{10}

\Line(260,-10)(260,50)
\Photon(250,72.5)(260,50){2}{6}
\Line(260,50)(300,50)
\GOval(260,10)(2,4)(90){0}
\Gluon(260,30)(300,30){-2}{10}

\Line(375,-10)(375,50)
\Photon(365,72.5)(375,50){2}{6}
\Line(375,50)(415,50)
\GlueArc(375,20)(10,270,450){2}{8}
\Gluon(387,20)(425,20){-2}{8}
\GCirc(387,20){1}{1}

\end{picture}

\begin{picture}(100,100)(0,-15)

\Text(20,-8)[c]{\footnotesize{(i)}}
\Text(135,-8)[c]{\footnotesize{(l)}}
\Text(250,-8)[c]{\footnotesize{(m)}}
\Text(365,-8)[c]{\footnotesize{(n)}}

\Text(70,70)[l]{\scriptsize{$p_1$}}
\Text(55,70)[l]{\scriptsize{$\mu$}}
\Text(160,70)[l]{\scriptsize{$p_1$}}
\Text(145,70)[l]{\scriptsize{$\mu$}}
\Text(310,32)[r]{\scriptsize{$p_1$}}
\Text(310,45.5)[r]{\scriptsize{$\mu$}}
\Text(425,12)[r]{\scriptsize{$p_1$}}
\Text(425,28)[r]{\scriptsize{$\mu$}}

\Line(30,0)(30,50)
\Photon(20,72.5)(30,50){2}{6}
\Line(30,50)(80,50)
\GlueArc(30,50)(20,270,360){2}{8}
\Gluon(65,72.5)(65,50){-2}{6}

\Line(145,0)(145,50)
\Photon(135,72.5)(145,50){2}{6}
\Line(145,50)(195,50)
\GlueArc(145,50)(20,270,360){2}{8}
\Gluon(155,72.5)(155,50){-2}{6}

\Line(260,0)(260,50)
\Photon(250,72.5)(260,50){2}{6}
\Line(260,50)(310,50)
\GlueArc(260,50)(20,270,360){2}{8}
\Gluon(260,40)(270,40){2}{3}
\Gluon(283,40)(310,40){2}{6}

\Line(375,0)(375,50)
\Photon(365,72.5)(375,50){2}{6}
\Line(375,50)(425,50)
\GlueArc(375,50)(20,270,360){2}{8}
\Gluon(375,20)(425,20){-2}{14}

\end{picture}

\begin{picture}(100,100)(0,-15)

\Text(20,-8)[c]{\footnotesize{(o)}}
\Text(135,-8)[c]{\footnotesize{(p)}}
\Text(250,-8)[c]{\footnotesize{(q)}}

\Text(65,70)[l]{\scriptsize{$p_1$}}
\Text(50,70)[l]{\scriptsize{$\mu$}}
\Text(195,17)[r]{\scriptsize{$p_1$}}
\Text(195,31)[r]{\scriptsize{$\mu$}}
\Text(310,28)[r]{\scriptsize{$p_1$}}
\Text(310,44)[r]{\scriptsize{$\mu$}}

\Line(30,0)(30,50)
\Photon(20,72.5)(30,50){2}{6}
\Line(30,50)(80,50)
\GOval(30,25)(2,4)(90){0}
\Gluon(60,72.5)(60,50){2}{6}

\Line(145,0)(145,50)
\Photon(135,72.5)(145,50){2}{6}
\Line(145,50)(195,50)
\GOval(170,50)(2,4)(0){0}
\Gluon(145,25)(195,25){2}{10}

\Line(260,0)(260,50)
\Photon(250,72.5)(260,50){2}{6}
\Line(260,50)(310,50)
\GlueArc(260,50)(20,270,360){2}{8}
\Gluon(277,35)(310,35){2}{8}
\GCirc(276.5,35.5){1}{1}

\end{picture}

\vspace{1.5cm}

\caption{The one loop two particle phase space appearing in the two loop 
QCD absorptive
PT construction.  Diagrams (a),$\dots$,(d) define the $s$-channel amplitude 
${\cal T}^{[2]}_{{\frak a}\, s}$,   while all the others define the $t$-channel
amplitude  ${\cal T}^{[2]}_{{\frak a}\, t}$.} 

\label{fig5}

\end{figure}

\newpage

\begin{figure}

\begin{picture}(100,100)(-60,0)

\Text(20,18)[c]{\footnotesize{(a)}}
\Text(135,18)[c]{\footnotesize{(b)}}
\Text(250,18)[c]{\footnotesize{(c)}}

\Text(50,65)[r]{\scriptsize{$p_1$}}
\Text(70,77)[c]{\scriptsize{$\nu$}}
\Text(75,65)[l]{\scriptsize{$p_2$}}
\Text(55,77)[c]{\scriptsize{$\mu$}}
\Text(180,65)[r]{\scriptsize{$p_1$}}
\Text(185,77)[c]{\scriptsize{$\mu$}}
\Text(205,65)[l]{\scriptsize{$p_2$}}
\Text(205,77)[c]{\scriptsize{$\nu$}}
\Text(303,81)[l]{\scriptsize{$p_1$}}
\Text(326,81)[c]{\scriptsize{$\mu$}}
\Text(303,55)[l]{\scriptsize{$p_2$}}
\Text(326,55)[c]{\scriptsize{$\nu$}}

\Line(20,27.5)(30,50)
\Photon(20,72.5)(30,50){2}{6}
\Line(30,50)(90,50)
\Gluon(70,72.5)(70,50){-2}{6}
\Gluon(55,72.5)(55,50){-2}{6}

\Line(135,27.5)(145,50)
\Photon(135,72.5)(145,50){2}{6}
\Line(145,50)(205,50)
\Gluon(185,72.5)(185,62){-2}{2}
\Gluon(185,55)(185,50){-2}{1}
\Gluon(205,72.5)(170,50){-2}{8}

\Line(250,27.5)(260,50)
\Photon(250,72.5)(260,50){2}{6}
\Line(260,50)(320,50)
\GlueArc(300,50)(18,90,180){2}{6}
\Gluon(300,68)(320,55){-2}{5}
\Gluon(300,68)(320,81){-2}{5}
\GCirc(300,68){1}{1}

\end{picture}

\begin{picture}(100,100)(-60,-25)

\Text(20,-8)[c]{\footnotesize{(d)}}
\Text(135,-8)[c]{\footnotesize{(e)}}
\Text(250,-8)[c]{\footnotesize{(f)}}

\Text(60,43)[r]{\scriptsize{$p_1$}}
\Text(75,35)[c]{\scriptsize{$\mu$}}
\Text(60,12)[r]{\scriptsize{$p_2$}}
\Text(75,20)[c]{\scriptsize{$\nu$}}
\Text(180,40)[r]{\scriptsize{$p_1$}}
\Text(190,35)[c]{\scriptsize{$\mu$}}
\Text(180,15)[r]{\scriptsize{$p_2$}}
\Text(190,20)[c]{\scriptsize{$\nu$}}
\Text(283,40.5)[l]{\scriptsize{$p_1$}}
\Text(306,40.5)[c]{\scriptsize{$\mu$}}
\Text(283,14.5)[l]{\scriptsize{$p_2$}}
\Text(306,14.5)[c]{\scriptsize{$\nu$}}

\Line(30,0)(30,50)
\Photon(20,72.5)(30,50){2}{6}
\Line(30,50)(70,50)
\Gluon(30,20)(70,20){2}{9}
\Gluon(30,35)(70,35){2}{9}

\Line(145,0)(145,50)
\Photon(135,72.5)(145,50){2}{6}
\Line(145,50)(185,50)
\Gluon(145,35)(185,20){2}{9}
\Gluon(145,20)(158,24.88){2}{3}
\Gluon(185,35)(172,30.13){-2}{3}

\Line(260,0)(260,50)
\Photon(250,72.5)(260,50){2}{6}
\Line(260,50)(300,50)
\Gluon(260,27.5)(280,27.5){2}{4}
\Gluon(280,27.5)(300,40.5){-2}{5}
\Gluon(280,27.5)(300,14.5){-2}{5}
\GCirc(280,27.5){1}{1}

\end{picture}

\begin{picture}(100,100)(-60,-25)

\Text(20,7)[c]{\footnotesize{(g)}}
\Text(135,7)[c]{\footnotesize{(h)}}

\Text(60,70)[r]{\scriptsize{$p_1$}}
\Text(74,76)[c]{\scriptsize{$\mu$}}
\Text(60,43)[r]{\scriptsize{$p_2$}}
\Text(75,35)[c]{\scriptsize{$\nu$}}
\Text(180,43)[r]{\scriptsize{$p_1$}}
\Text(190,35)[c]{\scriptsize{$\mu$}}
\Text(172,26)[r]{\scriptsize{$p_2$}}
\Text(177,23)[c]{\scriptsize{$\nu$}}

\Line(30,15)(30,50)
\Photon(20,72.5)(30,50){2}{6}
\Line(30,50)(70,50)
\Gluon(30,35)(70,35){2}{9}
\Gluon(70,72.5)(50,50){-2}{8}

\Line(145,15)(145,50)
\Photon(135,72.5)(145,50){2}{6}
\Line(145,50)(185,50)
\Gluon(145,35)(165,35){2}{4}
\Gluon(175,35)(185,35){2}{2}
\Gluon(175,27.5)(155,50){-2}{8}

\end{picture}

\caption{The tree level three  particle phase space appearing in the two loop
QCD  absorptive PT construction.  Diagrams (a),$\dots$,(c) define the
$s$-channel amplitude  ${\cal T}^{[3]}_{{\frak b}\, s}$,  while all the others
define the $t$-channel one ${\cal T}^{[3]}_{{\frak b}\, t}$.} 

\label{fig6}

\end{figure}

\begin{figure}

\begin{center}

\begin{picture}(100,100)(0,0)

\Text(-80,50)[r]{$\Big( $}
\Text(-10,50)[l]{$\Big)^{-1}\ = \ \Big( $}
\Line(-75,50)(-15,50)
\GCirc(-45,50){4}{0}

\Line(37,50)(97,50)
\Text(102,50)[l]{$\Big)^{-1}\ -\ $}

\Line(140,50)(210,50)
\PhotonArc(175,50)(18,0,180){2}{7.5}
\GCirc(175,50){4}{0}
\GCirc(175,68){4}{0}
\GCirc(193,50){4}{0}

\end{picture}

\end{center}

\caption{The SD equation for the electron propagator $S$.}

\label{fig7}

\end{figure}

\end{document}